\documentclass{aa}
\usepackage{graphicx}
\usepackage[varg]{txfonts}
\usepackage{amsmath}
\usepackage{amssymb}
\usepackage{bm}
\usepackage{multicol}
\usepackage{multirow}
\usepackage{tabularx}
\usepackage{subcaption}
\usepackage{lscape}
\usepackage{placeins}
\usepackage{float}
\usepackage{algorithm}
\usepackage{algpseudocode}
\usepackage{hyperref}
\usepackage{orcidlink}
\usepackage{dblfloatfix}
\usepackage{etoolbox}
\usepackage{xcolor}
\usepackage[normalem]{ulem}
\usepackage{lineno}
\definecolor{revisionorange}{RGB}{230,110,0}

\newcommand{\TransitNet}{TransitNet\xspace}
\newcommand{\SNR}{\ensuremath{\mathrm{S/N}}}
\newcommand{\snr}{S/N}
\newcommand{\orcid}[1]{\,\orcidlink{#1}}
\title{\TransitNet: A Compact Attention-Augmented Deep Learning Framework for Low-\snr{} Transit Blind Searches}
\titlerunning{\TransitNet}
\authorrunning{Yan et al.}
\author{
  Xingchen~Yan\inst{1,2}\corrauth{yanxingchen0@gmail.com}\orcid{0009-0009-3252-6043}
  \and Jian~Ge\inst{1}\corrauth{jge@shao.ac.cn}\orcid{0000-0002-9144-5712}
  \and Qingtian~Liu\inst{1,2}\email{qtxplorer@shao.ac.cn}\orcid{0009-0009-3642-8993}
  \and Kevin~Willis\inst{3}\email{kevin.w.willis@gmail.com}
  \and Quanquan~Hu\inst{1,2}\email{huqq1917@outlook.com}\orcid{0009-0009-9761-8678}
  \and Jiapeng~Zhu\inst{1}\email{jpzhu@smail.nju.edu.cn}
}
\institute{Shanghai Astronomical Observatory, Shanghai 200030, China
  \and University of Chinese Academy of Sciences, Yanqi Lake Campus, East Road 1, Huairou, Beijing 101408, China
  \and Science Talent Training Center, Gainesville, FL 32606, USA
}
\date{Received 16 June 2026}
\abstract
{The observational incompleteness of intermediate-to-long-period Earth-size planets motivates the development of more sensitive transit-search methods for low signal-to-noise ratio (\snr{}) regimes.}
{We present ``\TransitNet'', a compact attention-augmented deep-learning framework for low-\snr{} transit blind searches, and evaluate its performance in recovering Earth-size and sub-Earth-size planetary transits.}
{To enable realistic method development and objective threshold calibration under blind-search conditions, we developed a unified framework for dataset construction, benchmarking, and threshold selection. We evaluated \TransitNet on recovery benchmarks constructed from unseen Kepler targets and compared its performance with TLS and BLS. Additional validation was performed on selected real Kepler targets.}
{On recovery benchmarks, \TransitNet achieved an accuracy of 95.2\% in the $\SNR{}=6$--8 regime, with ROC-AUC and PR-AP values of 0.974 and 0.982, respectively, exceeding the performance of both TLS and BLS. In injected Earth-size and sub-Earth-size transit recovery experiments, it recovered 93.0\% of signals, compared with 63.1\% for TLS and 60.0\% for BLS. The model additionally provides attention-based estimates of transit windows and midpoints. Applied to real Kepler observations, it recovered all 34 selected confirmed planets with a 1.24~h mean absolute error of the transit-midpoint estimates.}
{\TransitNet provides an accurate, computationally efficient, and scalable framework for low-\snr{} transit blind searches and supports future searches for longer-period Earth-size and sub-Earth-size planets.}
\keywords{
planets and satellites: detection --
(stars:) planetary systems --
techniques: photometric --
methods: data analysis --
methods: statistical
}

\begin{document}
% Keep the column boxes at \textheight (\flushbottom, as in the journal).
% article.cls sets \parskip to 0pt plus 1pt; size10.clo gives \floatsep,
% \intextsep and \dblfloatsep a plus component; aa.cls gives \section a
% before-skip of 16dd plus 4pt. \flushbottom stretches whichever of these
% is present, so a short column opens one gap between paragraphs, after a
% float, or above a heading. Those skips are rigid (shrink only). The only
% remaining stretch is a fraction of a point on \baselineskip, so a small
% shortfall is shared by every line instead of collecting in one place.
% \clubpenalty/\widowpenalty are lowered from aa.cls's 10000 so a column
% can still take one line from the next paragraph.
\makeatletter
\g@addto@macro\normalsize{%
  \addtolength{\baselineskip}{0pt plus 0.15pt}%
  \abovedisplayskip 8.55\p@ \@minus5\p@
  \abovedisplayshortskip \z@
  \belowdisplayshortskip 6\p@ \@minus3\p@
  \belowdisplayskip \abovedisplayskip
}
\g@addto@macro\small{%
  \addtolength{\baselineskip}{0pt plus 0.12pt}%
  \abovedisplayskip 8.5\p@ \@minus4\p@
  \abovedisplayshortskip \z@
  \belowdisplayshortskip 4\p@ \@minus2\p@
  \belowdisplayskip \abovedisplayskip
}
\normalsize
\setlength{\parskip}{0pt}
\setlength{\floatsep}{12pt minus 2pt}
\setlength{\textfloatsep}{18pt minus 4pt}
\setlength{\intextsep}{12pt minus 2pt}
\setlength{\dblfloatsep}{12pt minus 2pt}
\setlength{\dbltextfloatsep}{18pt minus 4pt}
\patchcmd{\section}%
  {-16dd\@plus-4\p@ \@minus-4\p@}{-16dd}%
  {}{\typeout{WARNING: could not freeze section skip}}%
\renewcommand\@listi{%
  \leftmargin\leftmargini
  \parsep\z@
  \topsep 8\p@ \@minus4\p@
  \itemsep\parsep
}
\let\listI\@listi
\clubpenalty=4000
\@clubpenalty\clubpenalty
\widowpenalty=4000
\flushbottom
% \normalsize restores the article list spacing (itemsep 4pt + parsep 4pt,
% both stretchable). aa.cls sets \bibsep to 0 but never applies it, so
% \flushbottom opens a blank line between references. Keep entries on the
% bibliography leading.
\let\TN@thebibliography\thebibliography
\renewcommand\thebibliography[1]{%
  \TN@thebibliography{#1}%
  \setlength{\itemsep}{0pt}%
  \setlength{\parsep}{0pt}%
  \setlength{\parskip}{0pt}%
  \setlength{\@topsep}{0pt}%
  \setlength{\@topsepadd}{0pt}%
}
\makeatother
% \linenumbers
\maketitle
\nolinenumbers

\section{Introduction}
\label{sec:intro}
The transit method has emerged as the most productive technique for exoplanet detection.
Leveraging both ground-based surveys such as WASP \citep{Pollacco2006a} and KELT \citep{pepperKELTSouthTelescope12012}, 
and space-based missions including CoRoT \citep{auvergneCoRoTSatelliteFlight2009}, Kepler/K2 \citep{kochKEPLERMISSIONDESIGN2010, Howell2014}, and
TESS \citep{rickerTransitingExoplanetSurvey2014},
these methods have yielded over 6000 confirmed exoplanets to date\footnote{\label{fn:exoplanet_archive}\url{https://exoplanetarchive.ipac.caltech.edu/docs/counts_detail.html}}.
Upcoming missions such as PLATO \citep{catalaPLATOPLAnetaryTransits2009, rauerPLATOMission2014, Rauer2025} 
and ET (the Earth 2.0 space mission; \citealt{Ge2022a, Ge2022b, Ge2022c, Ge2024a, Ge2024b}) will further expand our capability to detect and characterize Earth-size exoplanets.
Among these missions, the Kepler space telescope has been particularly transformative: 
its 4-year primary mission discovered over 2700 confirmed exoplanets and identified an additional \textasciitilde1900 candidates awaiting confirmation$^{\ref{fn:exoplanet_archive}}$.

\begin{figure}[!t]
  \centering
  \includegraphics[width=\hsize]{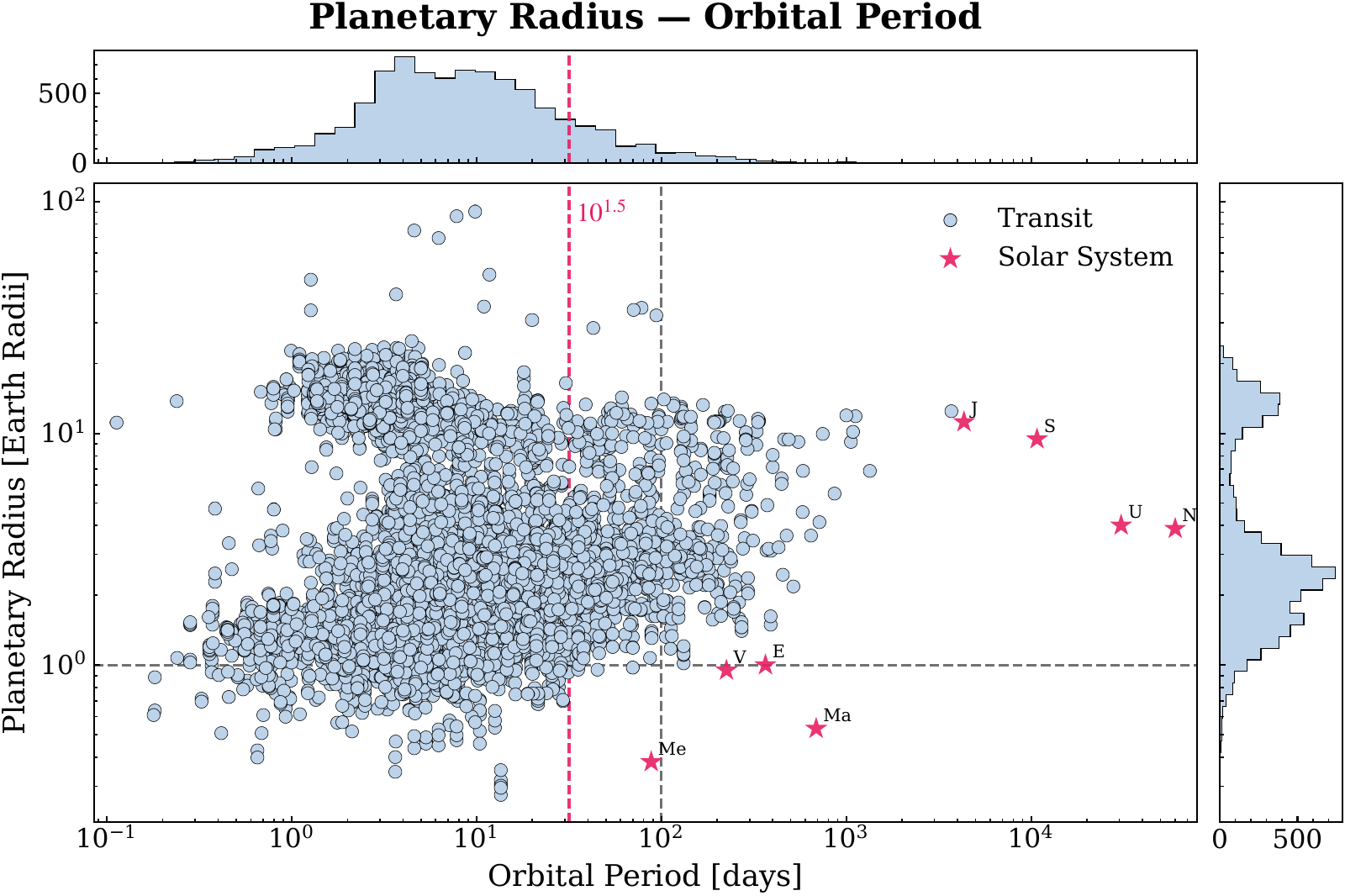}
  \caption{Confirmed exoplanets on the $R_{\rm p}$--$P$ plane (log scale), with marginal histograms of $P$ (top) and $R_{\rm p}$ (right). 
  Solar System planets are shown as labeled magenta stars; gray dashed lines mark $R_{\rm p}=1\,R_\oplus$, $P=100$~days, and red vertical dashed line marks $P=10^{1.5}$~days ($\approx31.6$~days). 
  }
  \label{fig:dist_of_CP}
\end{figure}

Despite the rapid growth of the exoplanet population, Earth-size planets on relatively long-period orbits remain underrepresented. 
In practice, the observed exoplanet population is dominated by larger ($R > 2R_\oplus$) planets on short-period orbits, with the small-radius, intermediate-to-long-period (I2LP) regime ($R \leq R_\oplus$, $P > 10^{1.5}\approx 31.6$ days; Fig.~\ref{fig:dist_of_CP}) remaining largely devoid of confirmed planets.

Nevertheless, previous occurrence-rate studies suggest that Earth-size planets are common around Sun-like stars. 
For example, \citet{Fressin2013} found that $16.5 \pm 3.6$ percent of FGK stars host at least one Earth-size planet ($0.8$--$1.25R_\oplus$) with $P \leq 85$ days, while \citet{Petigura2013} found that $26 \pm 3$ percent of Sun-like stars host a $1$--$2R_\oplus$ planet with $P=5$--$100$ days, with the occurrence remaining approximately constant with increasing period up to $\sim200$ days.
Using the final Kepler DR25 catalog, \citet{Bryson2021} estimated a conservative habitable-zone occurrence of $0.37$--$0.60$ rocky planets per Solar-like star.

Moreover, Kepler was designed to achieve $\lesssim20$~ppm photometric precision for detecting the 84~ppm transit signal of an Earth--Sun analog (\citealt{kochKEPLERMISSIONDESIGN2010}), and a subset of targets achieved a 6-hr Combined Differential Photometric Precision (CDPP) $\lesssim30$~ppm in practice \citep{Christiansen2012}. 
Therefore, photometric precision alone is unlikely to fully explain the scarcity of Earth-size planets at I2LP, motivating consideration of low-SNR transit-search sensitivity.

Transit-search pipelines first search light curves for transit-like signals, which are then subjected to detailed candidate vetting \citep{Jenkins2012}. 
The Kepler pipeline, for example, uses a wavelet-based, adaptive matched filter to search for periodic transit signatures \citep{jenkinsTransitingPlanetSearch2010,jenkinsOVERVIEWKEPLERSCIENCE2010}.
In addition, periodogram-based methods are widely used in the community, 
among which the Box Least Squares (BLS; \citealt{kovacsBoxfittingAlgorithmSearch2002}) algorithm models transits as box-shaped signals, 
while the Transit Least Squares (TLS; \citealt{hippkeOptimizedTransitDetection2019}) method improves sensitivity by incorporating more realistic transit shapes with limb darkening. 
Additional approaches include the Transit Comb Filter (TCF; \citealt{gondhalekarTCFPeriodogramHigh2023}) combined with ARIMA noise modeling \citep{Melton2024a}, 
and the Quasi-Periodic Automated Transit Search (QATS; \citealt{Carter2013}) algorithm for detecting transits with timing variations.

Machine learning is increasingly used for transit detection and validation.
Examples include dimensionality reduction and similarity-based classification \citep{thompsonMACHINELEARNINGTECHNIQUE2015}, 
Bayesian model selection \citep{Mullally2016}, self-organizing maps for transit shape classification \citep{armstrongTransitShapesSelforganizing2017}, 
and time-series outlier rejection techniques such as TSARDI \citep{Mislis2018a}.
Gaussian process classifiers have been applied to probabilistic planet validation \citep{Armstrong2021a}, 
while random forests have been used in automated vetting pipelines such as Autovetter \citep{mccauliffAUTOMATICCLASSIFICATIONKEPLER2015}. 
Ensemble and boosting methods, including XGBoost and GBDT, have further improved performance in identifying rare transit signals 
\citep{Mislis2016,malikExoplanetDetectionUsing2021,pratyushAutomationTransitingExoplanet,panahiDetectionTransitingExoplanets2022,Melton2024a}.

A landmark study by \citet{shallueIdentifyingExoplanetsDeep2018} introduced AstroNet, 
a CNN model that classifies Kepler Threshold Crossing Events (TCEs) and demonstrated high accuracy in distinguishing planetary signals from false positives.
Subsequent studies have confirmed the strong performance of CNNs compared to traditional machine learning methods \citep{Pearson2018}, 
and extended AstroNet-like architectures to K2 and TESS data \citep{dattiloIdentifyingExoplanetsDeep2019,Yu2019,Tey2023}. 
Improvements such as ExoNet incorporate additional domain knowledge, including stellar parameters and centroid information, 
yielding significant gains in recall for low-\snr{} signals \citep{Ansdell2018}.
More recent architectures, including ExoMiner and related models, 
have achieved high precision and recall in automated vetting and enabled the validation of new exoplanets \citep{valizadeganExoMinerHighlyAccurate2022}. 
These models have since been extended to TESS data and full-frame image pipelines, 
further improving performance and scalability \citep{Valizadegan2023a,Valizadegan2025,Martinho2026ExoMinerPP20}.

A number of studies have leveraged simulated or synthetic data to construct more diverse training sets.
\citet{Zucker2018} tested CNNs for detecting transit signals based on simulated data.
\citet{osbornRapidClassificationTESS2020} created a raw TESS artificial dataset using the Lilith simulator \citep{Li2019} to train ExoNet 
and validate it on real TESS data.
\citet{Alvarez2023} also tested their CNNs for detecting transit signals based on simulated data.
These studies further highlight the importance of synthetic data in addressing data scarcity 
and improving model generalization across realistic observational domains \citep{Zucker2018,yehSearchingPossibleExoplanet2021,  Alvarez2023}.
Most recently, the GPU phase folding and CNN (GPFC) system of \citet{Wang2024a} 
combined semi-synthetic Kepler training with GPU-accelerated blind search, discovering ultra-short-period planet candidates at the shortest periods and smallest radii reported to date for such objects, with transit \snr{} as low as $\sim 6$-7 \citep{Wang2024}.

CNN-based transit searches have explored a range of input representations and model architectures. Some encode the period--epoch search space as two-dimensional maps of phase-folded light-curve segments \citep{chintarungruangchaiDetectingExoplanetTransits2019a,cuellarDeepLearningExoplanets2022}, while others combine CNNs with recurrent, attention, or Transformer-based architectures for transit detection and candidate classification \citep{Paetzold2025TRANSCENDENCE,Thomas2025KeplerHybrid,salinasDistinguishingPlanetaryTransit2023,Salinas2025,Choudhary2025ViTGAFRP}. 
Related developments in candidate vetting include attention-augmented residual networks and generative models coupled to conventional classifiers \citep{G2023,Xie2025RAAKeplerTESS,Fiscale2025CFM}. 
Beyond conventional periodic searches, deep-learning methods have increasingly been explored to identify challenging signals, particularly shallow, low-S/N, and single-transit events \citep{Dvash2022,Vivien2025,cuiIdentifyLightcurveSignals2022,aydoganExoplanetDetectionMachine2022}.

In this work, motivated by the scarcity of I2LP and Earth-size planets, the limitations of existing search methods, and the need for scalable data-driven approaches, we introduce ``\TransitNet'', a compact attention-augmented deep-learning framework for low-\snr{} transit blind searches using phase-folded light curves.
\TransitNet retains the \snr{} gain of phase folding while learning low-\snr{} transit morphology directly from data, targeting improved sensitivity to Earth- and sub-Earth-size planets at intermediate periods.
Coupled with GPU-based phase folding, it efficiently scans trial periods and epochs while remaining lightweight for large-scale surveys.
It also provides an attention-informed estimate of the transit midpoint, offering auxiliary prior information for subsequent candidate vetting and related downstream analyses.
We demonstrate improved detection sensitivity over BLS and TLS in the low-\snr{}, $P\in[30,60]$ days regime, providing a basis for extending blind searches to longer-period terrestrial planets with fewer observable transits.

The remainder of this paper is organized as follows.
Section~\ref{sec:method} describes the dataset construction pipeline and methods, including Periodic Chunk Permutation (PCP) for non-transit controls, and the \TransitNet{} architecture.
Section~\ref{sec:exp} presents ablation studies quantifying the contributions of Multi-Head Attention (MHA) and Fully Convolutional Network (FCN), and benchmarks \TransitNet{} against BLS and TLS on unseen Kepler targets, including recovery sensitivity, cross-target generalization, Earth-size transit recovery, inference efficiency, and transit-window estimation.
Section~\ref{sec:discussion} summarizes our findings and outlines future work.

\section{Materials and methods}

\label{sec:method}
Following previous CNN-based transit searches that use simulated or semi-synthetic light curves for model training \citep{Zucker2018,osbornRapidClassificationTESS2020,Alvarez2023,Wang2024a, Wang2024}, we do not construct the training set from confirmed transits. In particular, the number of reliably labeled low-\snr{} events is insufficient to cover the $(P,T_{14},\delta)$ domain targeted in this work, where $P$ is the orbital period, $T_{14}$ is the transit duration, and $\delta$ is the transit depth. We therefore construct a semi-synthetic Kepler dataset of phase-folded light curves (Sect.~\ref{sec:data}) and use it to train \TransitNet (Sect.~\ref{sec:archi}).

This section first describes the construction of the training data, including the generation of semi-synthetic light curves and PCP for non-transit controls, and then discusses the design rationale underlying \TransitNet.

\subsection{Benchmark dataset generation for transit detection}
\label{sec:data}
Through extensive experimentation, we have established a standardized and principled pipeline for constructing datasets for deep learning-based exoplanet detection models:

\textit{Observational settings and transit parameter space:}
Determine the sampling cadence of the telescope data, compute an appropriate number of phase bins for folding and binning, 
and specify the values and ranges of key transit parameters to be explored in the exoplanet search, including the orbital period $P$, transit depth $\delta$, 
and transit duration $T_{14}$.

\textit{Synthetic light-curve generation:}
Simulated transit signals are constructed using parameter combinations sampled from the predefined ranges specified in the previous step. 
In the fully synthetic case, these signals are inserted into Gaussian Noise Light Curves (GNLC) or other Non-Transiting Light Curves (NTLC) to generate Artificial Transiting Light Curves (ATLC). 
In the semi-synthetic case, they are inserted into real Transit-Masked Light Curves (TMLC). 
A third option is to use real light curves that already contain confirmed exoplanet signals.

\textit{Transit and non-transit samples:}
Transit signal samples are derived from ATLCs obtained in the previous step after folding and binning. 
Non-transit signal samples are generated by folding and binning NTLC, ensuring that no transit features are present.

\textit{Dataset specifications:}
Ensure that the dataset contains a sufficient number of samples, 
that the injected transit signals cover the target parameter space as comprehensively as possible.

\subsubsection{Distributions of parameters}
\label{sec:dist_params}
To simulate transit signals that resemble the observed exoplanet population as closely as possible,
we first define the target parameter space for sample construction by focusing on orbital periods $P\in[30,60]$~days and $\SNR{}\in[6,15]$, corresponding to the onset of the I2LP and low-\snr{} regime considered in this work, with 30~days adopted as the lower-period boundary for computational convenience (Fig.~\ref{fig:dist_of_CP}).

Using the KOI tables available from the NASA Exoplanet Archive \citep{Thompson_2018, Christiansen2025}, 
we select confirmed Kepler planets within this period range as the reference population, from which the admissible ranges $P \in [30, 60]$ days, $T_{14} \in [1.99, 7.99]$ hours, and $\delta \in [79, 1262]$ ppm are derived.

\subsubsection{Generation of artificial transiting and non-transiting light curves}
\label{sec:alc}
We incorporate TMLCs extracted from real Kepler observations to generate ATLCs rather than adopt an idealized noise model, 
thereby preserving the authentic noise characteristics of space-based photometry (semi-synthetic approach).

\textit{Sources of TMLCs:}
We randomly select $\sim$700 Kepler Input Catalog (KIC; \citealt{Brown_2011}) targets corresponding to confirmed planetary systems identified in the Kepler Object of Interest (KOI; \citealt{Thompson_2018}) catalog, excluding known false positives, to serve as the source pool for constructing TMLCs. 
The corresponding KIC identifiers are then used to retrieve the Pre-search Data Conditioning Simple Aperture Photometry (PDCSAP; \texttt{PDCSAP\_FLUX}) light curves at a long cadence of 29.4\,min from the Mikulski Archive for Space Telescopes (MAST)\footnote{\url{https://archive.stsci.edu/kepler/}}.
We then concatenate all available quarterly segments and apply the detrending and normalization procedures described above \citep{Wang2024a, Wang2024, liu2026delosdetectingshallowtransits}. Confirmed transit signals are masked only when constructing TMLCs; the independent benchmark and confirmed-planet light curves (Table~\ref{tab:benchmarks}) undergo the same preprocessing without signal masking. This treatment suppresses long-timescale trends that could distort the injected signals while retaining realistic short-timescale and correlated Kepler residual noise.

\textit{ATLC generation:} 
We adopt a basic geometric trapezoidal transit model with distinct ingress and egress phases, neglecting limb-darkening effects \citep{Wang2024a, Wang2024}. This approximation is adequate for our low-\snr{} regime, where photometric noise dominates and the transit duration is typically much shorter than the orbital period, $T_{14}\ll P$, while enabling efficient large-scale signal generation.
For each TMLC, we inject a single transit signal into the light curve to produce the corresponding ATLC, 
with parameters ($P$, $T_{14}$, and $\delta$) uniformly sampled from predefined ranges.
The transit epoch $T_0$, defined as the reference transit epoch, 
is drawn from $\mathcal{U}(0, P)$ and determines the phase offset of the injected transit signal.
The randomly drawn parameters do not guarantee that all generated ATLCs fall within the target \snr{} range.  
 We thus calculated the \snr{} of each candidate ATLC following the formula in \citet{kovacsBoxfittingAlgorithmSearch2002}:
\begin{equation}
  \label{eq:snr}
  \SNR{} = \frac{\delta}{\sigma} \sqrt{n\frac{T_{14}}{P}},
\end{equation}
where $\sigma$ denotes the standard deviation of the out-of-transit flux measurements, $n$ is the total number of observations in the light curve. 
The quantity $n T_{14} / P$ corresponds to the effective number of in-transit data points.
This strategy ensures broad coverage of the joint target parameter space considered in this study ($P\in[30,60]$~days and $\SNR{}\in[6,15]$) and fills gaps in the distribution of confirmed exoplanet parameters that cannot be achieved using real observational data alone.
The injected transit signals are combined with TMLCs to form ATLCs, preserving the intrinsic photometric trends and noise properties, 
and yielding data that closely resemble real observations while enabling controlled training sample generation.
Since TMLCs may contain undetected weak transit signals from additional planetary companions, 
this hybrid strategy better mimics realistic survey conditions and exposes the model to genuine observational systematics and residual noise.

\textit{Preparation of transit and non-transit signal samples:} 
Each ATLC is subsequently phase-folded at the injected orbital period and resampled into $N=4096$ uniform phase bins.
The choice of $N$ is motivated by the requirement that even the shortest transit events be adequately resolved. 
Consider the limiting case of the longest orbital period ($P_{\rm max} = 60$~days) and the shortest transit duration ($T_{14,min} \approx 2$~hours). 
The temporal width of each phase bin is
\begin{equation}
  \Delta t = \frac{P_{\rm max}}{N} = \frac{60 \times 24 \times 60}{4096} \approx 21.1~\mathrm{min}.
\end{equation}
Under this configuration, a 2-hour transit spans approximately 6 phase bins, 
providing sufficient sampling to characterize the transit morphology. 
This bin count corresponds to a power of two ($2^{12}$), 
which is favorable for binary-based computation and memory alignment, 
and may offer implementation advantages in neural network architectures.
Note that, after phase folding and binning, transit signals may appear at any position within the phase window, 
including cases where they extend across window boundaries, consistent with realistic exoplanet search conditions.
Considering that non-transit signal samples should ideally contain no transit features, 
the non-transit signal samples are constructed from two complementary sources to ensure both label purity and realistic noise characteristics.

\textit{Gaussian-noise controls:}
GNLCs are generated and phase-folded over specified period ranges to provide an idealized non-transit case. 
The noise amplitude (standard deviation) of the GNLCs is sampled from the empirical distribution of the selected TMLCs.

\textit{Periodic Chunk Permutation:}
To preserve realistic instrumental and astrophysical noise characteristics, 
we implement a PCP strategy applied to each TMLC. 
The procedure is inspired by the quarter-shuffling framework of \citep{Telesco2026}. 
Each light curve is partitioned and reorganized using randomly selected periods uniformly drawn from $[30, 60]$~days, effectively disrupting residual periodic structures that may arise from masked or undetected transit signals (see Appendix \ref{app:pcp} for details).

The final dataset for model training comprises $\sim$71000 unique transit signal samples derived from ATLCs and $\sim$85000 non-transit signal samples from both GNLCs and PCP TMLCs, 
with all samples binned to a length of $N$.
Given the large scale of this dataset and the need to phase fold each light curve at multiple trial periods, 
this process utilizes GPU-accelerated phase folding \citep{Wang2024a, Wang2024, HuGPUTLS} 
for all phase folding operations, which significantly reduces the computational time required for processing large numbers of light curves.

\subsection{\TransitNet Architecture and Design Rationale}
\label{sec:archi}
\TransitNet comprises three modules with complementary roles: the Filtering Module (FM) suppresses noise and extracts local temporal features from the phase-folded light curve; 
the MHA module captures long-range dependencies and enhances sensitivity to low-\snr{} transit morphologies, 
in which the attention weight matrix is further leveraged to inform subsequent transit midpoint estimation, while the FCN module fuses and refines the representation while preserving spatial structure to produce a scalar detection score.

The input to the model is the phase-folded, binned sequence constructed in Section~\ref{sec:data}:
$X = [f_1, f_2, \ldots, f_{L_0}]$, where $L_0 = N= 4096$ is the number of phase bins adopted in Section~\ref{sec:alc}.

\subsubsection{Filtering module}
\label{sec:cnn}
The FM acts as a learnable, data-driven front-end that suppresses noise and extracts initial local temporal features from the phase-folded light curve.

Low-S/N transit signals often exhibit shallow depths and may have either short or relatively long durations. These weak signals can be easily obscured by photometric noise, stellar variability, and instrumental systematics, making them challenging to detect using conventional template-matching and threshold-based methods.
Some correlated noise and instrumental systematics exhibit concentrated spectral features, in contrast to the relatively flat spectrum of white noise, potentially enabling more effective frequency-selective suppression.

One-dimensional convolution performs local weighted sums using kernels learned through back-propagation, where a kernel is a small set of learnable weights that slides across the input sequence to extract local features. From a signal-processing perspective, these learned kernels act as data-driven filters. The convolution theorem relates convolution in the time domain to pointwise multiplication in the frequency domain,
\begin{equation}
\mathcal{F}\{f(t) * g(t)\} = F(\omega) \cdot G(\omega).
\end{equation}
where $\mathcal{F}$ denotes the Fourier transform, $*$ represents
convolution, and $F(\omega)$ and $G(\omega)$ are the Fourier transforms
of $f(t)$ and $g(t)$, respectively.
 Thus, gradient-based training allows the FM to learn adaptive filters that selectively suppress noise-dominated components while preserving transit-related features, providing a signal-processing motivation for using CNNs in the FM.

The design of the FM is inspired by the convolutional stack used for global-view feature extraction in AstroNet \citep{shallueIdentifyingExoplanetsDeep2018}, 
where multiple one-dimensional convolutional layers perform hierarchical feature extraction on the time series. 
The FM comprises five one-dimensional convolutional layers with channel dimensions increasing in powers of two. Each layer consists of one-dimensional convolution, ReLU activation, max pooling, and dropout. 
The channel dimension is the number of kernels applied in parallel, each producing its own filtered sequence, or feature map; ReLU is the pointwise nonlinearity $\max(0,x)$, without which a stack of convolutions would reduce to a single linear filter; max pooling retains the largest activation within each local window, reducing the sequence length while preserving strong local responses. Transit-related activations can therefore propagate through successive pooling stages; dropout randomly sets a fraction of the features to zero at each training step, so that the network cannot rely on a few individual features. With our default hyperparameters, the input $X$ is transformed by the FM into a feature tensor $F \in \mathbb{R}^{L_1 \times 256}$, where $L_1=128$ is determined by the kernel sizes, strides, and pooling factors of the convolutional layers. The 256-dimensional feature representation is retained in light of the demonstrated effectiveness of the global-view architecture in AstroNet.

The FM extracts hierarchical local features from the phase-folded light curve, but local responses alone may be insufficient to distinguish low-\snr{} transit signals from stochastic or systematic variations. We therefore use a relatively shallow convolutional stack for local feature extraction and preliminary filtering, while MHA captures global dependencies between phase bins and identifies regions that jointly support a transit-like pattern.

\subsubsection{Multi-head attention module}
\label{sec:mha}
\begin{figure*}[!tp]
  \centering
  \includegraphics[width=\textwidth]{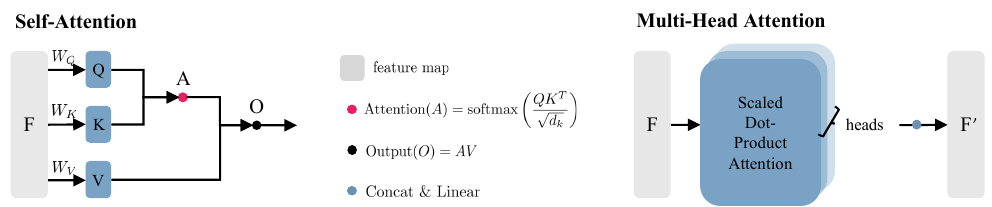}
  \caption{
    Schematic of the MHA computation. 
    The input feature matrix $F$, encoding flux patterns from the folded transit profile, is linearly projected into queries $Q$ (search templates for transit-like features), 
    keys $K$ (reference catalog of flux patterns), and values $V$ (actual flux information) using learned weights $W_Q, W_K,$ and $W_V$. 
    Attention weights $A$ are computed via scaled dot-product attention, $\mathrm{softmax}\!\left(QK^{\top}/\sqrt{d_K}\right)$, 
    assigning adaptive weights to different phase bins based on their contribution to transit detection, and applied to the values to produce per-head outputs $O = AV$. 
    Outputs from all heads, each potentially specializing in different aspects of transit morphology, 
    are concatenated and passed through a final linear projection to yield the enhanced feature matrix $F'$.}
  % Alt text: Flowchart of multi-head attention showing linear projections to queries, keys, and values, scaled dot-product attention, per-head outputs, concatenation, and final projection to refined features.
  \label{fig:mha}
\end{figure*}
The MHA module provides a global receptive field over the folded transit profile and enhances sensitivity to low-\snr{} transit features with consistent morphology.

MHA was introduced in the Transformer architecture by \citet{vaswaniAttentionAllYou2017} to capture long-range dependencies through direct interactions between sequence positions. It comprises $h$ parallel self-attention heads, whose outputs are concatenated and linearly projected.

Let $F \in \mathbb{R}^{L_1 \times d_f}$ denote the feature matrix from the FM, where $L_1$ is the downsampled sequence length and $d_f$ is the feature dimension. Self-attention operates on this representation, with $F$ serving as the source for the query, key, and value matrices. These matrices are obtained through learned linear projections:
\begin{equation}
Q = F W^Q,\quad K = F W^K,\quad V = F W^V\,
\end{equation}
where $W^Q, W^K, W^V \in \mathbb{R}^{d_f \times d_{\mathrm{model}}}$ are learnable weight matrices, yielding
$Q, K, V \in \mathbb{R}^{L_1 \times d_{\mathrm{model}}}$;
$d_{\mathrm{model}}$ is the attention embedding dimension.
$W^Q$ maps $F$ to a query space encoding what to look for;
$W^K$ to a key space encoding the patterns available for matching;
and $W^V$ to a value space encoding the flux information to be aggregated.
For each attention head, the projected $Q$, $K$, and $V$ representations are split along the feature dimension into $h$ subspaces, each of dimension $d_Q=d_K=d_V=d_{\mathrm{model}}/h$.
Denoting the $i$-th head by subscript $i$, we have
$Q_i, K_i, V_i \in \mathbb{R}^{L_1 \times d_K}$ with $d_Q=d_K=d_V$.
The scaled dot-product attention for head $i$, 
\begin{equation}
\mathrm{Attention}(Q_i, K_i, V_i)
= \mathrm{softmax}\!\left( \frac{Q_i K_i^{\top}}{\sqrt{d_K}} \right) V_i,
\label{eq:attn}
\end{equation}
is then modeling pairwise interaction between phase bins over the folded transit profile. The similarity $Q_i K_i^{\top}$ encourages morphologically consistent transit features across phases:
bins that are symmetrically distributed around the transit center may receive higher attention weights when they exhibit similar flux dips, while other bins receive lower weights, enabling low-\snr{} transit signals to be emphasized and unrelated fluctuations to be suppressed.
The scale factor $1/\sqrt{d_K}$ keeps the dot products from growing with $d_K$ and stabilizes the softmax.
The resulting attention matrix
$A_i = \mathrm{softmax}(Q_i K_i^{\top}/\sqrt{d_K}) \in \mathbb{R}^{L_1 \times L_1}$
encodes pairwise relationships between phase bins:
each entry $A_{i,j}$ represents the attention weight assigned to phase bin $j$ (the source) when constructing the output representation for phase bin $i$ (the target),
where both $i$ and $j$ index positions along the phase-folded sequence ($i, j \in \{1, 2, \ldots, L_1\}$).
The output $A_i V_i$ is the corresponding weighted combination of value vectors, where each row of $A_i V_i$ aggregates information from all phase bins according to the attention weights in $A_i$.

MHA computes self-attention in parallel across $h$ heads. Their outputs are concatenated and linearly projected to obtain the final representation:
\begin{equation}
\begin{aligned}
  \mathrm{MultiHead}(F) &= \mathrm{Concat}(\mathrm{head}_1, \ldots, \mathrm{head}_h)\, W^O, \\
  \mathrm{head}_i &= \mathrm{Attention}(Q_i, K_i, V_i).
\end{aligned}
\end{equation}
where $W^O \in \mathbb{R}^{h d_V \times d_{\mathrm{model}}}$. The overall computation flow of MHA is illustrated in Fig.~\ref{fig:mha}.

In our implementation, $F \in \mathbb{R}^{L_1 \times 256}$ is first projected to $Q,K,V \in \mathbb{R}^{L_1 \times 16}$, then processed by MHA with $h=8$ heads and per-head dimensions $d_K=d_V=d_{\mathrm{model}}/h=2$. 
This choice balances parallel learning of diverse transit-related patterns, such as flux dips and ingress/egress structure, against the limited dimensionality of each attention head. 

\subsubsection{Fully convolutional network module}
\label{sec:fcn}
The FCN module maps the MHA-refined representation to a scalar detection score while preserving the spatial and channel structure of the transit profile. 

In many CNN-based models, the final mapping to scalar outputs is implemented with fully connected (dense) layers. 
In our experiments, this design did not make full use of the multi-channel features produced by MHA: flattening the feature matrix and passing it through dense layers discards the spatial structure of the folded transit profile and the coupling between feature channels. 
To preserve this structure, we use a FCN module. 
The FCN \citep{Longa} employs pointwise ($1\times1$) convolutions, analogous to the channel-mixing stage of depthwise-separable architectures \citep{Howard2017, Chollet2017, Tan2020}. These layers fuse information across channels at each phase bin while preserving the sequence dimension.

This design refines and fuses the MHA features while preserving transit morphology, yielding a compact scalar representation for subsequent classification. As shown in Fig.~\ref{fig:model}, the FCN consists of several lightweight convolutional blocks with channel widths $(128, 64, 32, 16, 8, 1)$, progressively reducing the feature map to a single element through convolution, followed by a sigmoid activation. The resulting scalar score indicates the presence or absence of a transit signal in the input sequence.

\subsubsection{Overall architecture and relation to classical methods}
\begin{figure*}[!tp]
  \centering
  \includegraphics[width=\textwidth]{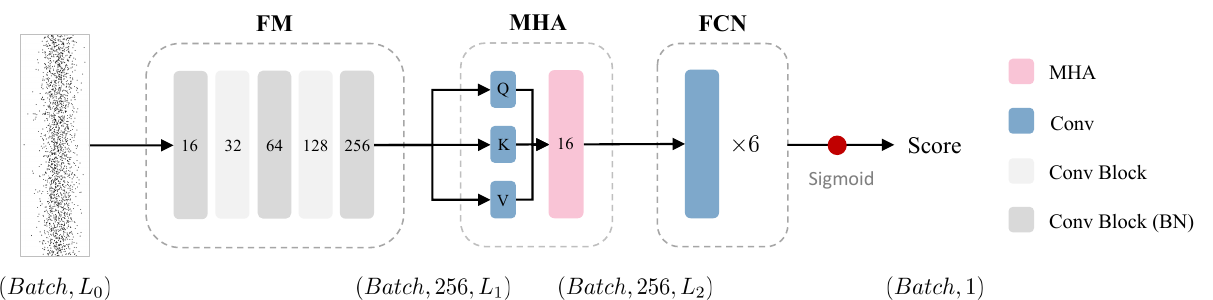}
  \caption{Architecture of \TransitNet after multiple optimizations. Our network processes a batch of 1D sequences $(\mathrm{Batch}, L_0)$ through three stages: FM, MHA, and FCN. 
  In the FM, five convolution blocks (CB) expand the channel width $(16, 32, 64, 128, 256)$ and produce features of shape $(\text{Batch}, 256, L_1)$. 
  Two variants of CBs are used: one with Batch Normalization (BN) and one without, both of which consist of Conv1D $\rightarrow$ ReLU $\rightarrow$ MaxPool $\rightarrow$ Dropout. BN rescales the intermediate features to zero mean and unit variance within each training batch, which stabilizes the optimization.
  Numbers on the CBs denote channel dimensions; the label $16$ on the MHA block denotes $d_{\mathrm{model}}$.
  The MHA projects the FM features to $Q$, $K$, and $V$ with attention embedding dimension $d_{\mathrm{model}}=16$, applies $h=8$-head attention with per-head dimension $d_K=d_V=2$, preserving the sequence length $L_1$. 
  The FCN applies a lightweight Conv repeated $\times 6$; the channel widths follow $(128, 64, 32, 16, 8, 1)$. 
  A final pointwise projection with a sigmoid yields a scalar score per sequence, giving the output shape $(\mathrm{Batch}, 1)$.}
  \label{fig:model}
\end{figure*}

The complete architecture, consisting of the FM, MHA, and FCN modules, is referred to as \TransitNet (Fig.~\ref{fig:model}).

In terms of both detection principle and computational mechanism, \TransitNet differs from classical transit search methods such as BLS and TLS. These methods explicitly evaluate transit models while optimizing over parameters such as phase and duration at each trial period, whereas \TransitNet maps each fixed-length, phase-folded profile directly to a detection score through a learned forward pass. This data-driven approach avoids imposing a specific analytic transit template, which can be restrictive when low-\snr{} transit signals are affected by stochastic or temporally correlated noise and exhibit deviations from idealized transit morphology.

The convolutional layers in \TransitNet learn hierarchical local features that capture transit morphology and small-scale variations, while MHA integrates information across the full phase domain in a content-dependent manner, allowing features from different phase regions to jointly contribute to the detection score. Once trained, the model replaces the per-period parameter optimization with a fixed sequence of convolution and attention operations. For a fixed folded-sequence length, the computational cost of each forward pass is therefore independent of any additional template-parameter search and can be efficiently batched across trial periods on GPUs. 

\begin{figure*}[!tp]
  \centering
  \begin{subfigure}{0.495\textwidth}
    \centering
    \includegraphics[width=\linewidth]{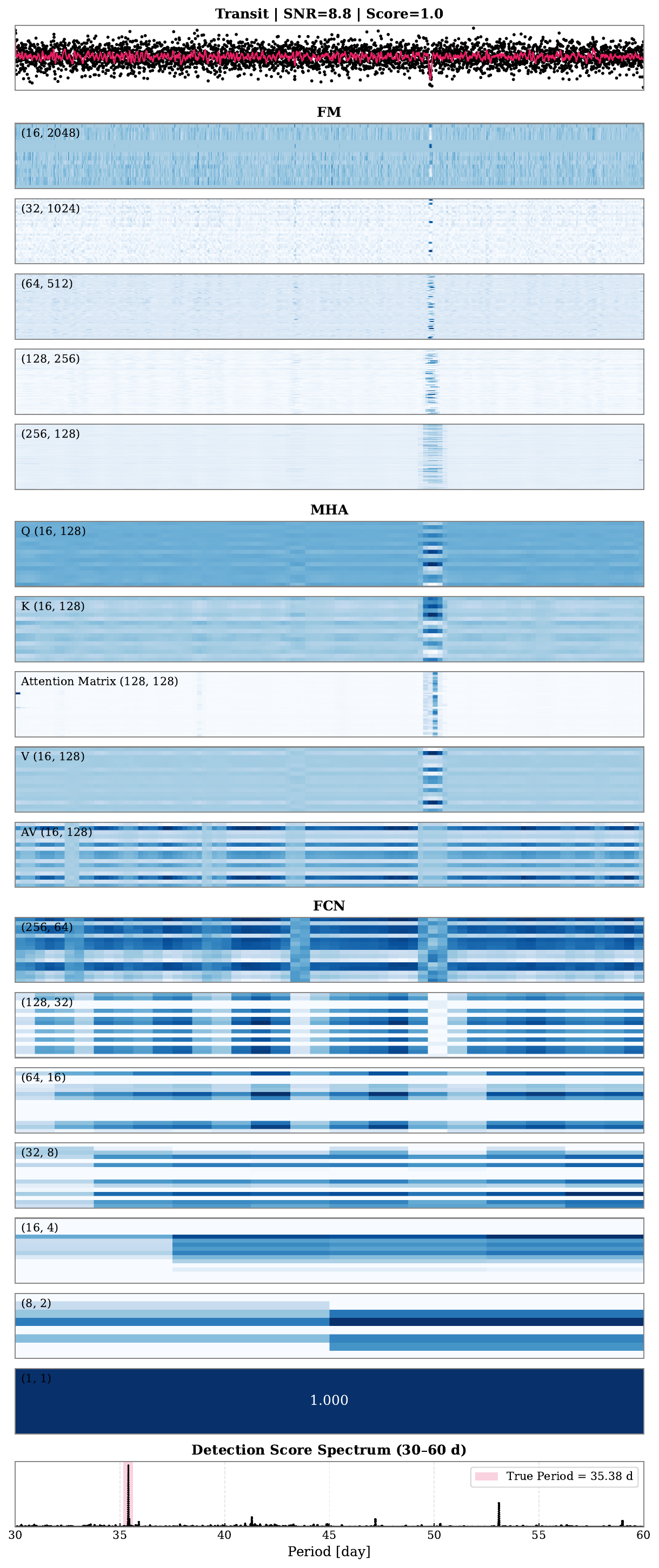}
  \end{subfigure}
  % \hfill
  \begin{subfigure}{0.495\textwidth}
    \centering
    \includegraphics[width=\linewidth]{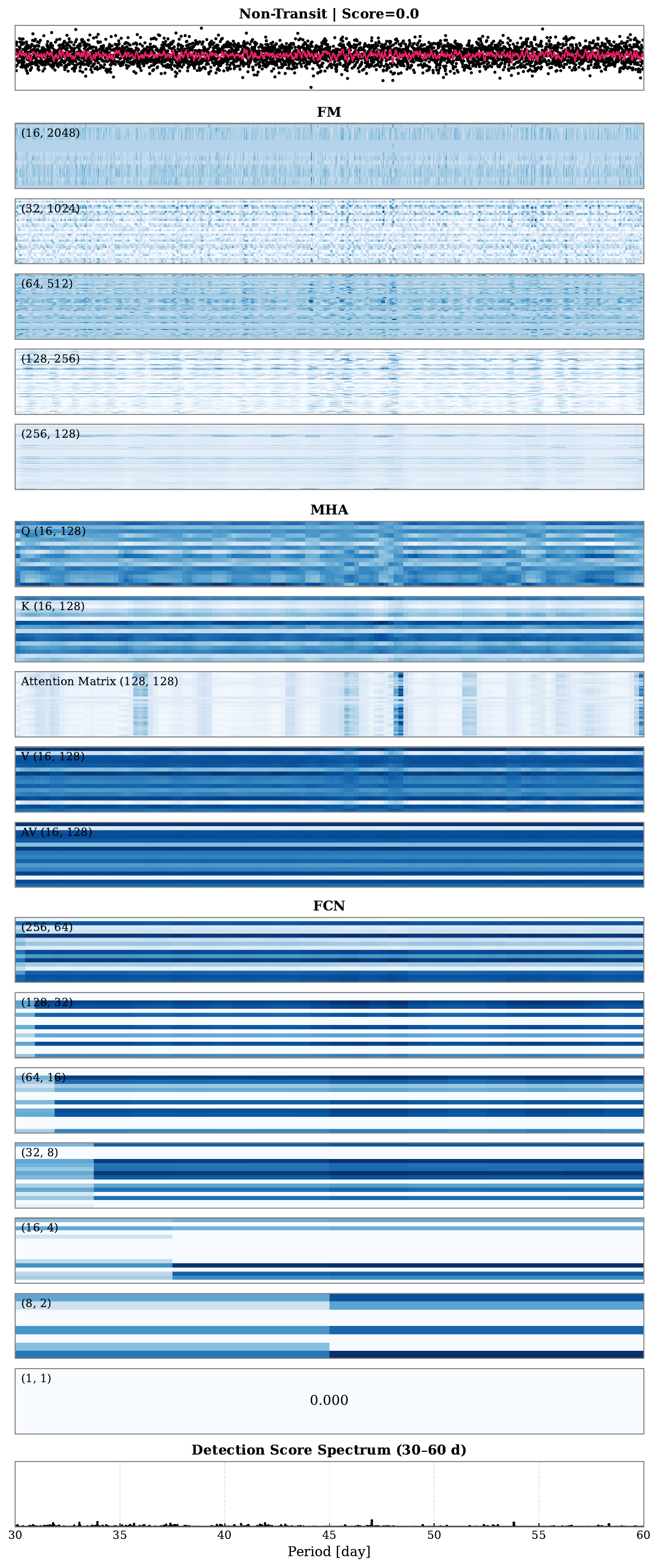}
  \end{subfigure}
  \caption{
  End-to-end feature-map visualization of \TransitNet for a transit (left) and a non-transit (right) semi-synthetic ATLC pair from Section~\ref{sec:alc}.
  From top to bottom: input light curves; FM (rows 2--6); MHA --- $Q$, $K$, attention matrix $A = \mathrm{softmax}(Q K^{\top}/\sqrt{d_K})$, $V$, and $AV$ (rows 7--11); FCN (rows 12--18); detection score spectrum (bottom).}
  % Alt text: Side-by-side stacked panels for a transit input on the left and a non-transit input on the right, tracing feature maps through filtering, attention matrices, convolutional layers, and final detection scores.
  \label{fig:model_output}
\end{figure*}

Fig.~\ref{fig:model_output} illustrates the complete pipeline from a folded light curve to the final detection score for transit and non-transit examples.
For the transit example, the convolutional feature maps show a clear progression with depth: stochastic fluctuations are progressively suppressed, while the transit signature becomes sharper and more spatially coherent in phase. This behavior is consistent with the intended role of the FM in extracting and enhancing transit morphology from noisy folded light curves. 

In contrast, the non-transit example remains diffuse across layers, without a stable localized structure indicative of a periodic dimming event.
The MHA representations further accentuate this distinction. For transit signals, the learned query-key-value interactions produce attention maps that assign higher weight to a small set of phase bins jointly consistent with a transit pattern, while down-weighting noise-dominated regions. For non-transit inputs, the attention remains diffuse and fragmented, with no coherent phase pattern.

Taken together, these feature “fingerprints” demonstrate that \TransitNet learns to couple local convolutional filtering with global, content-dependent reweighting, enabling the final FCN module to separate transit from non-transit sequences based on high-level, physically interpretable representations rather than on low-level noise fluctuations.
The FCN rows in the figure show this refinement, and the detection score spectrum at the bottom yields a sharp peak at the true period for the transit case and a flat response for the non-transit.

\section{Experiments and results}
\label{sec:exp}
This section presents comprehensive evaluations of TransitNet (Secs.~\ref{sec:compar}, \ref{sec:speed_comparison}, and \ref{sec:t0_estimation}) and comparisons with TLS and BLS (Secs.~\ref{sec:sens}--\ref{sec:era}).

\subsection{Ablation studies}
\label{sec:compar}
\begin{table}[!t]
  \centering
  \caption{
    Description of the modules used in model construction.
    % The design of these modules aims to investigate the capacity of different architectural building blocks to learn data features.
  }
  \label{tab:modules}
  \begin{tabular}{ll}
  \hline
  \hline
  \textbf{Module} & \textbf{Description} \\
  \hline
  FM   & Filtering module. \\
  MHA  & Multi-Head Attention. \\
  FCN  & Fully Convolutional Network. \\
  FCNN & Fully Connected Neural Network. \\
  \hline
  \hline
  \end{tabular}
\end{table}
Ablation studies are first conducted to quantify the contributions of the MHA and FCN modules to low-\snr{} transit feature extraction, training stability, and classification performance (Section~\ref{sec:compar}), using the training set constructed in Section~\ref{sec:data}.
To better understand the role of each component in \TransitNet, we perform ablation studies in which selected modules are individually removed or replaced, and the resulting changes in model performance are evaluated.

Each model is constructed from four distinct building blocks: FM, MHA, FCN, and Fully-Connected Neural Network (FCNN). 
We compare four architectures, denoted (a)--(d) in Table~\ref{tab:tp_results}, including \TransitNet as the full model and three ablated variants:
(a) FM~+~FCN,
(b) FM~+~FCNN,
(c) FM~+~MHA~+~FCNN, and
(d) FM~+~MHA~+~FCN (\TransitNet).
Table~\ref{tab:modules} provides annotations and descriptions for the individual components.

The design rationale for the MHA and FCN modules has been discussed in Sections~\ref{sec:mha} and \ref{sec:fcn}. Here, it is restated as two falsifiable expectations, which the measurements in Section~\ref{sec:metrics} are designed to test:
(i) delegating global temporal context to MHA improves sensitivity to low-\snr{} transits; and
(ii) an FCN output head, matched to the structure of the MHA output, separates transit from non-transit signals more effectively than an FCNN while requiring substantially fewer parameters.

These expectations are evaluated by comparing the four architectures in terms of low-\snr{} transit feature extraction, training stability, and classification performance.

\subsubsection{Training configurations}
To address the class imbalance between transit and non-transit signals, we adopt weighted random sampling with replacement during training. Each sample is assigned a weight inversely proportional to the number of samples in its class: 
$w_i = 1/N_{c_i}$
where $c_i$ denotes the ground-truth class of sample $i$, and $N_{c_i}$ is the total number of samples belonging to class $c_i$. The sampling probability of each sample is proportional to $w_i$, so that minority-class samples are more likely to be selected during training.

To mitigate the variability inherent in stochastic deep learning training, we adopt five-fold cross-validation that promotes a fair and objective comparison, 
in which the training and test sets are mutually exclusive, with a training-to-test ratio of 4:1 in each fold.
Each model underwent multiple training sessions with carefully tuned hyperparameters to reach near-optimal settings. 

% To ensure objective five-fold cross-validation and mitigate training variability, comparison and visualization used the median-performing model from a representative fold.
To further enhance the generalization ability of the model and alleviate overfitting, 
dropout layers \citep{Srivastava2014} are added after each convolutional layer in all models. 
\citet{Ioffe2015} introduced internal covariate shift and proposed BN to mitigate its adverse effects, which motivates our use of BN to improve training stability.
Model optimization is performed using AdamW \citep{loshchilovDecoupledWeightDecay2019}, which decouples weight decay from the adaptive gradient updates of Adam \citep{Kingma2017}, 
improving regularization and generalization. A binary cross-entropy loss function is used. After empirical tuning, the final configuration adopts a learning rate of $8\times10^{-4}$ and a weight decay of $1\times10^{-6}$. 
Dropout is applied at rates of $0.08$ and $0.05$ in both convolutional and attention layers, respectively, to reduce overfitting. 
The model is trained with a batch size of $3072$, and early stopping (patience = 5) is employed to halt training when no further improvement in validation loss is observed.

\subsubsection{Experimental results}
\label{sec:metrics}
Table~\ref{tab:tp_results} presents the evaluation metrics on the test set for architectures (a)--(d), reported as mean~$\pm$~standard deviation over five-fold cross-validation.
Among these, \TransitNet achieves the best overall performance, attaining an average F1-score of 99.1\% on the test set after five-fold cross-validation. 
From the table, we observe two consistent effects.

\textit{Role of MHA:}
Comparing architectures (a) with (d) and (b) with (c), introducing MHA enables the model to capture global temporal dependencies among data points, 
enhancing sensitivity to low-\snr{} transit signatures and improving the modeling of both transit events and complex photometric noise.

\textit{Role of the FCN output:}
Comparing (a) with (b) and (c) with (d), 
using an FCN as the output mapping layer improves the modeling of spatial structure in the feature matrices 
while substantially reducing the parameter count by approximately 98\% (from 17522K to 350K) and 79\% (from 1807K to 380K), respectively.
This is because FCN preserves spatial topology and exploits weight sharing, 
whereas FCNN flattens the features and requires significantly more parameters for the mapping.

These paired comparisons are obtained from our own ablation experiments and confirm the two expectations 
stated at the beginning of Section~\ref{sec:compar}: 
MHA improves sensitivity to low-\snr{} transit signals, and the FCN output mapping outperforms an FCNN 
while substantially reducing model complexity.

\begin{table*}
  \centering
  \caption{
      Evaluation results on transit samples for four model architectures, 
      evaluated on the test sets and averaged over five-fold cross-validation:
      (a) FM + FCN, 
      (b) FM + FCNN, 
      (c) FM + MHA + FCNN, and 
      (d) FM + MHA + FCN (\TransitNet). 
      All metrics are reported as percentages (\%), and parameter counts are given in thousands (K, $\times 1000$). 
      Operating-point metrics are the true positive rate (TPR) at a false positive rate (FPR) of $1\%$ and the recall ($\mathcal{R}$) at a precision ($\mathcal{P}$) of $98\%$.
      Best-performing values are highlighted in bold, while second-best values are shown in italics. 
      Comparative analysis indicates that replacing FCNN with FCN improves performance with substantially fewer parameters (a vs.\ b; c vs. d), 
      incorporating MHA enhances representation learning (b vs.\ c; a vs. d).
  }
  \label{tab:tp_results} 
  \begin{tabular}{lccccccl}
  \hline
  \hline
  \textbf{Model} & \textbf{Accuracy} & \textbf{Precision} & \textbf{Recall} & \textbf{F1-Score} & \textbf{TPR @ $\mathrm{FPR}=1\%$} & \textbf{$\mathcal{R}$ @ $\mathcal{P}=98\%$} & \textbf{Num. of Params. (K)} \\
  \hline
  (a) & $98.0 \pm 0.4$ & $98.7 \pm 0.9$ & $97.4 \pm 0.2$ & $98.0 \pm 0.4$ & $96.9 \pm 1.3$ & $97.9 \pm 0.8$ & \textbf{350} \\
  (b) & $97.8 \pm 0.3$ & $99.2 \pm 0.2$ & $96.4 \pm 0.6$ & $97.8 \pm 0.3$ & $96.8 \pm 0.4$ & $97.5 \pm 0.4$ & 17,522 \\
  % (c) & $98.5 \pm 0.1$ & $99.3 \pm 0.2$ & $97.6 \pm 0.35$ & $98.5 \pm 0.1$ & $98.0 \pm 0.2$ & $98.5 \pm 0.2$ & 17,959 \\
  (c) & ${\textit{98.9}} \pm 0.2$ & ${\textbf{99.5}} \pm 0.3$ & ${\textit{98.4}} \pm 0.3$ & ${\textit{98.9}} \pm 0.2$ & ${\textit{98.8}} \pm 0.4$ & ${\textit{99.2}} \pm 0.3$ & 1,807 \\
  (d) & ${\textbf{99.1}} \pm 0.1$ & ${\textit{99.4}} \pm 0.2$ & ${\textbf{98.8}} \pm 0.1$ & ${\textbf{99.1}} \pm 0.1$ & ${\textbf{99.1}} \pm 0.1$ & ${\textbf{99.4}} \pm 0.1$ & ${\textit{376}}$ \\
  \hline
  \hline
  \end{tabular}
\end{table*}

\subsubsection{Stability and generalization}
\label{sec:results}
Stability and generalization are both relevant when evaluating a model: the former indicates whether 
comparable performance can be reproduced across training runs under fixed hyperparameters, 
while the latter relates to performance on held-out data and thus to potential applicability in real-world exoplanet searches.

Table~\ref{tab:tp_results} reports each metric as the mean~$\pm$~standard deviation over five-fold cross-validation. 
The standard deviations suggest that the joint use of FCN and MHA is associated with improved training stability. In particular, model (d) exhibits the smallest fold-to-fold variance across all evaluation metrics.
Generalization performance is evaluated using the standard test-set classification metrics as well as operating-point statistics derived from the ROC and PR curves, reported as the TPR at an FPR of $1\%$ and the recall at a precision of $98\%$, as listed in the table. 
Meanwhile, \TransitNet achieves the highest TPR of $99.1\%$ at an FPR of $1\%$ and the highest recall of $99.4\%$ at a precision of $98\%$ among all models. 

\begin{figure}[!t]
  \centering
  \includegraphics[width=\hsize]{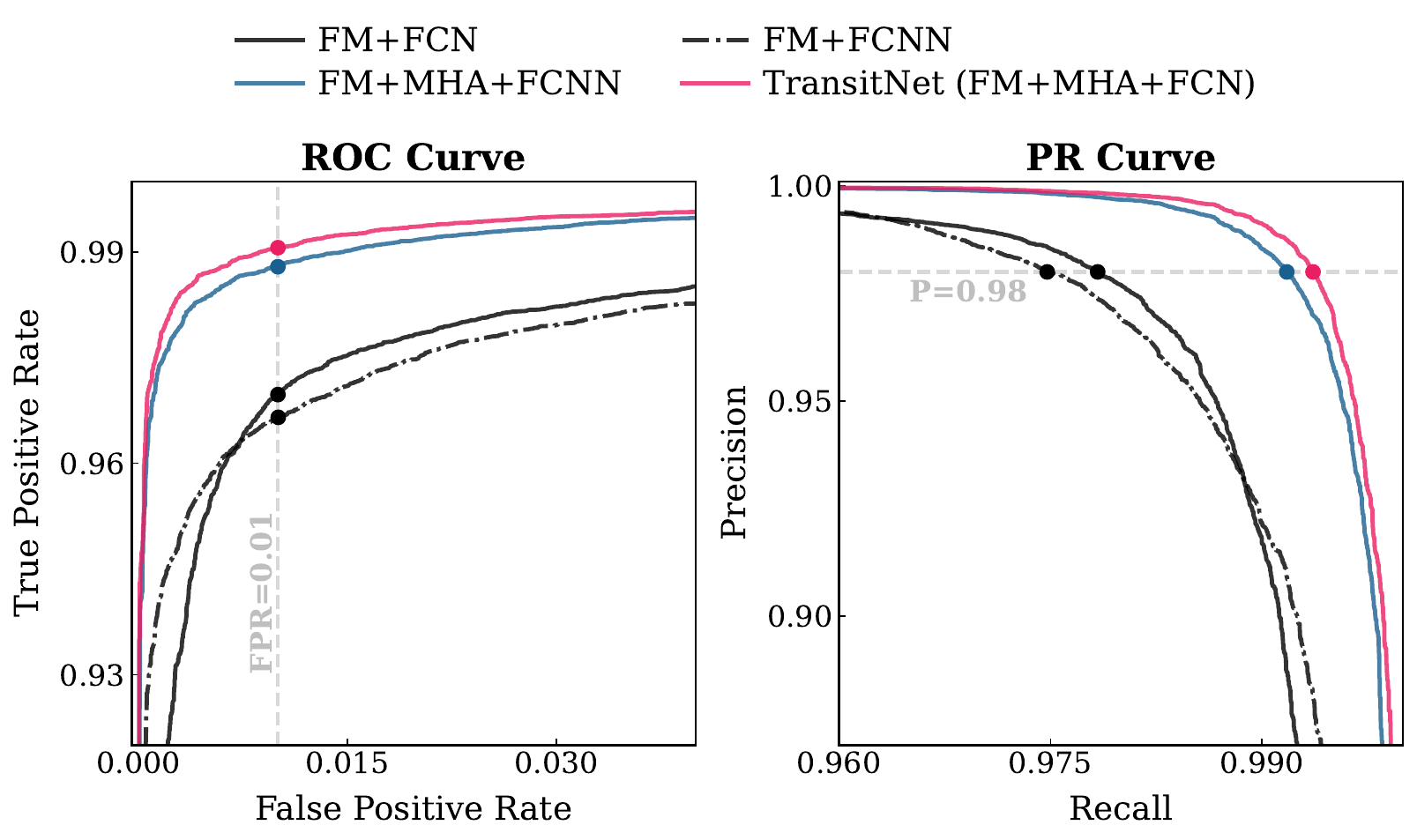}
  \caption{ROC and PR curves on the test set for four model architectures. Curves are averaged over five-fold cross-validation. 
  Filled circles indicate the operating points at a fixed FPR of $1\%$ on the ROC curves and a fixed precision of $98\%$ on the PR curves (Table~\ref{tab:tp_results}), 
  as highlighted by vertical dashed lines.}
  \label{fig:pr_roc}
\end{figure}

These operating-point results are consistent with the global trends shown in Fig.~\ref{fig:pr_roc}.
The combination of \TransitNet achieves the best performance, demonstrating the importance of tailoring the architecture to the characteristics of low-\snr{} transit signals. MHA establishes global interactions across the phase domain and enhances sensitivity to low-\snr{} transit signatures, while the FCN extracts local features and progressively refines their mapping to the classification boundary.

\subsection{Comparative evaluations of BLS, TLS and \TransitNet{} in low-\snr{} exoplanet searches}
\label{sec:apply}
In this section, we benchmark \TransitNet against BLS and TLS in the search for low-\snr{} transit signals under realistic transit blind-search conditions, using Kepler targets that were never seen during training.

All experiments share the same source targets, the same sample-construction pipeline, and the same search configuration (Sect.~\ref{sec:exp_config}). 
The three semi-synthetic benchmarks then address complementary questions: sensitivity as a function of \snr{} within a single target (Sect.~\ref{sec:sens}); robustness across heterogeneous targets, together with the calibration of a single global operating threshold (Sect.~\ref{sec:cross_kic}); and the recovery of Earth-size and sub-Earth-size planets (Sect.~\ref{sec:era}).
Sections~\ref{sec:speed_comparison} and \ref{sec:t0_estimation} finally report inference efficiency and transit-midpoint estimation.

\subsection{Experimental configuration}
\label{sec:exp_config}

\subsubsection{Benchmark datasets}
\label{sec:eval_sets}
All three semi-synthetic benchmarks are constructed from a common pool of 60 KIC targets randomly selected from Kepler stars that contributed no samples to either the training or validation sets, ensuring complete target-level separation from the $\sim$700 KIC targets used to construct the training TMLCs. Each target is processed using the same light-curve construction procedure described in Section~\ref{sec:alc}, yielding one TMLC per target.

These TMLCs then serve as the basis for benchmark sample generation. Injecting a single trapezoidal transit signal produces an ATLC, while applying PCP produces multiple PCP TMLCs. The injected signals span $P\in[30,60]$~days, $\SNR{}\in[6,15]$, and $T_{14}\in[1.991,7.987]$~h (Sect.~\ref{sec:dist_params}). The three benchmarks differ only in their target subsets and constraints on the injected parameters.

\textit{Low-\snr{} Transit Recovery Set} (Sect.~\ref{sec:sens}): one KIC, chosen from the pool for its minimal residual systematics after detrending. Repeated independent injections into its TMLC, each inserting one signal with an independently sampled \snr{}, give 1000 ATLCs, and 1000 PCP TMLCs of the same light curve provide the non-transit samples. With the host noise fixed, this set isolates the dependence of detectability on \snr{}.

\textit{Cross-KIC Recovery Set} (Sect.~\ref{sec:cross_kic}): all 60 TMLCs. Each target receives 20 independent injection runs on the same light curve, every run inserting a single signal with its own orbital and transit parameters and hence its own \snr{}, and 25 PCP TMLCs are constructed from the same light curve. A random subset of 1000 ATLCs and 1000 PCP TMLCs is used for scoring. These injection realizations are entirely disjoint from those of the previous set, and the 60 hosts span diverse stellar variability and noise environments.

\textit{Earth-size and Sub-Earth-size Recovery Set} (Sect.~\ref{sec:era}):
all 60 TMLCs serve as candidate backgrounds for transit injection.
For each background, the Earth-size reference depth is defined as
$\delta_\oplus=(R_\oplus/R_\star)^2$.
Candidate transits are generated by uniformly sampling $P$, $T_{14}$,
and \snr{} from their prescribed ranges, with the injection depth
$\delta$ derived from Eq.~\ref{eq:snr}; candidates with
$\delta>\delta_\oplus$ are rejected.
This procedure is repeated until 1000 ATLCs are obtained.

\textit{Confirmed-Planet Set}: it comprises the detrended light curves of 34 confirmed Kepler planets whose $P$ and \snr{} fall within the predefined target ranges, enabling end-to-end evaluation on genuine limb-darkened transits without synthetic injection (Appendix~\ref{app:real}).
Table~\ref{tab:benchmarks} summarizes the composition, purpose, and headline result of all four sets.

\begin{table*}[!t]
  \centering
  \caption{Overview of the four evaluation sets. The ATLC / PCP column gives the numbers of injected transit and PCP non-transit samples. The last column reports BLS, TLS, and \TransitNet, with the best value in bold; accuracy and recall are percentages. Operating points are BLS SDE = 6.76, TLS SDE = 9.80, and \TransitNet score = 0.54. Low-\snr{} accuracy is measured at \snr{} $=6$--8, while ROC-AUC is threshold-independent and averaged over KICs; the corresponding mean PR-AP values are 0.967, 0.975, and 0.982. The first three sets share 60 KICs unused in training; the Low-\snr{} set uses one target selected for low residual systematics. The Earth-size set reports recall only because it contains no non-transit samples. Cross-KIC and Earth-size injections are independent. The Confirmed-Planet Set uses real detrended light curves without injection; only \TransitNet was applied, with mean and median midpoint errors of 1.24 and 0.62~h.
  }
  \label{tab:benchmarks}
  \small
  \setlength{\tabcolsep}{3.5pt}
  \renewcommand{\arraystretch}{1.1}
  \begin{tabular*}{\textwidth}{@{\extracolsep{\fill}}lcccl@{}}
  \hline
  \hline
  \textbf{Evaluation set} & \textbf{Targets} & \textbf{ATLC / PCP} & \textbf{Question addressed} & \textbf{Main result} \\
  \hline
  \textit{Low-\snr{} Transit Recovery Set}
    & 1 KIC
    & $1000$ / $1000$
    & S/N sensitivity at fixed host noise
    & Accuracy: $73.9$ / $86.9$ / $\mathbf{95.2}$ \\
  \textit{Cross-KIC Recovery Set}
    & 60 KICs
    & $1000$ / $1000$
    & Robustness across hosts
    & ROC-AUC: $0.947$ / $0.958$ / $\mathbf{0.974}$ \\
  \textit{Earth-size and Sub-Earth-size Recovery Set}
    & 60 KICs
    & $1000$ / \ldots
    & Recovery of Earth-size planets
    & Recall: $60.0$ / $63.1$ / $\mathbf{93.0}$ \\
  \textit{Confirmed-Planet Set}
    & 34 planets
    & $34$ / \ldots
    & End-to-end test on genuine transits
    & Recovery: \ldots{} / \ldots{} / $\mathbf{100}$ \\
  \hline
  \hline
  \end{tabular*}
\end{table*}

\subsubsection{Search configuration and score extraction}
\label{sec:exp_setting}
To emulate realistic transit blind-search scenarios as closely as possible while ensuring a fair comparison, a common search configuration is adopted for all methods and for the first three benchmarks of Section~\ref{sec:eval_sets}.

BLS, TLS, and \TransitNet operate on the same detrended, full-length light curves and produce detection-score spectra evaluated over $P\in[30,60]$~days on an identical period grid.
The common period grid was generated using \texttt{period\_grid} method in  \href{https://transitleastsquares.readthedocs.io/en/latest/}{\textsc{transitleastsquares}} package \citep{hippkeOptimizedTransitDetection2019}, assuming $M_\star=1.0\,M_\odot$, $R_\star=1.0\,R_\odot$, a minimum of two observed transits, and an oversampling factor of 3.0, evaluated at the Kepler baseline $T_{\mathrm{span}}\approx4$~yr. 
This yielded approximately $1.15\times10^4$ trial periods. 
The BLS trial-duration grid was chosen to bracket the injected-duration range of Section~\ref{sec:eval_sets}, whereas TLS internally constructs its own duration grid from the stellar and orbital parameters.

BLS was implemented using the \texttt{power} method of the \texttt{BoxLeastSquares} class in \href{https://www.astropy.org/}{\textsc{Astropy}} package \citep{astropy:2013,astropy:2018,astropy:2022}.
In the call to \texttt{power()}, we set \texttt{objective}=\texttt{"snr"} and \texttt{method}=\texttt{"fast"}.
When per-point uncertainties were unavailable, we approximated the photometric noise level using the standard deviation of the flux time series.
TLS was run using GTLS \citep{HuGPUTLS}, a GPU-accelerated implementation of TLS, with identical period limits and grid settings. 
Additional GTLS parameters included \texttt{T0\_fit\_margin}=0.125 and \texttt{fast}=\texttt{True};
all other GPU-TLS settings were kept at their default values. The TLS detection statistic is the raw Signal Detection Efficiency (SDE) returned by the code. 
\TransitNet used the same period grid and produced a detection score spectrum over trial periods.
For BLS, the standardized detection strength was evaluated by converting the raw BLS power to an SDE:
\begin{equation}
{\rm SDE}_{\rm BLS}(p_i)=
\frac{p_i-\widetilde{p}_{\rm bg}}{\sigma_{\rm bg}},
\label{eq:bls_sde}
\end{equation}
where $p_i$ is the BLS power at the $i$-th trial period, $\widetilde{p}_{\rm bg}$ is the median of the background BLS power distribution after applying $4\sigma$ clipping for three iterations, and $\sigma_{\rm bg}=1.4826\,{\rm median}(|p-\widetilde{p}_{\rm bg}|)$ is the robust scatter estimated from the clipped background.

To ensure a consistent and fair comparison across methods, we adopt different score-selection strategies for transit and non-transit samples. 
For ATLCs containing injected transit signals, the detection score is evaluated at the true injected period for all methods, and a signal is considered recovered if this score exceeds the method-specific operating threshold (BLS: $\mathrm{SDE}=6.76$, TLS: $\mathrm{SDE}=9.80$, and \TransitNet: $\mathrm{Score}=0.54$; Section~\ref{sec:thres_select}). 

This protocol scans a common blind-search period grid and evaluates the score at the injected period, thereby measuring transit detectability independently of period-selection accuracy. For PCP TMLCs, the score is the maximum SDE for BLS and TLS, and the most significant peak above the local background for \TransitNet, reflecting the candidate-selection behavior of a blind search.

\subsection{Single-KIC transit blind-search sensitivity in the low-\snr{} regime}
\label{sec:sens}
The objective of this experiment is to evaluate the sensitivity of each algorithm to transit signals with $\SNR{}=6$--15 within a single target system, using the Low-\snr{} Transit Recovery Set of Section~\ref{sec:eval_sets}.

\subsubsection{Detection score distributions}

\begin{figure}[!t]
  \centering
  \includegraphics[width=\hsize]{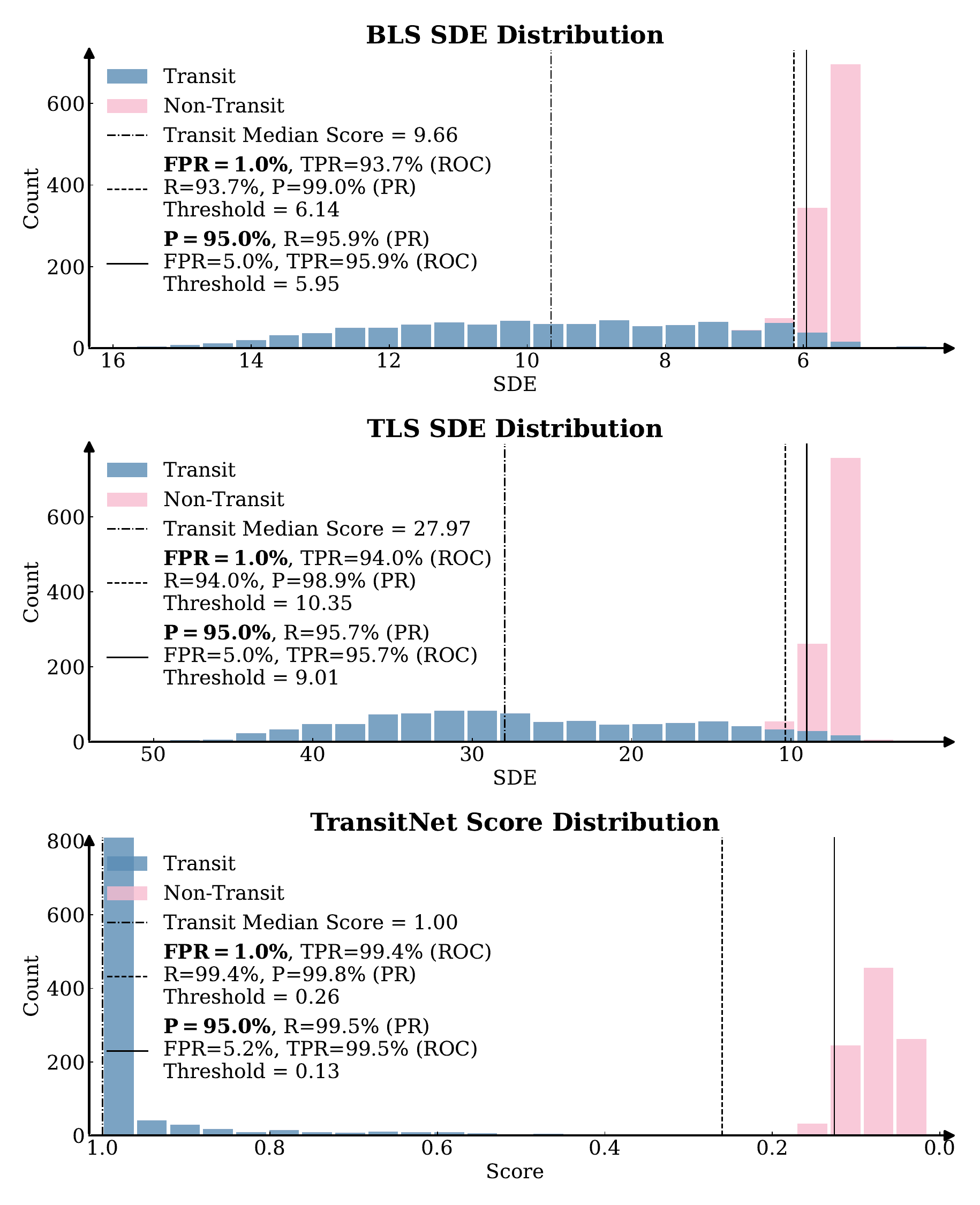}
  \caption{
    Detection score distributions of transit (blue) and non-transit (pink) light curves on the Low-\snr{} Transit Recovery Set for BLS, TLS, and \TransitNet.
    BLS and TLS are shown as SDE, and \TransitNet{} as its sigmoid score on $(0,1)$; the axes are not commensurate, and higher scores lie to the left.
    The dashed, dotted, and solid lines mark the transit median, the threshold at $\mathrm{FPR}=1\%$, and the threshold at a precision of $95\%$.
    }

  \label{fig:score_dist}
\end{figure}

Following the score-extraction procedure described in Section~\ref{sec:exp_setting}, we obtain detection score spectra for all ATLCs and PCP TMLCs, yielding the score distributions of BLS, TLS, and \TransitNet shown in Fig.~\ref{fig:score_dist}. 

For each method, the histograms show the score distributions of transit-containing ATLCs and non-transit PCP TMLCs, respectively. A smaller overlap between the two distributions indicates a stronger ability to distinguish transit signals from non-transit variability.
The figure also highlights the operating thresholds corresponding to a fixed FPR of 1\% and a fixed precision of 95\%, together with the associated performance metrics. 

Among the three methods, \TransitNet exhibits the clearest separation between the transit and non-transit score distributions, indicating stronger discrimination. Because the injected and non-transiting samples are constructed from the same KIC backgrounds, this separation reflects improved recovery of low-\snr{} transit signals under matched stellar noise conditions. Consequently, \TransitNet achieves a higher transit recovery rate and a lower FPR, indicating greater sensitivity while reducing false detections in realistic transit-search scenarios.

\subsubsection{Transit blind-search sensitivity across low-\snr{} regimes}
This experiment evaluates the performance of each method across different \snr{} regimes. 

\begin{figure*}[!tp]
  \centering
  \includegraphics[width=\textwidth]{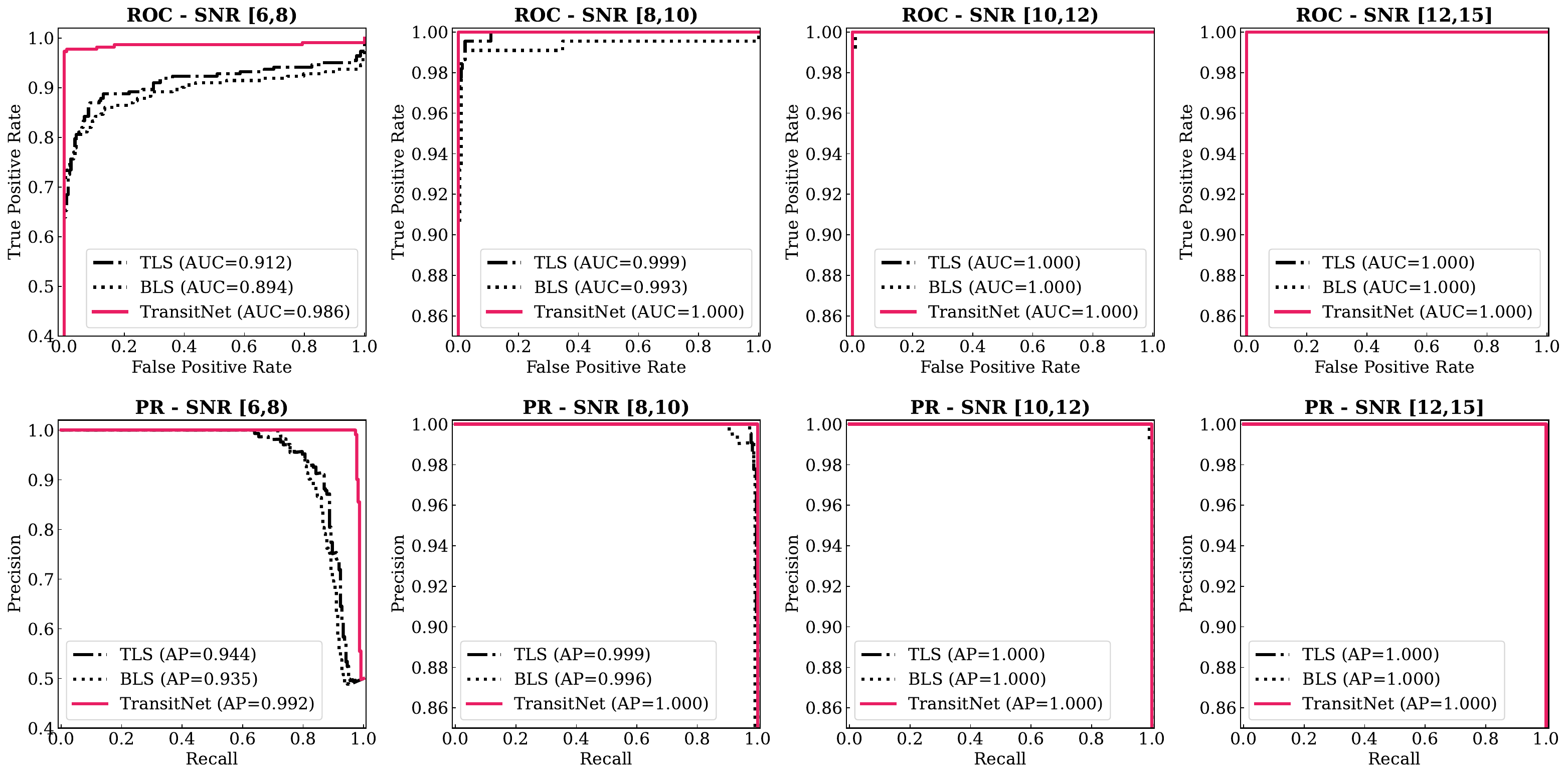}
  \caption{
    ROC and PR curves for BLS, TLS, and \TransitNet on the Low-\snr{} Transit Recovery Set, evaluated in four \snr{} bins: $[6,8)$, $[8,10)$, $[10,12)$, and $[12,15]$. Each bin contains equal numbers of transit and non-transit samples. 
    TransitNet achieves substantially higher detection performance in the low-SNR regime, while the performance differences diminish as \snr{} increases.
    }
  % Alt text: Grid of receiver operating characteristic and precision-recall curves in four signal-to-noise bins comparing BLS, TLS, and TransitNet, with the largest TransitNet advantage at the lowest signal-to-noise levels.
  \label{fig:roc_pr_lc}
\end{figure*}

\begin{figure}[!t]
  \centering
  \includegraphics[width=0.95\hsize]{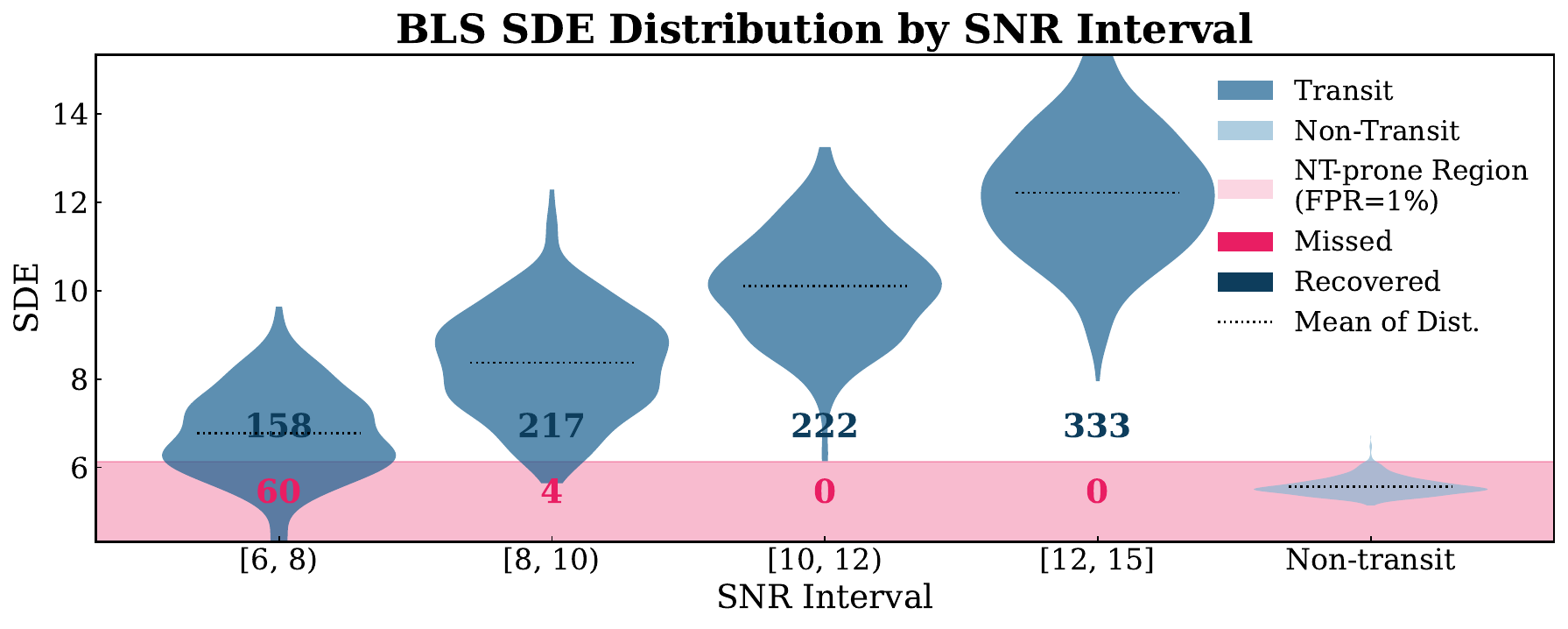}
  \includegraphics[width=0.95\hsize]{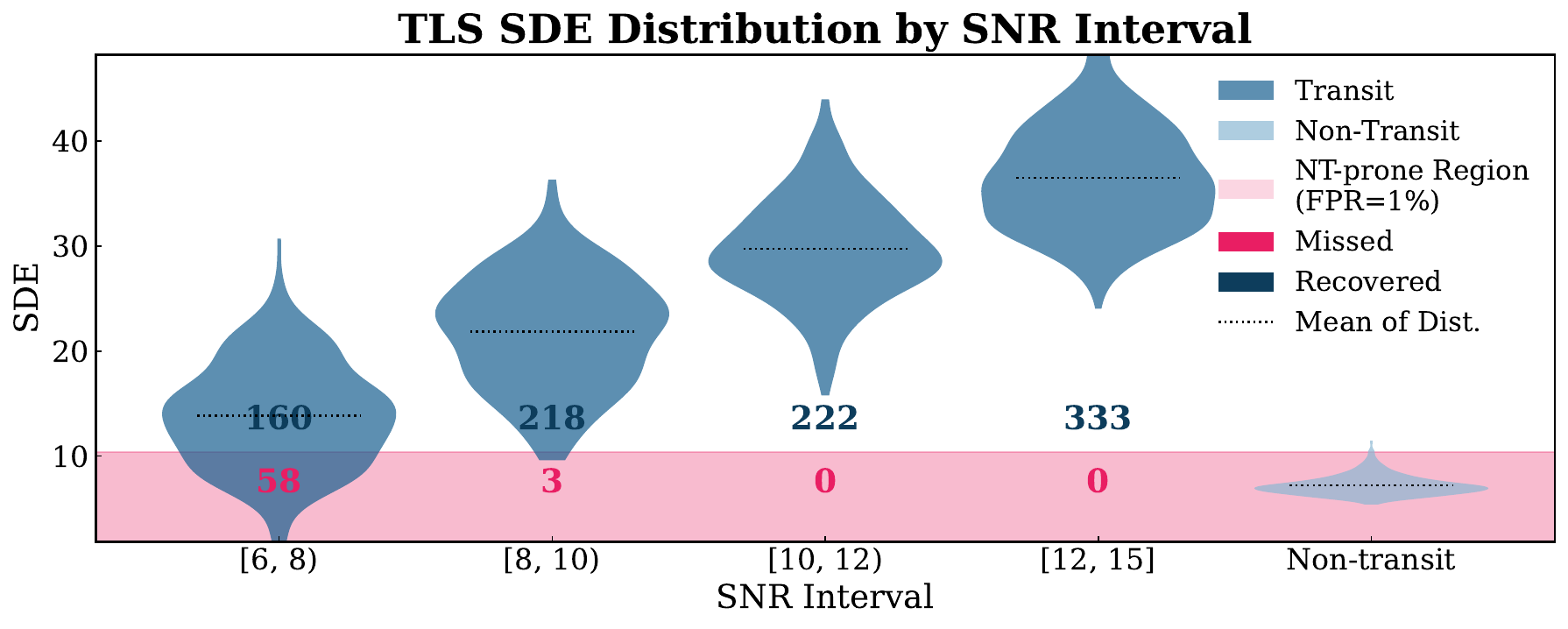}
  \includegraphics[width=0.95\hsize]{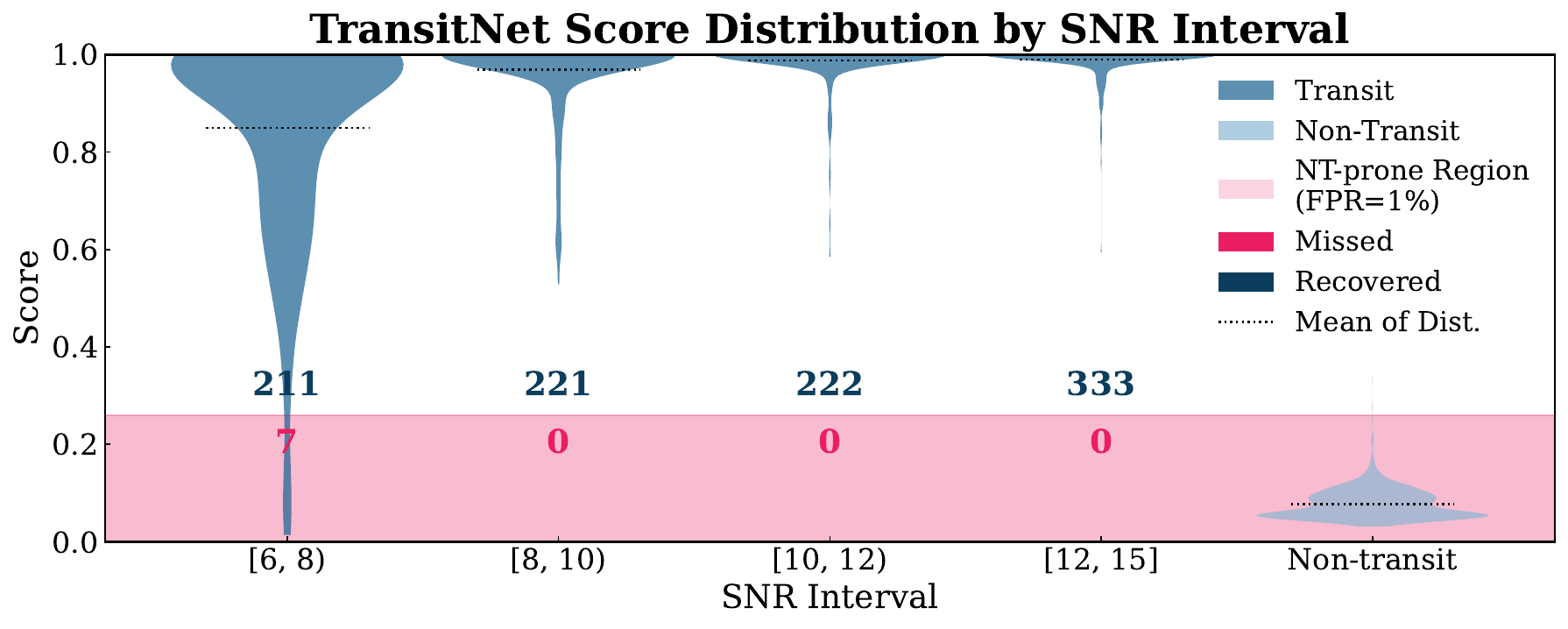}
  \caption{
    Violin plots of detection scores on the Low-\snr{} Transit Recovery Set for BLS, TLS, and \TransitNet, 
    stratified by \snr{} bins $[6,8)$, $[8,10)$, $[10,12)$, and $[12,15]$, 
    with an additional column aggregating all non-transit light curves (rightmost). 
    The light pink shaded band, termed the non-transit-prone (NT-prone) region, 
    is bounded above by the score threshold corresponding to a FPR of 1\%.
    \TransitNet consistently places a larger fraction of low-\snr{} transit signals above the NT-prone region, 
    indicating superior recoverability under realistic false-alarm constraints.  
  }
  % Alt text: Three rows of violin plots showing transit and non-transit detection scores across signal-to-noise bins for BLS, TLS, and TransitNet, with a shaded low-score region marking likely missed detections.
  \label{fig:dist_comp}
\end{figure}

Fig.~\ref{fig:roc_pr_lc} shows the ROC and PR curves of BLS, TLS, and \TransitNet on the Low-\snr{} Transit Recovery Set. \TransitNet consistently achieves superior performance, with the largest advantage observed in the lowest-\snr{} bins. The performance gap gradually narrows as the \snr{} increases.

Fig.~\ref{fig:dist_comp} complements this analysis by showing the score distributions of transit and non-transit samples within each \snr{} bin. The overlap between the two distributions reflects the difficulty of distinguishing low-\snr{} transit signals from noise. 
The shaded non-transit-prone (NT-prone) region below the threshold corresponding to a FPR of 1\% highlights signals most susceptible to missed detections, as well as noise fluctuations that may be incorrectly identified as transit candidates. 
Consistent with the ROC and PR results, \TransitNet exhibits substantially less overlap between the two distributions, indicating stronger sensitivity to low-\snr{} transit signals.

\subsection{Cross-KIC generalization and performance evaluation}
\label{sec:cross_kic}
The objective of this experiment is to evaluate the generalization capability and overall transit-search performance of each algorithm across multiple unseen target systems, using the Cross-KIC Recovery Set of Section~\ref{sec:eval_sets}.

\subsubsection{Overall detection performance and threshold optimization}
\label{sec:thres_select}
\begin{figure}[!t]
  \centering
  \includegraphics[width=0.98\hsize]{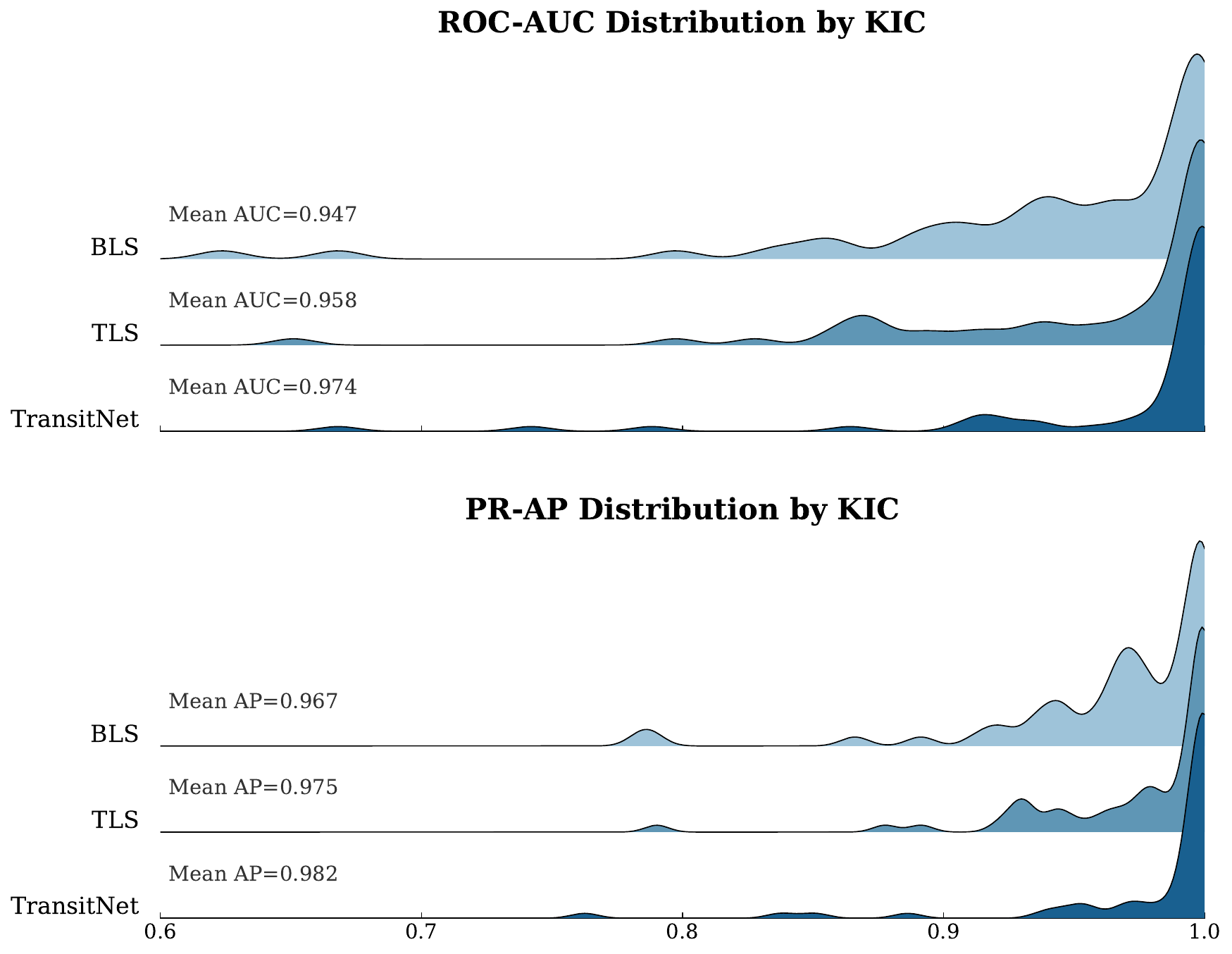}
  \caption{Distribution of per-KIC ROC-AUC and PR-AP values on the Cross-KIC Recovery Set. 
    For each KIC, the metrics are computed from the corresponding transit and non-transit detection scores spectrum. 
    These distributions evaluate the transit recovery performance of different algorithms across previously unseen KICs. 
    Distributions that are more concentrated near 1 indicate better and more stable performance under varying stellar noise conditions.
  }
  % Alt text: Ridge density plots of per-target ROC area under the curve and average precision for BLS, TLS, and TransitNet across sixty unseen Kepler targets, with TransitNet most concentrated near unity.
  \label{fig:per_kic_roc_pr_ridge}
\end{figure}

For each KIC and algorithm, the transit and non-transit samples belonging to that target yield a target-specific detection-score distribution. ROC-AUC and PR-AP computed from this distribution provide one performance-metric pair per target. Repeating the procedure across all 60 KIC targets yields the cross-target distributions of these metrics shown in Fig.~\ref{fig:per_kic_roc_pr_ridge}, while the underlying per-KIC detection-score distributions for individual methods are provided in Figs.~\ref{fig:kic_score_ridge_dlm}--\ref{fig:kic_score_ridge_bls}.

A stronger concentration of these metrics near unity indicates better cross-target generalization, greater robustness to the diverse noise characteristics across KIC targets, and more stable sensitivity to low-\snr{} transit signals. Results on the Cross-KIC Recovery Set further demonstrate that \TransitNet consistently retains the low-\snr{} transit detection capability observed on individual KIC targets across multiple KIC targets. \TransitNet achieves the strongest overall performance, with mean ROC-AUC and PR-AP values of 0.974 and 0.982, respectively, outperforming both BLS and TLS.

\begin{figure}[!t]
  \centering
  \includegraphics[width=\hsize]{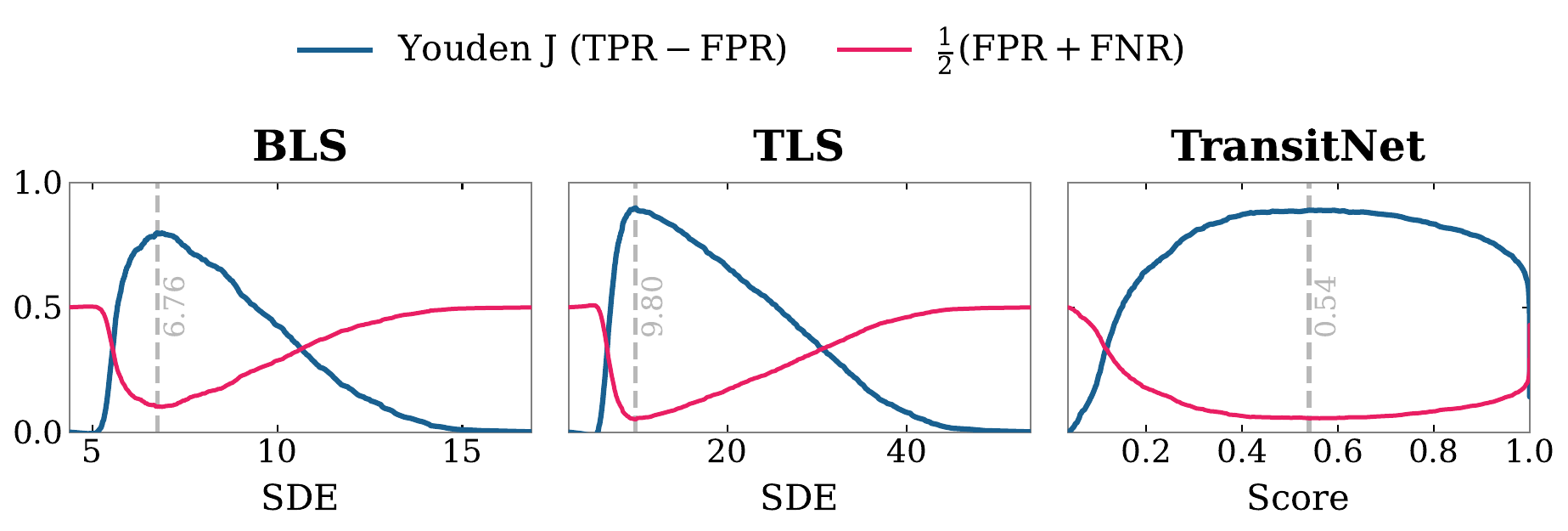}
  \caption{
    Threshold selection for BLS, TLS, and \TransitNet on the Cross-KIC Recovery Set based on maximizing Youden's statistic, $J=\mathrm{TPR}-\mathrm{FPR}$, 
    and equivalently minimizing the mean classification error, $\tfrac{1}{2}(\mathrm{FPR}+\mathrm{FNR})$. 
    Metrics are computed independently for each KIC and then macro-averaged across all KICs. 
    Vertical dashed lines indicate the adopted operating thresholds (BLS: 6.76, TLS: 9.80, and \TransitNet: 0.54), 
    which can serve as practical global operating thresholds in transit blind-searches. 
    The narrow optima of BLS and TLS indicate strong sensitivity to threshold selection, 
    whereas \TransitNet maintains near-optimal performance over a broad range of thresholds, 
    suggesting greater robustness across diverse stellar noise environments.
    }
  % Alt text: Macro-averaged true positive rate minus false positive rate versus detection threshold for BLS, TLS, and TransitNet, with vertical markers at the selected operating thresholds.
  \label{fig:threshold_macro_metrics_sweep}
\end{figure}

To enable a fair and reproducible comparison across methods, we introduce a unified threshold-selection procedure for transit blind-search evaluation. For each algorithm, all detection scores obtained from the Cross-KIC Recovery Set are collected, and a common binary decision rule is applied. Candidate thresholds are evaluated using macro-averaged TPR and FPR computed across KIC targets, from which the Youden statistic \citep{youden1950} is derived.
\begin{equation}
J(\theta)=\overline{\mathrm{TPR}}(\theta)-\overline{\mathrm{FPR}}(\theta),
\end{equation}
The threshold $\theta$ that maximizes $J(\theta)$ is adopted as the global operating point for each algorithm.
The corresponding definition of the average classification error is given by
\begin{equation}
    E(\theta) = \tfrac{1}{2}\bigl[\overline{\mathrm{FPR}}(\theta) + \overline{\mathrm{FNR}}(\theta)\bigr],
    \label{eq:macro-error}
\end{equation}
where the equivalence between the maximization of the Youden index and the minimization of $E(\theta)$ is provided in Appendix~\ref{app:thres_select}, along with the corresponding pseudocode for the threshold-selection procedure.
The optimal operating threshold for each method is selected by maximizing the difference between the TPR and FPR, equivalent to maximizing Youden's J statistic.
Fig.~\ref{fig:threshold_macro_metrics_sweep} shows the resulting threshold-sweep curves.

The optimal operating thresholds are $\mathrm{SDE}=6.76$ for BLS, $\mathrm{SDE}=9.80$ for TLS, and $\mathrm{Score}=0.54$ for \TransitNet. While BLS and TLS exhibit sharp maxima, indicating strong sensitivity to threshold selection, \TransitNet maintains near-optimal performance over a substantially broader range of thresholds. This behavior suggests greater robustness to threshold perturbations and more stable detection performance across heterogeneous stellar noise environments. The selected thresholds are subsequently fixed for the remaining experiments.

\begin{figure*}[!t]
  \centering
  \includegraphics[width=0.95\textwidth]{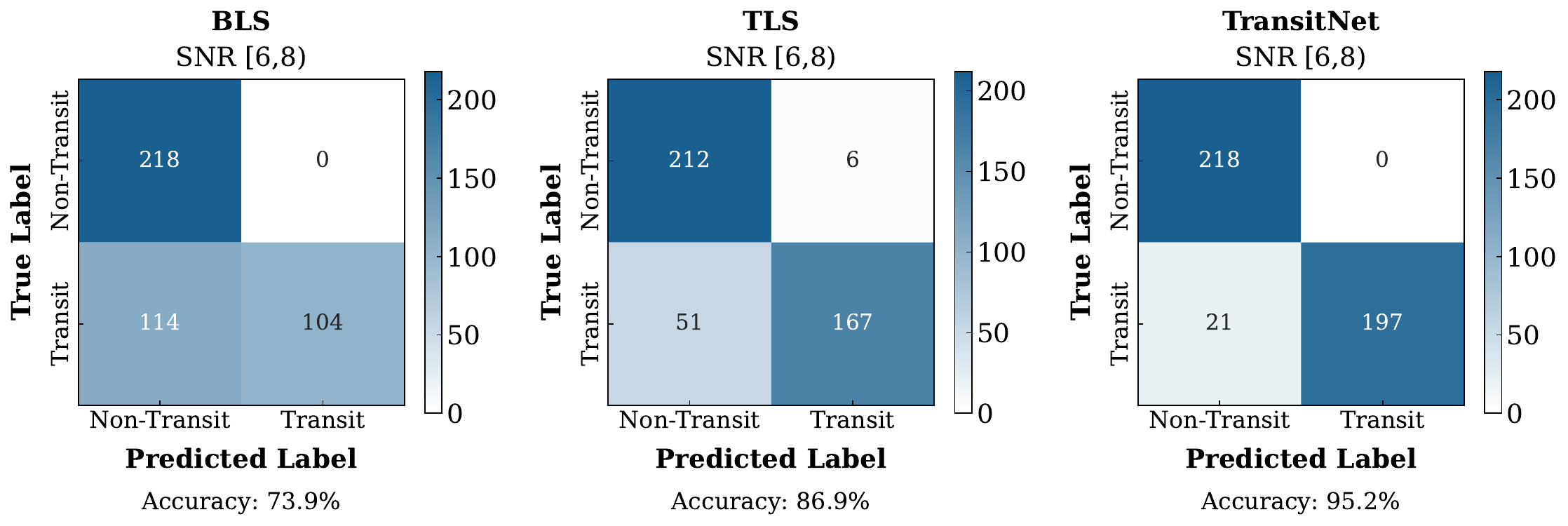}
  \caption{
    Confusion matrices for BLS, TLS, and \TransitNet on the Low-\snr{} Transit Recovery Set with \snr{} $\in [6,8)$. 
    Classification uses the operating thresholds from Fig.~\ref{fig:threshold_macro_metrics_sweep}. 
    Among the three methods, \TransitNet demonstrates the best overall performance, achieving the highest accuracy (95.2\%), compared to TLS (86.9\%) and BLS (73.9\%). 
    The overall confusion matrix for the complete dataset is shown in Fig.~\ref{fig:cnf_m_overall}.
  }
  % Alt text: Three confusion matrices for BLS, TLS, and TransitNet on the lowest signal-to-noise bin, showing counts of true and false positives and negatives and overall accuracies of 73.9, 86.9, and 95.2 percent.
  \label{fig:cnf_m}
\end{figure*}

\begin{figure*}[!t]
  \centering
  \includegraphics[width=0.98\textwidth]{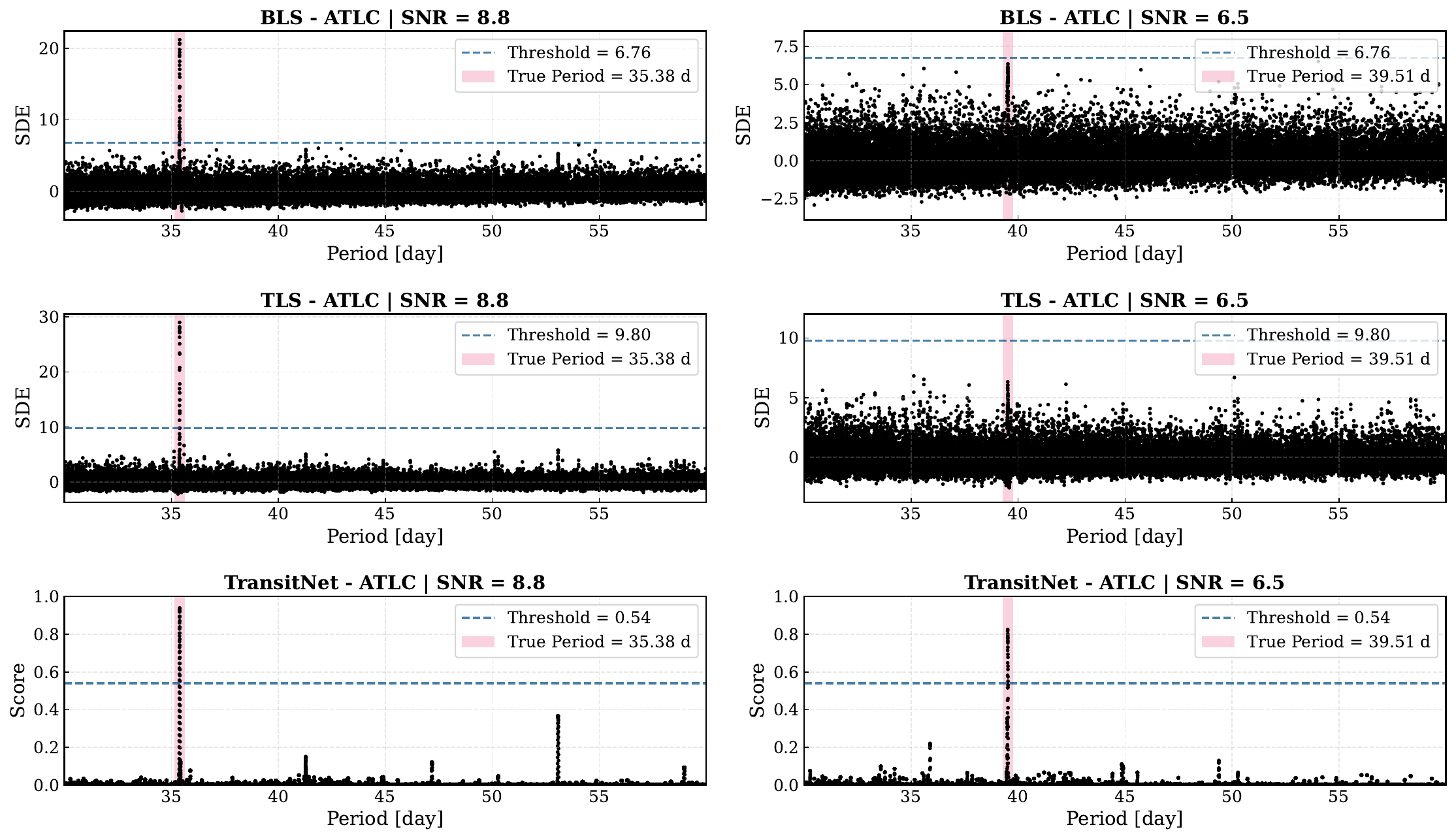}
  \caption{
    Detection score spectra for BLS, TLS, and \TransitNet on semi-synthetic ATLCs. 
    Left column: injected transit with \snr{} = 8.8 and $P=35.38$~d. 
    Right column: injected lower-\SNR{} transit with \snr{} = 6.5 and $P=39.51$~d, where the target transit is recovered only by TransitNet. 
    Horizontal lines mark the operating thresholds from Fig.~\ref{fig:threshold_macro_metrics_sweep}.
    This comparison directly illustrates \TransitNet’s enhanced sensitivity, enabling detection of low-\SNR{} transit that would be missed by BLS or TLS.
  }
  % Alt text: Two-column comparison of detection score versus trial period for a recovered high-signal injected transit and a low-signal case detected only by TransitNet, with horizontal threshold lines for each method.
  \label{fig:score_spec}
\end{figure*}

Using the selected operating thresholds, we further visualize the confusion matrices of all methods on the Low-\snr{} Transit Recovery Set. For each injected transit, the score at the injected period, obtained from the common blind-search spectrum, is compared with the corresponding threshold to yield a binary recovery decision. \TransitNet consistently outperforms TLS and BLS, achieving the highest overall accuracy of 98.9\% on the complete dataset (Fig.~\ref{fig:cnf_m_overall}). Its advantage is particularly pronounced in the challenging low-\snr{} regime (\snr{} $= 6$--$8$), where it attains an accuracy of 95.2\% (Fig.~\ref{fig:cnf_m}).

\subsubsection{Case studies on semi-synthetic low-\snr{} transits}
We further examine the detection-score spectra produced by all three methods on semi-synthetic ATLCs with injected transits at different \snr{} levels, highlighting the practical implications of the threshold-based classification and illustrating the superior sensitivity of \TransitNet.

As is shown in Fig.~\ref{fig:score_spec}, for an ATLC with an injected transit signal of \snr{} = 8.8 and a true period of 35.38 days (left column), all three algorithms successfully recover the primary signal, with detection scores exceeding their respective thresholds. Notably, \TransitNet additionally identifies a secondary peak above the threshold at approximately 53 days, which may correspond to a period alias or a potential secondary signal.
More significantly, for an ATLC with a lower \snr{} of 6.5 and a true period of 39.51 days (right column), \TransitNet successfully detected the signal with a score well above its threshold of 0.54, 
while BLS and TLS failed to detect the signal as their scores at the true period remained below their respective thresholds.

This observation directly demonstrates \TransitNet's enhanced sensitivity to low-\snr{} transit signals that are missed by TLS and BLS under the adopted thresholds in this example, highlighting a key advantage for detecting signals near the detection limit in real transit blind searches.

\subsection{Earth-size and sub-Earth-size transit recovery analysis}
\label{sec:era}
The objective of this experiment is to quantify the fraction of Earth-size and sub-Earth-size transit signals ($R_{\mathrm{p}} \leq 1\,R_\oplus$) that each algorithm can recover under realistic stellar and instrumental noise conditions, using the Earth-size and Sub-Earth-size Recovery Set of Section~\ref{sec:eval_sets}. 

\begin{figure*}[!tp]
  \centering
  \includegraphics[width=\textwidth]{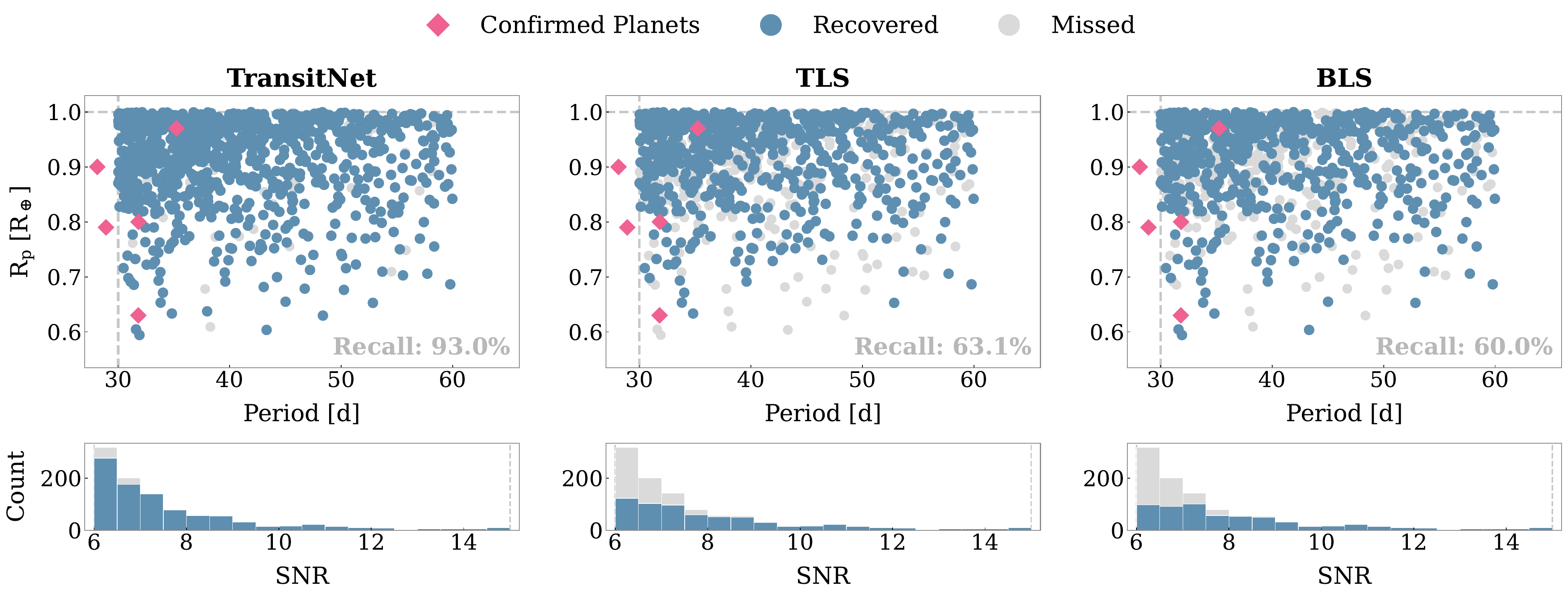}
  \caption{
  Recovery results on the Earth-size and Sub-Earth-size Recovery Set.
  \textit{Top:} planet radius versus period; blue and gray circles denote injected transits recovered and missed at algorithm-specific operating thresholds (Fig.~\ref{fig:threshold_macro_metrics_sweep}), and pink diamonds mark confirmed Kepler planets.
  Dashed guides indicate $R_{\mathrm{p}}=1\,R_{\oplus}$ and $P=30~\mathrm{d}$.
  \textit{Bottom:} stacked histograms of the injected sample as a function of \snr{}.
  \TransitNet achieves $93.0\%$ recall, substantially exceeding TLS ($63.1\%$) and BLS ($60.0\%$).
  }
  \label{fig:earth_sim_scatter}
\end{figure*}

The recovery rate is defined as $N_{\rm recovered}/N_{\rm total}$, representing the fraction of injected Earth-size and sub-Earth-size transits successfully recovered under realistic Kepler noise conditions. \TransitNet achieves a recovery rate of 93.0\%, substantially outperforming TLS at 63.1\% and BLS at 60.0\%. The corresponding recovery results across the injected parameter space are shown in Fig.~\ref{fig:earth_sim_scatter}.

Notably, the performance gain is concentrated primarily in the \snr{} range of 6--8, consistent with the results obtained in the previous experiments. 
The markedly higher recovery rate indicates that our deep-learning approach is significantly more sensitive than conventional search algorithms, highlighting its potential for improving the completeness of Earth-size or even terrestrial exoplanet detections. 

\subsection{Speed and inference efficiency}
\label{sec:speed_comparison}
Beyond detection accuracy, computational efficiency is a critical consideration for large-scale transit blind-search surveys.

To evaluate the runtime performance of \TransitNet, we conducted a benchmark comparison against both CPU and GPU implementations of BLS and TLS, as well as GPU-accelerated BLS \citep[GPU-BLS;][]{cuvarbase}.
All GPU benchmarks were performed on an NVIDIA RTX 4090 GPU with 24,564 MiB of memory. The CPU baselines were run on an AMD Ryzen 9 7950X processor (16 cores, 32 threads). We randomly selected a real Kepler light curve from a confirmed exoplanet host star. Following the grid-generation scheme proposed by \citet{hippkeOptimizedTransitDetection2019}, period grids were generated with oversampling factors (OS) ranging from 1 to 10, as implemented in the \textsc{transitleastsquares} Python package\footnote{\url{https://pypi.org/project/transitleastsquares/}} (v1.32). For each OS, the corresponding set of trial periods covering 30--60 days was constructed, and the runtime of each method was evaluated. 

To ensure a fair comparison, each algorithm was first subjected to one warm-up run, after which the execution time was measured over three consecutive trials. 
All methods were evaluated using the same input light curve. For \TransitNet, the reported runtime includes both the GPU-based folding stage and the subsequent neural-network inference. This protocol helps ensure that TLS, BLS, and \TransitNet are compared under identical search ranges and period-grid configurations.

\begin{figure}[!t]
  \centering
  \includegraphics[width=\hsize]{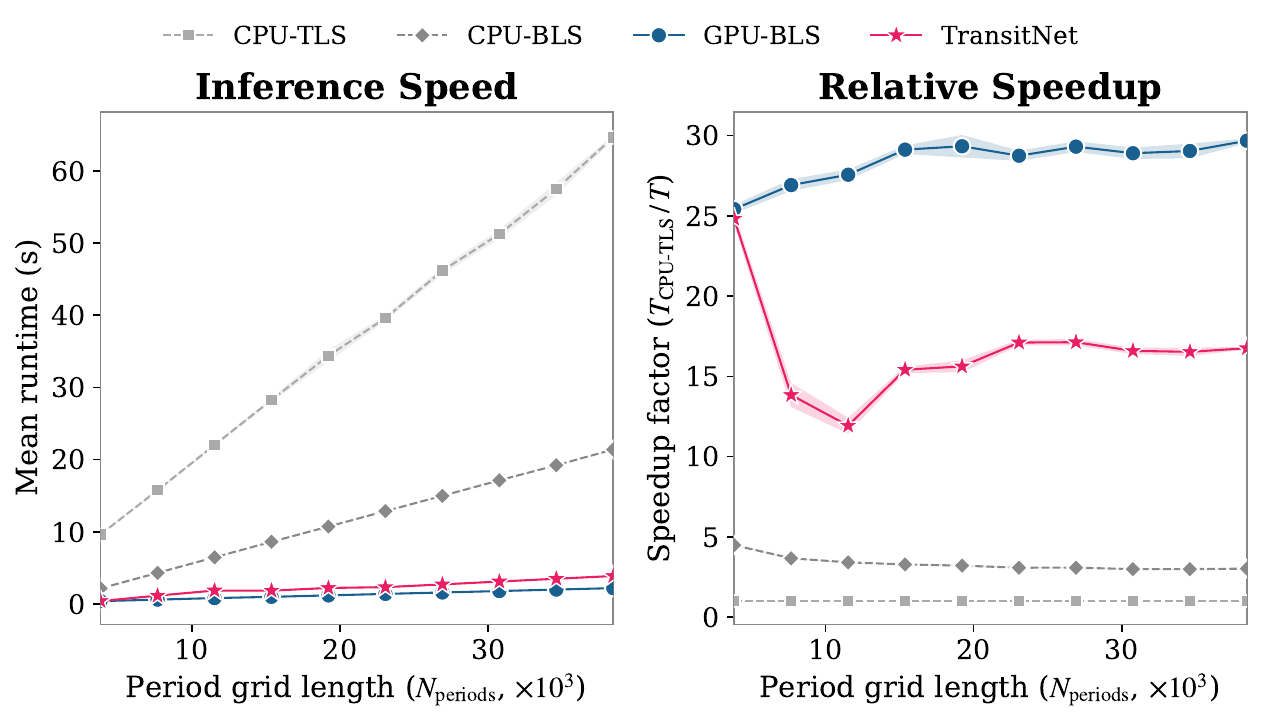}
  \caption{
Runtime scaling of transit-search algorithms as a function of period-grid size.
Dashed lines denote CPU-based implementations, while solid lines denote GPU-accelerated implementations.
The left panel shows the mean wall-clock inference time; the right panel shows the speed-up factor relative to CPU-TLS, $T_{\mathrm{CPU\text{-}TLS}}/T$.
Curves compare CPU-TLS, CPU-BLS, GPU-BLS, and \TransitNet.
The horizontal axis gives $N_{\mathrm{periods}}$ in units of $10^{3}$, controlled by sweeping the period-grid oversampling factor from 1 to 10.
  }
  % Alt text: Left panel shows mean runtime versus number of trial periods for CPU and GPU versions of BLS and TLS and for TransitNet; right panel shows speed-up relative to CPU TLS.
  \label{fig:speed_cmp}
\end{figure}
\TransitNet  combines superior low-\snr{} transit recovery with computational efficiency comparable to the fastest GPU-accelerated baseline. Over the tested range of oversampling factors, corresponding to $N_{\mathrm{periods}}\approx3.8\times10^{3}$--$3.8\times10^{4}$, inference completes within a few seconds, yielding speed-ups of $\sim$12--25$\times$ relative to CPU-TLS and $\sim$4--5$\times$ relative to CPU-BLS, while remaining within a factor of $\sim$2 of GPU-BLS in absolute runtime (Fig.~\ref{fig:speed_cmp}). 
These characteristics make \TransitNet well suited for future large-scale transit blind-search surveys targeting the poorly explored regime of $R_{\rm p}\leq 1\,R_\oplus$ and $P \gtrsim 30$ days, where improving detectability requires both high sensitivity and computational efficiency.

\subsection{Transit midpoint estimation from multi-head attention}
\label{sec:t0_estimation}
Beyond binary transit detection, localizing the detected signal in phase space provides additional information for candidate vetting and subsequent transit characterization.

Transit window and midpoint estimation exploit the content-adaptive attention mechanism of the MHA described in Section~\ref{sec:mha}, 
in which phase segments consistent with transit morphology are assigned higher attention weights.
The one-head attention matrix $A \in\mathbb{R}^{L_1\times L_1}$ (Eq.~\ref{eq:attn}) is computed 
from query and key representations projected from the downsampled features produced 
by the FM, rather than directly on the original $L_0$-bin phase grid.
Here $L_0=4096$ and $L_1=128$ denote the lengths of the input sequence and the downsampled feature sequence after the FM,  
respectively.
Therefore, the MHA attention matrix ($h\times L_1\times L_1$) requires head aggregation and upsampling before being aligned with the transit window on the original $L_0$-bin input sequence.

\textit{Attention aggregation:} Let the MHA matrix be denoted as 
$\boldsymbol{A} \in \mathbb{R}^{h\times L \times L}$ (where $L=L_1$ is used in this work).
The outputs of the $h$ attention heads are averaged, 
followed by aggregation along the query dimension.
\begin{equation}
w_i = \frac{1}{h L}\sum_{m}^{h}\sum_{q}^{L} A_{q,i}^{(m)}.
\end{equation}
This yields the attention vector $\boldsymbol{w} \in \mathbb{R}^{L}$, 
which quantifies the overall importance assigned to each source position for transit morphology recognition under the global attention mechanism.

\textit{Temporal rescaling:}
Since $L_1 \ll L_0$, directly estimating the transit midpoint 
from the low-resolution $\boldsymbol{w}$ would introduce estimation errors.
Therefore, $\boldsymbol{w}$ is remapped onto the phase grid corresponding to the original input sequence 
to restore temporal resolution.
Specifically, $\boldsymbol{w}$ is first interpolated using cubic spline interpolation (\textsc{SciPy}, \citealt{Jones2001})
and resampled onto a uniform phase grid of length $L_0$, 
yielding the interpolated attention vector $\boldsymbol{w}' \in \mathbb{R}^{L_0}$.
Subsequently, the interpolated attention vector is smoothed with a five-bin moving-average filter 
to reduce possible interpolation artifacts, followed by renormalization to preserve the total attention mass.

\textit{Transit-window estimation:}
The phase bin corresponding to the global maximum of the attention distribution is identified as the reference peak. 
The algorithm then expands in both directions until the attention decreases 
to the full width at half maximum (FWHM), thereby defining the estimated transit window $[\hat{\tau}_a,\hat{\tau}_b]$.
Due to interpolation-induced shifts after upsampling, 
the attention peak location $\hat{\tau}_{\mathrm{peak}}$ is used only for transit-window localization 
rather than as the final transit-midpoint estimate.

\textit{Transit midpoint estimation:}
Within the estimated transit window $[\hat{\tau}_a,\hat{\tau}_b]$, 
the phase bin with the minimum normalized flux is selected as the estimate of the transit midpoint $\hat{\tau}_0$.
After obtaining $\hat{\tau}_0$, the transit window is re-centered on $\hat{\tau}_0$ while preserving its width, 
$\Delta\hat{\tau}=\hat{\tau}_b-\hat{\tau}_a$, thereby updating the interval $[\hat{\tau}_a',\hat{\tau}_b']$.
This strategy constrains the transit midpoint using the minimum of the normalized flux within the estimated transit window. 
Because the attention-derived window is obtained from a downsampled representation, its boundaries may not perfectly coincide with the true transit interval and are subject to interpolation uncertainty when projected back to the original sequence. 
Rather than relying on peaks in the attention vector $\boldsymbol{w}'$, the proposed method identifies the midpoint directly from the original photometric measurements within the estimated window. 
This design represents a practical compromise between localization precision and computational efficiency, combining coarse attention-based localization with refinement from higher-resolution flux measurements.
For applications requiring higher midpoint localization precision, 
this step may be further replaced by refined template matching within the estimated transit window, 
thereby yielding a more accurate estimate of the transit midpoint.

\begin{figure}[!t]
  \centering
  \includegraphics[width=\hsize]{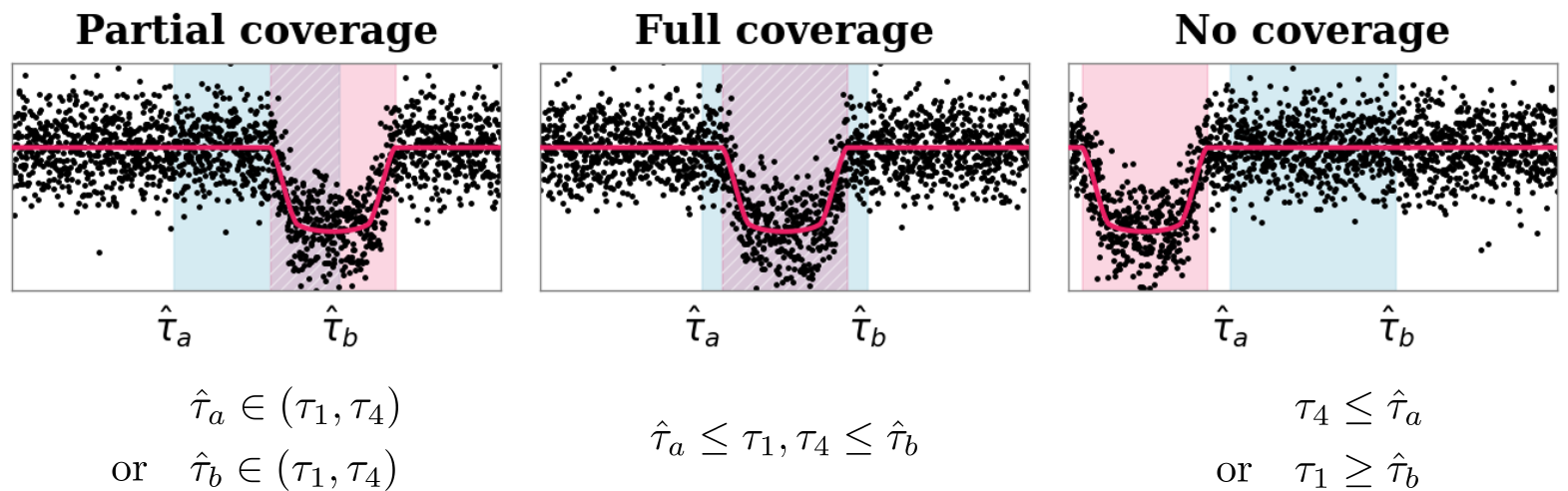}
  \caption{
    Schematic illustration of the transit-region scoring scheme comparing an estimated transit window, 
    $[\hat{\tau}_a,\hat{\tau}_b]$ (light blue), with the true in-transit interval, $[\tau_1,\tau_4]$ (magenta). 
    Black points show synthetic photometry generated from a \textsc{Batman} model (solid red curve), 
    and the hatched region denotes the covered transit interval.
    The three panels illustrate partial coverage (\textit{left}; $0<s<1$), 
    full coverage (\textit{center}; $[\tau_1,\tau_4]\subseteq[\hat{\tau}_a,\hat{\tau}_b]$, $s=1$), and no coverage (\textit{right}; $s=0$). 
    For partial coverage, the overlap duration is $D=\max\!\left(0,\min(\hat{\tau}_b,\tau_4)-\max(\hat{\tau}_a,\tau_1)\right)$, 
    with score $s=D/T_{14}$ and $T_{14}=\tau_4-\tau_1$.}
  % Alt text: Three schematic panels comparing an estimated transit window with the true in-transit interval for partial overlap, full overlap, and no overlap on synthetic folded light curves.
  \label{fig:t0_score}
\end{figure}

To quantitatively evaluate the accuracy of transit-window estimation, 
we introduce a transit-window coverage criterion (Fig.~\ref{fig:t0_score}).
Let the estimated transit window be denoted by $[\hat{\tau}_a,\hat{\tau}_b]$ and the true transit window by $[\tau_1,\tau_4]$.
The covered duration of the true transit interval is defined as
\begin{equation}
D=\max\left\{0,\,\min(\hat{\tau}_b,\tau_4)-\max(\hat{\tau}_a,\tau_1)\right\},
\label{eq:overlap_length}
\end{equation}
where the total true transit duration is given by $T_{14}=\tau_4-\tau_1$.
The transit-window coverage score is then defined as
\begin{equation}
s=\frac{D}{T_{14}},
\label{eq:overlap_score}
\end{equation}
which represents the fraction of the true transit interval covered by the estimated window and ranges from $0$ to $1$.

Specifically, $s=1$ when the true transit interval is completely covered by the estimated window, 
$0<s<1$ in cases of partial coverage, 
and $s=0$ when no part of the true transit interval is covered by the estimated window.
In addition, an independent evaluation dataset containing $\sim$7000 transit samples 
was constructed to assess the transit window and midpoint estimation capabilities of \TransitNet.

\begin{figure}[!t]
  \centering
  \includegraphics[width=\hsize]{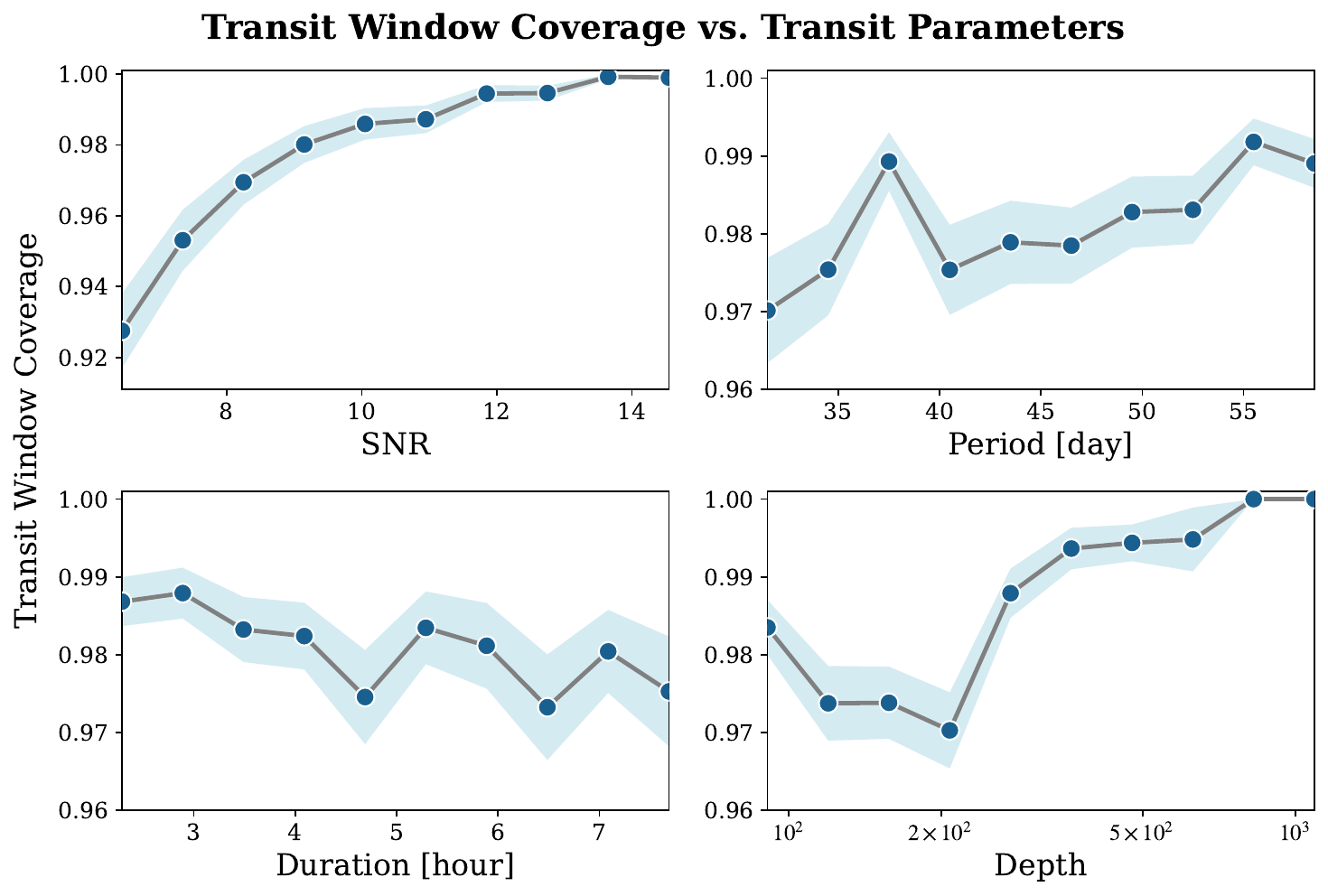}
  \caption{
  Transit window overlap versus transit and observational parameters on a separate evaluation set.
  Points show bin-averaged overlap scores; shaded bands indicate mean $\pm$ standard error of the mean (SEM).
  The score measures agreement between the estimated ingress-egress interval $[\hat{\tau}_a,\hat{\tau}_b]$ and the true transit window.
  \emph{Panels:} \snr{} (top left), orbital period in days (top right), transit duration in hours (bottom left), and transit depth (bottom right, log scale).
  }
  % Alt text: Four scatter panels with uncertainty bands showing mean transit-window overlap score versus signal-to-noise, orbital period, transit duration, and transit depth, remaining high across the parameter space.
  \label{fig:region_score_by_params}
\end{figure}
\begin{figure}[!t]
  \centering
  \includegraphics[width=\hsize]{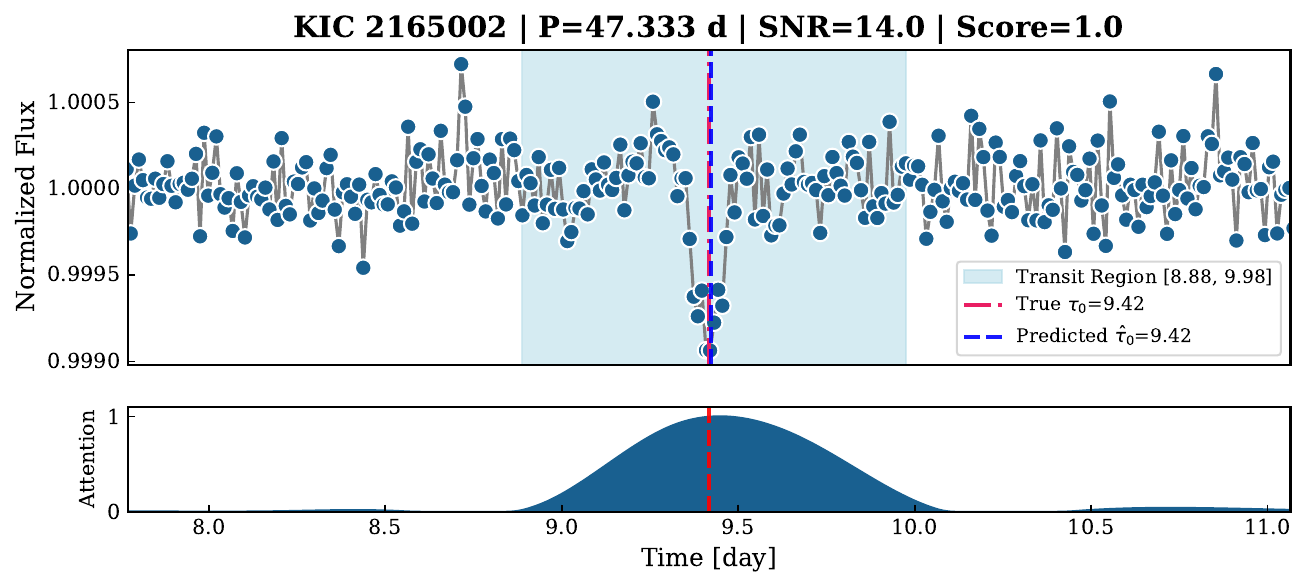}
  \includegraphics[width=\hsize]{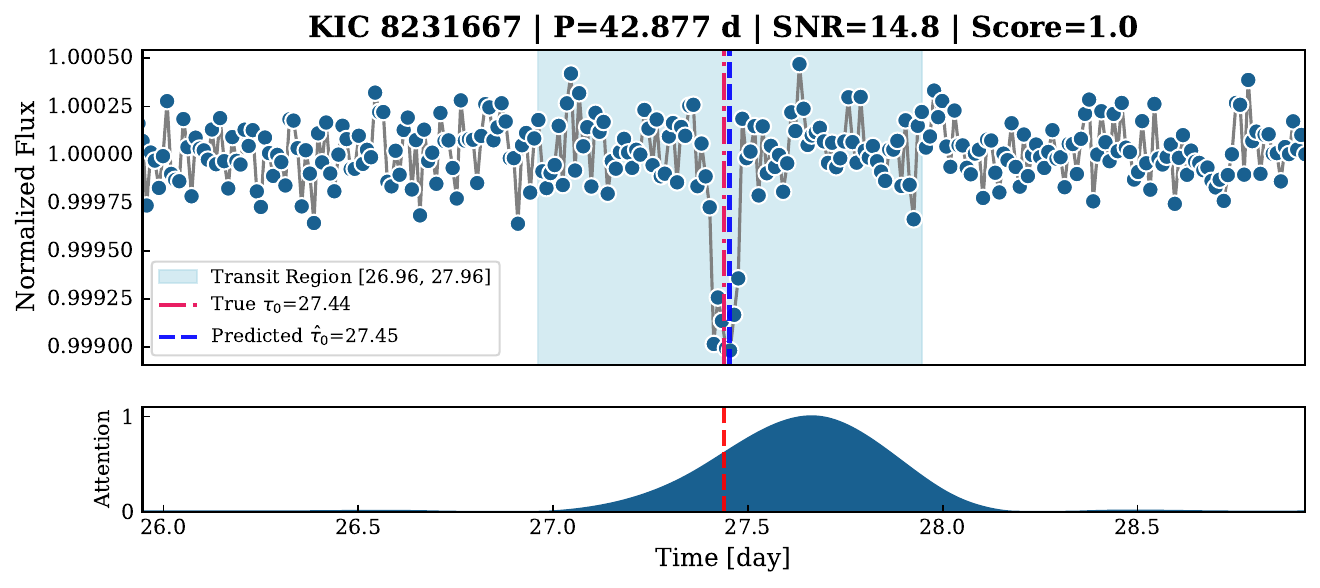}
  \caption{Analysis of two examples. The top panel in each figure shows the normalized flux over time. 
  The estimated transit midpoint ($\hat{\tau}_0$) is indicated by the red dashed line, 
  while the true $T_0$ is marked by the blue dashed line. 
  The bottom panel displays the attention weights, with peaks indicating the temporal segments most relevant for transit detection.}
  % Alt text: Two examples each with a top normalized flux time series marking estimated and true transit midpoints and a bottom attention-weight profile peaking near the transit event.
  \label{fig:t0_pred}
\end{figure}

Table~\ref{tab:coverage_stats} summarizes the proportions of the three coverage categories together with the mean overlap score achieved by \TransitNet on this dataset.
Across the investigated parameter space, the coverage score remains consistently high, 
with most bins satisfying $s>0.97$.
The coverage score depends more strongly on \snr{} and $\delta$, 
approaching nearly complete coverage for stronger transit signals, 
whereas only minor fluctuations are observed with orbital period and transit duration.
Fig.~\ref{fig:t0_pred} illustrates the estimated $\hat{\tau}_0$ for two genuine Kepler targets using \TransitNet,
along with their visualization of the processed attention weight sequence and the refined transit window.
Fig.~\ref{fig:region_score_by_params} further illustrates the variation of the coverage score $s$ as a function of \snr{}, $P$, $T_{14}$, and $\delta$.

All 34 confirmed Kepler planets in the Confirmed-Planet Set were recovered with scores above the detection threshold of 0.54. Their true transit midpoints, $\tau_0$, all fall within the corresponding estimated transit windows, $[\hat{\tau}_a,\hat{\tau}_b]$, supporting the use of the attention-based method for initial transit-window estimation and transit-epoch localization.

\TransitNet jointly performs transit detection and transit midpoint estimation in a single forward pass. 
This end-to-end formulation may support upstream and downstream survey tasks, 
including automated triage, transit-preserving detrending 
(by masking candidate transit regions prior to baseline removal), 
and the initialization of MCMC-based light-curve modeling (by providing initial $\tau_0$ estimates), 
% while the resulting midpoint estimates may also prove useful for transit timing variation (TTV) analyses 
and other large-scale follow-up studies. Further dedicated validation will be required 
to assess these potential applications.

\begin{table}[!t]
  \centering
  \caption{Transit-window coverage statistics and corresponding midpoint estimation errors on the independent evaluation dataset. 
  The coverage score quantifies the fraction of the true transit interval covered by the estimated window, 
  while the midpoint estimation error is defined as $|\tau_0-\hat{\tau}_0|$.
  The last two columns report the mean coverage score and mean midpoint estimation error within each category, respectively.
  The large errors in the no-coverage cases arise from complete misalignment with the true transit interval.
  Across all samples, the mean coverage score is $0.982\pm0.124$. 
  }
  \label{tab:coverage_stats}
  \begin{tabular}{lccc}
    \hline
    \hline
    Category & Fraction (\%) & Score & $|\tau_0 - \hat{\tau}_0|$ (h) \\
    \hline
    Full coverage    & 97.4 & 1.0 & 0.05 \\
    Partial coverage & 1.3  & 0.6 & 0.34\\
    No coverage      & 1.3  & 0.0 & 4.57 \\
    \hline
    \hline
  \end{tabular}
\end{table}

\section{Conclusion and discussion}
\label{sec:discussion}

\subsection{Conclusion}
Motivated by the observational incompleteness of I2LP Earth-size planets, we developed \TransitNet, a compact attention-based framework for low-\snr{} transit blind searches. Beyond the model architecture, we established a data construction and benchmarking pipeline tailored to realistic stellar variability and instrumental noise, including the PCP strategy for constructing reliable non-transit controls and recovery-oriented benchmarks for evaluating detection sensitivity.

Extensive experiments demonstrate that \TransitNet{} effectively recovers low-\snr{} transit signals across heterogeneous Kepler noise conditions and consistently outperforms TLS and BLS in the tested regimes. In the target I2LP range of $P\in[30,60]$~days and $\SNR{}\in[6,15]$, the model achieves 95.2\% accuracy at $\SNR{}=6$-8, with mean ROC-AUC and PR-AP of 0.974 and 0.982 on the cross-KIC evaluation, respectively. For injected Earth-size and sub-Earth-size transits, \TransitNet{} reaches a recovery rate of 93.0\%, compared with 63.1\% for TLS and 60.0\% for BLS. Together, these results demonstrate the effectiveness of the proposed architecture and data construction strategy for low-\snr{} transit recovery under realistic noise conditions.

Beyond detection, the learned attention patterns provide direct estimates of both the transit window and its midpoint phase in a single inference pass. The estimated transit windows achieve a mean coverage of $0.982\pm0.124$, while all 34 confirmed Kepler planets within the target parameter space are recovered with a mean absolute midpoint error of 1.24~h. These results show that \TransitNet{} can not only identify low-\snr{} transit signals but also provide physically meaningful temporal localization and initial transit timing information for subsequent candidate characterization. Combined with a compact 1.5~MB footprint and second-scale GPU inference, the framework is well suited to large-scale candidate searches.

As future transit surveys process millions of stellar light curves, practical blind-search pipelines will require a combination of sensitivity, scalability, and informative candidate outputs. By integrating low-\snr{} transit recovery, efficient inference, and physically meaningful transit localization and timing estimation, \TransitNet{} provides a practical framework for extending transit searches toward the challenging I2LP regime of Earth-size and sub-Earth-size planets.

\subsection{Generalization, overfitting, and scope of validity}
\label{sec:overfit}
Because \TransitNet is trained on semi-synthetic data, its evaluation warrants an explicit discussion of overfitting.
The evidence that the reported performance is not an artifact of the training distribution rests on four largely independent arguments.
Changing only the noise realization or the stellar target would not, by itself, guarantee generalization; the assessment below therefore does not rely on noise diversity alone.

First, the 60 KIC targets underlying every benchmark of Section~\ref{sec:apply} share no target with the $\sim$700 KICs used to build the training TMLCs.
The simulated transits are injected into real, detrended Kepler residual light curves rather than a parametric or purely Gaussian noise model. The training backgrounds therefore retain realistic instrumental systematics, short-timescale stellar variability, data gaps, and correlated residual noise. This target-level separation is more informative than changing only a random noise realization, and the per-target ROC-AUC and PR-AP distributions (Fig.~\ref{fig:per_kic_roc_pr_ridge}) show that performance is stable across this heterogeneous population rather than concentrated in a favorable subset. The common detrending suppresses long-timescale variability and reduces the distribution mismatch between training and application data, but it neither makes the residuals Gaussian nor guarantees insensitivity to all stellar activity.
Second, the non-transit class is dominated by PCP TMLCs, which retain genuine instrumental and astrophysical noise structure while destroying residual periodicity (Appendix~\ref{app:pcp});  
this reduces the risk that the classifier relies on an idealized noise model rather than transit morphology.

Third, and in our view most informative, the test of Appendix~\ref{app:real} involves no injection at all.
The signals used for training are basic geometric trapezoids without limb darkening (Section~\ref{sec:alc}), whereas the 34 confirmed Kepler planets of Table~\ref{tab:transitnet_results} have genuine limb-darkened transit profiles that are absent from the training distribution.
All 34 are recovered above the operating threshold with a mean absolute midpoint error of 1.24~h. This injection-free result provides direct empirical evidence that the model transfers from injected trapezoids to genuine Kepler transit morphologies in unseen targets, although it does not by itself establish unrestricted generalization.

Fourth, dropout, BN, decoupled weight decay, and early stopping are applied throughout, and the fold-to-fold dispersion of \TransitNet is at most $0.2\%$ on every metric in Table~\ref{tab:tp_results}.
Together, the target-disjoint benchmarks, injection-free confirmed-planet test, and regularization controls argue against simple memorization, while this conclusion remains limited to the Kepler-based noise, preprocessing, and parameter domain examined here.

These tests nonetheless leave the scope of validity bounded, and we state it explicitly rather than leaving it implicit.
\TransitNet is trained and validated for $P\in[30,60]$~days and $\SNR{}\in[6,15]$, and its quoted performance should be read as applying to that domain.
Known transit signals, including known strong signals, are masked before the targeted residual search. If an unknown higher-\snr{} signal remains, BLS and TLS provide an independent recovery channel. Within the range tested directly here, $10\leq\SNR{}\leq15$, the performance gap between \TransitNet and TLS narrows as \snr{} increases (Figs.~\ref{fig:roc_pr_lc} and \ref{fig:earth_sim_scatter}); performance for $\SNR{}>15$ has not been quantified.
Because \TransitNet operates on a phase-folded representation and does not receive the orbital period as an explicit input, a low-\snr{} signal outside the trained period interval may still produce an elevated score when a morphologically similar fold, alias, or harmonic lies within the searched grid. This out-of-distribution behavior has not been calibrated, however, and no completeness claim is made outside 30 to 60~days. However, these limitations can be addressed by retraining the model on newly generated datasets covering different orbital-period and \snr{} ranges, thereby adapting the search to other target domains.

In the intended residual-search workflow, \TransitNet serves as a candidate generator rather than the sole decision criterion. Candidate peaks will subsequently undergo harmonic analysis, odd--even transit-depth tests, and half-/double-period checks following \citet{Wang2024}; only candidates whose vetted periods lie in the target interval will be retained. These checks cannot recover a signal that the model fails to flag.
Extending the trained period range, together with the lower transit multiplicity and smaller transit duty cycle that this entails, is addressed in Section~\ref{sec:future}.

\subsection{Cross-survey transfer and generalization}
Compared to Kepler, other survey missions (e.g., TESS) differ in cadence, photometric precision, and target populations. Consequently, the cross-instrument generalization and transferability of deep-learning-based transit-search models merit discussion.
Depending on the transfer paradigm, generalization can be categorized into two types. 

The first is direct transfer, in which a model trained on Kepler data is applied to other survey data without additional modification. The second is indirect transfer, 
in which survey-specific datasets are constructed and the model is retrained. 

For direct transfer, its effectiveness depends on the following conditions: 
(i) the photometric precision and cadence of the light curves are sufficient to resolve the target exoplanet transit duration; 
and (ii) the period range of the target signals is covered by the period distribution of the training data.
For example, in the Kepler-based dataset adopted in this work, the cadence is 29.4\,min. 
The transit samples are generated from ATLCs containing injected transit signals with periods of 30--60\,days and \snr{} values between 6 and 15, whereas the non-transit samples are constructed from PCP TMLCs and GNLCs, with the noise standard deviations of the latter sampled from the corresponding source KICs.
Under these conditions, direct transfer may be feasible but requires dedicated validation.

Indirect transfer typically involves more complex data generation procedures 
and imposes higher demands on training efficiency and stability. 
The proposed model, with a lightweight architecture of approximately 1.5~MB and strong training stability (Section~\ref{sec:method}), 
as well as its demonstrated computational efficiency in inference (Section~\ref{sec:speed_comparison}), 
is well suited to survey-specific retraining and controlled direct-transfer tests.

\subsection{Future work}
\label{sec:future}
Two extensions follow from the domain of validity stated in Section~\ref{sec:overfit}.

The first is a quantitative test of cross-survey transfer.
Direct transfer, in which the Kepler-trained model is applied without retraining or with only limited fine-tuning, and indirect transfer, in which the training set is rebuilt for the target survey and the model is retrained, are set out in the preceding subsection; neither has yet been measured on an independent survey.
TESS is the appropriate near-term benchmark.
Its cadence, photometric precision, and stellar sample differ from the Kepler long-cadence data used here, and over most of the sky a single sector lasts about 27~days.
In the trained interval $P\in[30,60]$~d, that baseline admits at most one transit per sector; only the continuous-viewing zones, with a baseline of about one year, provide repeated events.
PLATO and ET are the relevant longer-baseline facilities for the same comparison once survey light curves, or simulations of comparable fidelity, become available.
Performance should be reported against cadence, photometric precision, the overlap between the searched periods and the training interval, and the difference in stellar population.
The operating threshold should be recalibrated on the target survey, so that completeness and the false-positive rate are assessed together.

The second extension moves the trained interval from $P\in[30,60]$~d to $[150,200]$~d.
These periods approach the habitable zones of K dwarfs.
They remain shorter than the orbital periods of the conservative habitable zone of solar-type stars \citep{Bryson2021}, and on a four-year Kepler-like baseline they still yield only a few transits.
They are therefore a controlled intermediate step toward Earth analogs.
Fewer co-added transits lower the phase-folded \snr{} and leave the transit covering a smaller fraction of the folded phase curve.
Completeness and the false-positive rate will need to be evaluated jointly in this regime before the Kepler threshold, and the performance quoted in this work, can be carried over.

In addition, we also plan to apply \TransitNet{} to a large-scale search of the Kepler data for Earth- and sub-Earth-sized exoplanets, with the aim of addressing part of the parameter-space gap associated with the I2LP population.

\begin{acknowledgements}
Funding for this study is provided by the Strategic Priority Program on Space Science of the Chinese Academy of Sciences (XDA15020600) and China's Space Origins Exploration Program (GJ11030405). 
JPZ is grateful for the support of the National Natural Science Foundation of China, Grant No. 12203087.
Software used includes: \textsc{Astropy} \citep{astropy:2013,astropy:2018,astropy:2022}, \textsc{PyTorch} \citep{Paszke2017}, \textsc{Jupyter} \citep{Kluyver2016}, \textsc{SciPy} \citep{Jones2001}, \textsc{Scikit-learn} \citep{scikit-learn,sklearn_api}, and \textsc{Matplotlib} \citep{Hunter2007}.
The authors thank Hui Zhang, Shiyin Shen, Bo Ma, and Jiwei Xie for their valuable comments and constructive suggestions, which helped improve the manuscript, and Zhenghong Liu for helpful discussions and assistance with the detrending procedure used in this work.
\end{acknowledgements}
\section*{Data availability}
The data used in this study are publicly available from the Kepler mission archive 
\href{https://archive.stsci.edu/kepler/}{https://archive.stsci.edu/kepler/}. 
The KOI and exoplanet parameters used in this work are obtained from the NASA Exoplanet Archive \citep{Christiansen2025} 
(\href{https://exoplanetarchive.ipac.caltech.edu/}{https://exoplanetarchive.ipac.caltech.edu/}), including the KOI table (doi:\href{https://doi.org/10.26133/NEA4}{10.26133/NEA4}). 
The ATLCs and PCP TMLCs derived from Kepler TMLCs are based on these public data and can be accessed upon request from the corresponding author.
\bibliographystyle{aa}
\bibliography{transitnet}

@misc{liu2026delosdetectingshallowtransits,
      title={DELOS: Detecting Shallow Transits in Kepler Photometry Using a Contrastive-Learning Framework}, 
      author={Qingtian Liu and Jian Ge and XingChen Yan and Kevin Willis and Xinyu Yao and QuanQuan Hu and Jiapeng Zhu},
      year={2026},
      eprint={2605.29428},
      archivePrefix={arXiv},
      primaryClass={astro-ph.EP},
      url={https://arxiv.org/abs/2605.29428}, 
}

@article{Thompson_2018,
    author = {Thompson, Susan E. and Coughlin, Jeffrey L. and Hoffman, Kelsey and Mullally, Fergal and Christiansen, Jessie L. and Burke, Christopher J. and Bryson, Steve and Batalha, Natalie and Haas, Michael R. and Catanzarite, Joseph and Rowe, Jason F. and Barentsen, Geert and Caldwell, Douglas A. and Clarke, Bruce D. and Jenkins, Jon M. and Li, Jie and Latham, David W. and Lissauer, Jack J. and Mathur, Savita and Morris, Robert L. and Seader, Shawn E. and Smith, Jeffrey C. and Klaus, Todd C. and Twicken, Joseph D. and Van Cleve, Jeffrey E. and Wohler, Bill and Akeson, Rachel and Ciardi, David R. and Cochran, William D. and Henze, Christopher E. and Howell, Steve B. and Huber, Daniel and Pr{\v s}a, Andrej and Ram{\'\i}rez, Solange V. and Morton, Timothy D. and Barclay, Thomas and Campbell, Jennifer R. and Chaplin, William J. and Charbonneau, David and Christensen-Dalsgaard, J{\o}rgen and Dotson, Jessie L. and Doyle, Laurance and Dunham, Edward W. and Dupree, Andrea K. and Ford, Eric B. and Geary, John C. and Girouard, Forrest R. and Isaacson, Howard and Kjeldsen, Hans and Quintana, Elisa V. and Ragozzine, Darin and Shabram, Megan and Shporer, Avi and Aguirre, Victor Silva and Steffen, Jason H. and Still, Martin and Tenenbaum, Peter and Welsh, William F. and Wolfgang, Angie and Zamudio, Khadeejah A and Koch, David G. and Borucki, William J.},
	doi = {10.3847/1538-4365/aab4f9},
	fjournal = {The Astrophysical Journal Supplement Series},
    journal = {ApJS},
	month = {apr},
	number = {2},
	pages = {38},
	publisher = {The American Astronomical Society},
	title = {Planetary Candidates Observed by Kepler. VIII. A Fully Automated Catalog with Measured Completeness and Reliability Based on Data Release 25},
	url = {https://doi.org/10.3847/1538-4365/aab4f9},
	volume = {235},
	year = {2018}
}

@article{youden1950,
    author = {Youden, W. J.},
    title = {Index for rating diagnostic tests},
    journal = {Cancer},
    volume = {3},
    number = {1},
    pages = {32-35},
    doi = {https://doi.org/10.1002/1097-0142(1950)3:1<32::AID-CNCR2820030106>3.0.CO;2-3},
    url = {https://acsjournals.onlinelibrary.wiley.com/doi/abs/10.1002/1097-0142%281950%293%3A1%3C32%3A%3AAID-CNCR2820030106%3E3.0.CO%3B2-3},
    eprint = {https://acsjournals.onlinelibrary.wiley.com/doi/pdf/10.1002/1097-0142%281950%293%3A1%3C32%3A%3AAID-CNCR2820030106%3E3.0.CO%3B2-3},
    year = {1950}
}

@article{Fressin2013,
  author = {Fressin, François and Torres, Guillermo and Charbonneau, David and Bryson, Stephen T. and Christiansen, Jessie and Dressing, Courtney D. and Jenkins, Jon M. and Walkowicz, Lucianne M. and Batalha, Natalie M.},
  doi = {10.1088/0004-637x/766/2/81},
  fjournal = {The Astrophysical Journal},
  issn = {1538-4357},
  journal = {ApJ},
  month = {March},
  number = {2},
  pages = {81},
  publisher = {American Astronomical Society},
  title = {THE FALSE POSITIVE RATE OFKEPLERAND THE OCCURRENCE OF PLANETS},
  url = {https://doi.org/10.1088/0004-637x/766/2/81},
  volume = {766},
  year = {2013}
}

@article{Bryson2021,
  author = {Bryson, Steve and Kunimoto, Michelle and Kopparapu, Ravi K. and Coughlin, Jeffrey L. and Borucki, William J. and Koch, David and Aguirre, Victor Silva and Allen, Christopher and Barentsen, Geert and Batalha, Natalie M. and Berger, Travis and Boss, Alan and Buchhave, Lars A. and Burke, Christopher J. and Caldwell, Douglas A. and Campbell, Jennifer R. and Catanzarite, Joseph and Chandrasekaran, Hema and Chaplin, William J. and Christiansen, Jessie L. and Christensen-Dalsgaard, Jørgen and Ciardi, David R. and Clarke, Bruce D. and Cochran, William D. and Dotson, Jessie L. and Doyle, Laurance R. and Duarte, Eduardo Seperuelo and Dunham, Edward W. and Dupree, Andrea K. and Endl, Michael and Fanson, James L. and Ford, Eric B. and Fujieh, Maura and Gautier III, Thomas N. and Geary, John C. and Gilliland, Ronald L and Girouard, Forrest R. and Gould, Alan and Haas, Michael R. and Henze, Christopher E. and Holman, Matthew J. and Howard, Andrew W. and Howell, Steve B. and Huber, Daniel and Hunter, Roger C. and Jenkins, Jon M. and Kjeldsen, Hans and Kolodziejczak, Jeffery and Larson, Kipp and Latham, David W. and Li, Jie and Mathur, Savita and Meibom, Søren and Middour, Chris and Morris, Robert L. and Morton, Timothy D. and Mullally, Fergal and Mullally, Susan E. and Pletcher, David and Prsa, Andrej and Quinn, Samuel N. and Quintana, Elisa V. and Ragozzine, Darin and Ramirez, Solange V. and Sanderfer, Dwight T. and Sasselov, Dimitar and Seader, Shawn E. and Shabram, Megan and Shporer, Avi and Smith, Jeffrey C. and Steffen, Jason H. and Still, Martin and Torres, Guillermo and Troeltzsch, John and Twicken, Joseph D. and Uddin, Akm Kamal and Van Cleve, Jeffrey E. and Voss, Janice and Weiss, Lauren M. and Welsh, William F. and Wohler, Bill and Zamudio, Khadeejah A},
  doi = {10.3847/1538-3881/abc418},
  fjournal = {The Astronomical Journal},
  issn = {1538-3881},
  journal = {AJ},
  month = {December},
  number = {1},
  pages = {36},
  publisher = {American Astronomical Society},
  title = {The Occurrence of Rocky Habitable-zone Planets around Solar-like Stars from Kepler Data},
  url = {https://doi.org/10.3847/1538-3881/abc418},
  volume = {161},
  year = {2020}
}

@article{Pollacco2006a,
  author = {Pollacco, D. L. and Skillen, I. and Cameron, A. Collier and Christian, D. J. and Hellier, C. and Irwin, J. and Lister, T. A. and Street, R. A. and West, R. G. and Anderson, D. and Clarkson, W. I. and Deeg, H. and Enoch, B. and Evans, A. and Fitzsimmons, A. and Haswell, C. A. and Hodgkin, S. and Horne, K. and Kane, S. R. and Keenan, F. P. and Maxted, P. F. L. and Norton, A. J. and Osborne, J. and Parley, N. R. and Ryans, R. S. I. and Smalley, B. and Wheatley, P. J. and Wilson, D. M.},
  doi = {10.1086/508556},
  fjournal = {Publications of the Astronomical Society of the Pacific},
  issn = {1538-3873},
  journal = {PASP},
  month = {October},
  number = {848},
  pages = {1407--1418},
  publisher = {IOP Publishing},
  title = {The WASP Project and the SuperWASP Cameras},
  url = {https://doi.org/10.1086/508556},
  volume = {118},
  year = {2006}
}

@article{pepperKELTSouthTelescope12012,
	author = {Pepper, Joshua and Kuhn, Rudolf B. and Siverd, Robert and James, David and Stassun, Keivan},
	doi = {10.1086/665044},
	fjournal = {Publications of the Astronomical Society of the Pacific},
    journal = {PASP},
	month = {mar},
	number = {913},
	pages = {230},
	publisher = {University of Chicago Press},
	title = {The KELT-South Telescope1},
	url = {https://doi.org/10.1086/665044},
	volume = {124},
	year = {2012}
}

@misc{Ge2022a,
  title={ET White Paper: To Find the First Earth 2.0}, 
  author={Jian Ge and Hui Zhang and Weicheng Zang and Hongping Deng and Shude Mao and Ji-Wei Xie and Hui-Gen Liu and Ji-Lin Zhou and Kevin Willis and Chelsea Huang and Steve B. Howell and Fabo Feng and Jiapeng Zhu and Xinyu Yao and Beibei Liu and Masataka Aizawa and Wei Zhu and Ya-Ping Li and Bo Ma and Quanzhi Ye and Jie Yu and Maosheng Xiang and Cong Yu and Shangfei Liu and Ming Yang and Mu-Tian Wang and Xian Shi and Tong Fang and Weikai Zong and Jinzhong Liu and Yu Zhang and Liyun Zhang and Kareem El-Badry and Rongfeng Shen and Pak-Hin Thomas Tam and Zhecheng Hu and Yanlv Yang and Yuan-Chuan Zou and Jia-Li Wu and Wei-Hua Lei and Jun-Jie Wei and Xue-Feng Wu and Tian-Rui Sun and Fa-Yin Wang and Bin-Bin Zhang and Dong Xu and Yuan-Pei Yang and Wen-Xiong Li and Dan-Feng Xiang and Xiaofeng Wang and Tinggui Wang and Bing Zhang and Peng Jia and Haibo Yuan and Jinghua Zhang and Sharon Xuesong Wang and Tianjun Gan and Wei Wang and Yinan Zhao and Yujuan Liu and Chuanxin Wei and Yanwu Kang and Baoyu Yang and Chao Qi and Xiaohua Liu and Quan Zhang and Yuji Zhu and Dan Zhou and Congcong Zhang and Yong Yu and Yongshuai Zhang and Yan Li and Zhenghong Tang and Chaoyan Wang and Fengtao Wang and Wei Li and Pengfei Cheng and Chao Shen and Baopeng Li and Yue Pan and Sen Yang and Wei Gao and Zongxi Song and Jian Wang and Hongfei Zhang and Cheng Chen and Hui Wang and Jun Zhang and Zhiyue Wang and Feng Zeng and Zhenhao Zheng and Jie Zhu and Yingfan Guo and Yihao Zhang and Yudong Li and Lin Wen and Jie Feng and Wen Chen and Kun Chen and Xingbo Han and Yingquan Yang and Haoyu Wang and Xuliang Duan and Jiangjiang Huang and Hong Liang and Shaolan Bi and Ning Gai and Zhishuai Ge and Zhao Guo and Yang Huang and Gang Li and Haining Li and Tanda Li and Yuxi and Lu and Hans-Walter Rix and Jianrong Shi and Fen Song and Yanke Tang and Yuan-Sen Ting and Tao Wu and Yaqian Wu and Taozhi Yang and Qing-Zhu Yin and Andrew Gould and Chung-Uk Lee and Subo Dong and Jennifer C. Yee and Yossi Shvartzvald and Hongjing Yang and Renkun Kuang and Jiyuan Zhang and Shilong Liao and Zhaoxiang Qi and Jun Yang and Ruisheng Zhang and Chen Jiang and Jian-Wen Ou and Yaguang Li and Paul Beck and Timothy R. Bedding and Tiago L. Campante and William J. Chaplin and Jørgen Christensen-Dalsgaard and Rafael A. García and Patrick Gaulme and Laurent Gizon and Saskia Hekker and Daniel Huber and Shourya Khanna and Yan Li and Savita Mathur and Andrea Miglio and Benoît Mosser and J. M. Joel Ong and Ângela R. G. Santos and Dennis Stello and Dominic M. Bowman and Mariel Lares-Martiz and Simon Murphy and Jia-Shu Niu and Xiao-Yu Ma and László Molnár and Jian-Ning Fu and Peter De Cat and Jie Su and the ET consortium},
  year={2022},
  eprint={2206.06693},
  archivePrefix={arXiv},
  primaryClass={astro-ph.IM},
  url={https://arxiv.org/abs/2206.06693}, 
}

@article{Ge2022b,
  author = {Ge, Jian and Zhang, Hui and Deng, Hongping and Howell, Steve B.},
  doi = {10.1016/j.xinn.2022.100271},
  fjournal = {The Innovation},
  issn = {2666-6758},
  journal = {The Innovation},
  month = {July},
  number = {4},
  pages = {100271},
  publisher = {Elsevier BV},
  title = {The ET mission to search for earth 2.0s},
  url = {https://doi.org/10.1016/j.xinn.2022.100271},
  volume = {3},
  year = {2022}
}

@inproceedings{Ge2022c,
  title        = {The Earth 2.0 Space Mission for Detecting Earth-like Planets Around Solar Type Stars},
  author       = {Ge, J. and others},
  year         = {2022},
  booktitle    = {Space Telescopes and Instrumentation 2022: Optical, Infrared, and Millimeter Wave},
  editor       = {Coyle, L. E. and Matsuura, S. and Perrin, M. D.},
  series       = {Proc. SPIE},
  volume       = {12180},
  pages        = {1218015},
  doi          = {10.1117/12.2630656},
  url          = {https://doi.org/10.1117/12.2630656},
  publisher    = {SPIE},
  organization = {SPIE},
  langid       = {english}
}

@article{Ge2024a,
  author  = {Ge, Jian and Chen, Wen and Chen, Yonghe and Song, Zongxi and Wang, Jian and Zhang, Hui and Li, Yan and Zang, Weicheng and Zhou, Dan and Zhang, Yongshuai and Chen, Kun and Yang, Yingquan and Mao, Shude and Huang, Chelsea and Yao, Xinyu and Li, Xinglong and Jiang, Haijiao and Yu, Yong and Tang, Zhenghong and Dong, Feng and Gao, Wei and Zhang, Hongfei and Shen, Chao and Wang, Fengtao and Wei, Chuanxin and Yang, Baoyu and Li, Yudong and Wen, Lin and Zhang, Pengjun and Zhang, Congcong and Xie, Jiwei and Ma, Bo and Deng, Hongping and Liu, Huigen and Duan, Xuliang and Wang, Haoyu and Huang, Jiangjiang and Gao, Yang and Wang, Yifei and Wang, Lei and Qin, Genjian and Liu, Xinyu and Gao, Jie},
  title   = {Search for a Second Earth -- the Earth 2.0 (ET) Space Mission (in Chinese)},
  journal = {Chinese Journal of Space Science},
  volume  = {44},
  number  = {3},
  pages   = {400--424},
  year    = {2024},
  doi     = {10.11728/cjss2024.03.yg05},
  url     = {https://www.cjss.ac.cn/en/article/doi/10.11728/cjss2024.03.yg05}
}

@inproceedings{Ge2024b,
  title        = {The Earth 2.0 Mission: Space Telescopes and Instrumentation 2024},
  author       = {Ge, J. and others},
  year         = {2024},
  booktitle    = {Space Telescopes and Instrumentation 2024: Optical, Infrared, and Millimeter Wave},
  editor       = {Coyle, L. E. and Matsuura, S. and Perrin, M. D.},
  series       = {Proc. SPIE},
  volume       = {13092},
  pages        = {1309218},
  doi          = {10.1117/12.3018669},
  url          = {https://doi.org/10.1117/12.3018669},
  publisher    = {SPIE},
  organization = {SPIE},
  langid       = {english}
}

@article{Rauer2025,
  author = {Rauer, Heike and Aerts, Conny and Cabrera, Juan and Deleuil, Magali and Erikson, Anders and Gizon, Laurent and Goupil, Mariejo and Heras, Ana and Walloschek, Thomas and Lorenzo-Alvarez, Jose and Marliani, Filippo and Martin-Garcia, César and Mas-Hesse, J. Miguel and O’Rourke, Laurence and Osborn, Hugh and Pagano, Isabella and Piotto, Giampaolo and Pollacco, Don and Ragazzoni, Roberto and Ramsay, Gavin and Udry, Stéphane and Appourchaux, Thierry and Benz, Willy and Brandeker, Alexis and Güdel, Manuel and Janot-Pacheco, Eduardo and Kabath, Petr and Kjeldsen, Hans and Min, Michiel and Santos, Nuno and Smith, Alan and Suarez, Juan-Carlos and Werner, Stephanie C. and Aboudan, Alessio and Abreu, Manuel and Acuña, Lorena and Adams, Moritz and Adibekyan, Vardan and Affer, Laura and Agneray, François and Agnor, Craig and Aguirre Børsen-Koch, Victor and Ahmed, Saad and Aigrain, Suzanne and Al-Bahlawan, Ashraf and Alcacera Gil, Ma de los Angeles and Alei, Eleonora and Alencar, Silvia and Alexander, Richard and Alfonso-Garzón, Julia and Alibert, Yann and Allende Prieto, Carlos and Almeida, Leonardo and Alonso Sobrino, Roi and Altavilla, Giuseppe and Althaus, Christian and Alvarez Trujillo, Luis Alonso and Amarsi, Anish and Ammler-von Eiff, Matthias and Amôres, Eduardo and Andrade, Laerte and Antoniadis-Karnavas, Alexandros and António, Carlos and Aparicio del Moral, Beatriz and Appolloni, Matteo and Arena, Claudio and Armstrong, David and Aroca Aliaga, Jose and Asplund, Martin and Audenaert, Jeroen and Auricchio, Natalia and Avelino, Pedro and Baeke, Ann and Baillié, Kevin and Balado, Ana and Ballber Balagueró, Pau and Balestra, Andrea and Ball, Warrick and Ballans, Herve and Ballot, Jerome and Barban, Caroline and Barbary, Gaële and Barbieri, Mauro and Barceló Forteza, Sebastià and Barker, Adrian and Barklem, Paul and Barnes, Sydney and Barrado Navascues, David and Barragan, Oscar and Baruteau, Clément and Basu, Sarbani and Baudin, Frederic and Baumeister, Philipp and Bayliss, Daniel and Bazot, Michael and Beck, Paul G. and Belkacem, Kevin and Bellinger, Earl and Benatti, Serena and Benomar, Othman and Bérard, Diane and Bergemann, Maria and Bergomi, Maria and Bernardo, Pierre and Biazzo, Katia and Bignamini, Andrea and Bigot, Lionel and Billot, Nicolas and Binet, Martin and Biondi, David and Biondi, Federico and Birch, Aaron C. and Bitsch, Bertram and Bluhm Ceballos, Paz Victoria and Bódi, Attila and Bognár, Zsófia and Boisse, Isabelle and Bolmont, Emeline and Bonanno, Alfio and Bonavita, Mariangela and Bonfanti, Andrea and Bonfils, Xavier and Bonito, Rosaria and Bonomo, Aldo Stefano and Börner, Anko and Boro Saikia, Sudeshna and Borreguero Martín, Elisa and Borsa, Francesco and Borsato, Luca and Bossini, Diego and Bouchy, Francois and Boué, Gwenaël and Boufleur, Rodrigo and Boumier, Patrick and Bourrier, Vincent and Bowman, Dominic M. and Bozzo, Enrico and Bradley, Louisa and Bray, John and Bressan, Alessandro and Breton, Sylvain and Brienza, Daniele and Brito, Ana and Brogi, Matteo and Brown, Beverly and Brown, David J. A. and Brun, Allan Sacha and Bruno, Giovanni and Bruns, Michael and Buchhave, Lars A. and Bugnet, Lisa and Buldgen, Gaël and Burgess, Patrick and Busatta, Andrea and Busso, Giorgia and Buzasi, Derek and Caballero, José A. and Cabral, Alexandre and Cabrero Gomez, Juan-Francisco and Calderone, Flavia and Cameron, Robert and Cameron, Andrew and Campante, Tiago and Campos Gestal, Néstor and Canto Martins, Bruno Leonardo and Cara, Christophe and Carone, Ludmila and Carrasco, Josep Manel and Casagrande, Luca and Casewell, Sarah L. and Cassisi, Santi and Castellani, Marco and Castro, Matthieu and Catala, Claude and Catalán Fernández, Irene and Catelan, Márcio and Cegla, Heather and Cerruti, Chiara and Cessa, Virginie and Chadid, Merieme and Chaplin, William and Charpinet, Stephane and Chiappini, Cristina and Chiarucci, Simone and Chiavassa, Andrea and Chinellato, Simonetta and Chirulli, Giovanni and Christensen-Dalsgaard, Jørgen and Church, Ross and Claret, Antonio and Clarke, Cathie and Claudi, Riccardo and Clermont, Lionel and Coelho, Hugo and Coelho, Joao and Cogato, Fabrizio and Colomé, Josep and Condamin, Mathieu and Conde García, Fernando and Conseil, Simon and Corbard, Thierry and Correia, Alexandre C. M. and Corsaro, Enrico and Cosentino, Rosario and Costes, Jean and Cottinelli, Andrea and Covone, Giovanni and Creevey, Orlagh L. and Crida, Aurelien and Csizmadia, Szilard and Cunha, Margarida and Curry, Patrick and da Costa, Jefferson and da Silva, Francys and Dalal, Shweta and Damasso, Mario and Damiani, Cilia and Damiani, Francesco and das Chagas, Maria Liduina and Davies, Melvyn and Davies, Guy and Davies, Ben and Davison, Gary and de Almeida, Leandro and de Angeli, Francesca and de Barros, Susana Cristina Cabral and de CastroLeão, Izan and de Freitas, Daniel Brito and de Freitas, Marcia Cristina and De Martino, Domitilla and de Medeiros, José Renan and de Paula, Luiz Alberto and de Pedraza Gómez, Álvaro and de Plaa, Jelle and De Ridder, Joris and Deal, Morgan and Decin, Leen and Deeg, Hans and Degl’Innocenti, Scilla and Deheuvels, Sebastien and del Burgo, Carlos and Del Sordo, Fabio and Delgado-Mena, Elisa and Demangeon, Olivier and Denk, Tilmann and Derekas, Aliz and Desert, Jean-Michel and Desidera, Silvano and Dexet, Marc and Di Criscienzo, Marcella and Di Giorgio, Anna Maria and Di Mauro, Maria Pia and Diaz Rial, Federico Jose and Díaz-García, José-Javier and Dima, Marco and Dinuzzi, Giacomo and Dionatos, Odysseas and Distefano, Elisa and do Nascimento, Jose-Dias and Domingo, Albert and D’Orazi, Valentina and Dorn, Caroline and Doyle, Lauren and Duarte, Elena and Ducellier, Florent and Dumaye, Luc and Dumusque, Xavier and Dupret, Marc-Antoine and Eggenberger, Patrick and Ehrenreich, David and Eigmüller, Philipp and Eising, Johannes and Emilio, Marcelo and Eriksson, Kjell and Ermocida, Marco and Escate Giribaldi, Riano Isidoro and Eschen, Yoshi and Espinosa Yáñez, Lucía and Estrela, Inês and Evans, Dafydd Wyn and Fabbian, Damian and Fabrizio, Michele and Faria, João Pedro and Farina, Maria and Farinato, Jacopo and Feliz, Dax and Feltzing, Sofia and Fenouillet, Thomas and Fernández, Miguel and Ferrari, Lorenza and Ferraz-Mello, Sylvio and Fialho, Fabio and Fienga, Agnes and Figueira, Pedro and Fiori, Laura and Flaccomio, Ettore and Focardi, Mauro and Foley, Steve and Fontignie, Jean and Ford, Dominic and Fornazier, Karin and Forveille, Thierry and Fossati, Luca and Franca, Rodrigo de Marca and Franco da Silva, Lucas and Frasca, Antonio and Fridlund, Malcolm and Furlan, Marco and Gabler, Sarah-Maria and Gaido, Marco and Gallagher, Andrew and Gallego Sempere, Paloma I. and Galli, Emanuele and García, Rafael A. and García Hernández, Antonio and Garcia Munoz, Antonio and García-Vázquez, Hugo and Garrido Haba, Rafael and Gaulme, Patrick and Gauthier, Nicolas and Gehan, Charlotte and Gent, Matthew and Georgieva, Iskra and Ghigo, Mauro and Giana, Edoardo and Gill, Samuel and Girardi, Leo and Giuliatti Winter, Silvia and Giusi, Giovanni and Gomes da Silva, João and Gómez Zazo, Luis Jorge and Gomez-Lopez, Juan Manuel and González Hernández, Jonay Isai and Gonzalez Murillo, Kevin and Gonzalo Melchor, Alejandro and Gorius, Nicolas and Gouel, Pierre-Vincent and Goulty, Duncan and Granata, Valentina and Grenfell, John Lee and Grießbach, Denis and Grolleau, Emmanuel and Grouffal, Salomé and Grziwa, Sascha and Guarcello, Mario Giuseppe and Gueguen, Loïc and Guenther, Eike Wolf and Guilhem, Terrasa and Guillerot, Lucas and Guillot, Tristan and Guiot, Pierre and Guterman, Pascal and Gutiérrez, Antonio and Gutiérrez-Canales, Fernando and Hagelberg, Janis and Haldemann, Jonas and Hall, Cassandra and Handberg, Rasmus and Harrison, Ian and Harrison, Diana L. and Hasiba, Johann and Haswell, Carole A. and Hatalova, Petra and Hatzes, Artie and Haywood, Raphaelle and Hébrard, Guillaume and Heckes, Frank and Heiter, Ulrike and Hekker, Saskia and Heller, René and Helling, Christiane and Helminiak, Krzysztof and Hemsley, Simon and Heng, Kevin and Herbst, Konstantin and Hermans, Aline and Hermes, JJ and Hidalgo Torres, Nadia and Hinkel, Natalie and Hobbs, David and Hodgkin, Simon and Hofmann, Karl and Hojjatpanah, Saeed and Houdek, Günter and Huber, Daniel and Huesler, Joseph and Hui-Bon-Hoa, Alain and Huygen, Rik and Huynh, Duc-Dat and Iro, Nicolas and Irwin, Jonathan and Irwin, Mike and Izidoro, André and Jacquinod, Sophie and Jannsen, Nicholas Emborg and Janson, Markus and Jeszenszky, Harald and Jiang, Chen and Jimenez Mancebo, Antonio José and Jofre, Paula and Johansen, Anders and Johnston, Cole and Jones, Geraint and Kallinger, Thomas and Kálmán, Szilárd and Kanitz, Thomas and Karjalainen, Marie and Karjalainen, Raine and Karoff, Christoffer and Kawaler, Steven and Kawata, Daisuke and Keereman, Arnoud and Keiderling, David and Kennedy, Tom and Kenworthy, Matthew and Kerschbaum, Franz and Kidger, Mark and Kiefer, Flavien and Kintziger, Christian and Kislyakova, Kristina and Kiss, László and Klagyivik, Peter and Klahr, Hubert and Klevas, Jonas and Kochukhov, Oleg and Köhler, Ulrich and Kolb, Ulrich and Koncz, Alexander and Korth, Judith and Kostogryz, Nadiia and Kovács, Gábor and Kovács, József and Kozhura, Oleg and Krivova, Natalie and Kuĉinskas, Arūnas and Kuhlemann, Ilyas and Kupka, Friedrich and Laauwen, Wouter and Labiano, Alvaro and Lagarde, Nadege and Laget, Philippe and Laky, Gunter and Lam, Kristine Wai Fun and Lambrechts, Michiel and Lammer, Helmut and Lanza, Antonino Francesco and Lanzafame, Alessandro and Lares Martiz, Mariel and Laskar, Jacques and Latter, Henrik and Lavanant, Tony and Lawrenson, Alastair and Lazzoni, Cecilia and Lebre, Agnes and Lebreton, Yveline and Lecavelier des Etangs, Alain and Lee, Katherine and Leinhardt, Zoe and Leleu, Adrien and Lendl, Monika and Leto, Giuseppe and Levillain, Yves and Libert, Anne-Sophie and Lichtenberg, Tim and Ligi, Roxanne and Lignieres, Francois and Lillo-Box, Jorge and Linsky, Jeffrey and Liu, John Scige and Loidolt, Dominik and Longval, Yuying and Lopes, Ilídio and Lorenzani, Andrea and Ludwig, Hans-Guenter and Lund, Mikkel and Lundkvist, Mia Sloth and Luri, Xavier and Maceroni, Carla and Madden, Sean and Madhusudhan, Nikku and Maggio, Antonio and Magliano, Christian and Magrin, Demetrio and Mahy, Laurent and Maibaum, Olaf and Malac-Allain, LeeRoy and Malapert, Jean-Christophe and Malavolta, Luca and Maldonado, Jesus and Mamonova, Elena and Manchon, Louis and Manjón, Andres and Mann, Andrew and Mantovan, Giacomo and Marafatto, Luca and Marconi, Marcella and Mardling, Rosemary and Marigo, Paola and Marinoni, Silvia and Marques, Rico and Marques, Joao Pedro and Marrese, Paola Maria and Marshall, Douglas and Martínez Perales, Silvia and Mary, David and Marzari, Francesco and Masana, Eduard and Mascher, Andrina and Mathis, Stéphane and Mathur, Savita and Martín Vodopivec, Iris and Mattiuci Figueiredo, Ana Carolina and Maxted, Pierre F. L. and Mazeh, Tsevi and Mazevet, Stephane and Mazzei, Francesco and McCormac, James and McMillan, Paul and Menou, Lucas and Merle, Thibault and Meru, Farzana and Mesa, Dino and Messina, Sergio and Mészáros, Szabolcs and Meunier, Nadége and Meunier, Jean-Charles and Micela, Giuseppina and Michaelis, Harald and Michel, Eric and Michielsen, Mathias and Michtchenko, Tatiana and Miglio, Andrea and Miguel, Yamila and Milligan, David and Mirouh, Giovanni and Mitchell, Morgan and Moedas, Nuno and Molendini, Francesca and Molnár, László and Mombarg, Joey and Montalban, Josefina and Montalto, Marco and Monteiro, Mário J. P. F. G. and Montoro Sánchez, Francisco and Morales, Juan Carlos and Morales-Calderon, Maria and Morbidelli, Alessandro and Mordasini, Christoph and Moreau, Chrystel and Morel, Thierry and Morello, Giuseppe and Morin, Julien and Mortier, Annelies and Mosser, Benoît and Mourard, Denis and Mousis, Olivier and Moutou, Claire and Mowlavi, Nami and Moya, Andrés and Muehlmann, Prisca and Muirhead, Philip and Munari, Matteo and Musella, Ilaria and Mustill, Alexander James and Nardetto, Nicolas and Nardiello, Domenico and Narita, Norio and Nascimbeni, Valerio and Nash, Anna and Neiner, Coralie and Nelson, Richard P. and Nettelmann, Nadine and Nicolini, Gianalfredo and Nielsen, Martin and Niemi, Sami-Matias and Noack, Lena and Noels-Grotsch, Arlette and Noll, Anthony and Norazman, Azib and Norton, Andrew J. and Nsamba, Benard and Ofir, Aviv and Ogilvie, Gordon and Olander, Terese and Olivetto, Christian and Olofsson, Göran and Ong, Joel and Ortolani, Sergio and Oshagh, Mahmoudreza and Ottacher, Harald and Ottensamer, Roland and Ouazzani, Rhita-Maria and Paardekooper, Sijme-Jan and Pace, Emanuele and Pajas, Miriam and Palacios, Ana and Palandri, Gaelle and Palle, Enric and Paproth, Carsten and Parro, Vanderlei and Parviainen, Hannu and Pascual Granado, Javier and Passegger, Vera Maria and Pastor-Morales, Carmen and Pätzold, Martin and Pedersen, May Gade and Pena Hidalgo, David and Pepe, Francesco and Pereira, Filipe and Persson, Carina M. and Pertenais, Martin and Peter, Gisbert and Petit, Antoine C. and Petit, Pascal and Pezzuto, Stefania and Pichierri, Gabriele and Pietrinferni, Adriano and Pinheiro, Fernando and Pinsonneault, Marc and Plachy, Emese and Plasson, Philippe and Plez, Bertrand and Poppenhaeger, Katja and Poretti, Ennio and Portaluri, Elisa and Portell, Jordi and Porto de Mello, Gustavo Frederico and Poyatos, Julien and Pozuelos, Francisco J. and Prada Moroni, Pier Giorgio and Pricopi, Dumitru and Prisinzano, Loredana and Quade, Matthias and Quirrenbach, Andreas and Rabanal Reina, Julio Arturo and Rabello Soares, Maria Cristina and Raimondo, Gabriella and Rainer, Monica and Ramón Rodón, Jose and Ramón-Ballesta, Alejandro and Ramos Zapata, Gonzalo and Rätz, Stefanie and Rauterberg, Christoph and Redman, Bob and Redmer, Ronald and Reese, Daniel and Regibo, Sara and Reiners, Ansgar and Reinhold, Timo and Renie, Christian and Ribas, Ignasi and Ribeiro, Sergio and Ricciardi, Thiago Pereira and Rice, Ken and Richard, Olivier and Riello, Marco and Rieutord, Michel and Ripepi, Vincenzo and Rixon, Guy and Rockstein, Steve and Rodón Ortiz, José Ramón and Rodrigo Rodríguez, María Teresa and Rodríguez Amor, Alberto and Rodríguez Díaz, Luisa Fernanda and Rodriguez Garcia, Juan Pablo and Rodriguez-Gomez, Julio and Roehlly, Yannick and Roig, Fernando and Rojas-Ayala, Bárbara and Rolf, Tobias and Rørsted, Jakob Lysgaard and Rosado, Hugo and Rosotti, Giovanni and Roth, Olivier and Roth, Markus and Rousseau, Alex and Roxburgh, Ian and Roy, Fabrice and Royer, Pierre and Ruane, Kirk and Rufini Mastropasqua, Sergio and Ruiz de Galarreta, Claudia and Russi, Andrea and Saar, Steven and Saillenfest, Melaine and Salaris, Maurizio and Salmon, Sebastien and Saltas, Ippocratis and Samadi, Réza and Samadi, Aunia and Samra, Dominic and Sanches da Silva, Tiago and Sánchez Carrasco, Miguel Andrés and Santerne, Alexandre and Santiago Pé, Amaia and Santoli, Francesco and Santos, Ängela R. G. and Sanz Mesa, Rosario and Sarro, Luis Manuel and Scandariato, Gaetano and Schäfer, Martin and Schlafly, Edward and Schmider, François-Xavier and Schneider, Jean and Schou, Jesper and Schunker, Hannah and Schwarzkopf, Gabriel Jörg and Serenelli, Aldo and Seynaeve, Dries and Shan, Yutong and Shapiro, Alexander and Shipman, Russel and Sicilia, Daniela and Sierra sanmartin, Maria Angeles and Sigot, Axelle and Silliman, Kyle and Silvotti, Roberto and Simon, Attila E. and Simoyama Napoli, Ricardo and Skarka, Marek and Smalley, Barry and Smiljanic, Rodolfo and Smit, Samuel and Smith, Alexis and Smith, Leigh and Snellen, Ignas and Sódor, Ádám and Sohl, Frank and Solanki, Sami K. and Sortino, Francesca and Sousa, Sérgio and Southworth, John and Souto, Diogo and Sozzetti, Alessandro and Stamatellos, Dimitris and Stassun, Keivan and Steller, Manfred and Stello, Dennis and Stelzer, Beate and Stiebeler, Ulrike and Stokholm, Amalie and Storelvmo, Trude and Strassmeier, Klaus and Strøm, Paul Anthony and Strugarek, Antoine and Sulis, Sophia and Švanda, Michal and Szabados, László and Szabó, Róbert and Szabó, Gyula M. and Szuszkiewicz, Ewa and Talens, Geert Jan and Teti, Daniele and Theisen, Tom and Thévenin, Frédéric and Thoul, Anne and Tiphene, Didier and Titz-Weider, Ruth and Tkachenko, Andrew and Tomecki, Daniel and Tonfat, Jorge and Tosi, Nicola and Trampedach, Regner and Traven, Gregor and Triaud, Amaury and Trønnes, Reidar and Tsantaki, Maria and Tschentscher, Matthias and Turin, Arnaud and Tvaruzka, Adam and Ulmer, Bernd and Ulmer-Moll, Solène and Ulusoy, Ceren and Umbriaco, Gabriele and Valencia, Diana and Valentini, Marica and Valio, Adriana and Valverde Guijarro, Ángel Luis and Van Eylen, Vincent and Van Grootel, Valerie and van Kempen, Tim A. and Van Reeth, Timothy and Van Zelst, Iris and Vandenbussche, Bart and Vasiliou, Konstantinos and Vasilyev, Valeriy and Vaz de Mascarenhas, David and Vazan, Allona and Vela Nunez, Marina and Velloso, Eduardo Nunes and Ventura, Rita and Ventura, Paolo and Venturini, Julia and Vera Trallero, Isabel and Veras, Dimitri and Verdugo, Eva and Verma, Kuldeep and Vibert, Didier and Vicanek Martinez, Tobias and Vida, Krisztián and Vigan, Arthur and Villacorta, Antonio and Villaver, Eva and Villaverde Aparicio, Marcos and Viotto, Valentina and Vorobyov, Eduard and Vorontsov, Sergey and Wagner, Frank W. and Walton, Nicholas and Walton, Dave and Wang, Haiyang and Waters, Rens and Watson, Christopher and Wedemeyer, Sven and Weeks, Angharad and Weingrill, Jörg and Weiss, Annita and Wendler, Belinda and West, Richard and Westerdorff, Karsten and Westphal, Pierre-Amaury and Wheatley, Peter and White, Tim and Whittaker, Amadou and Wickhusen, Kai and Wilson, Thomas and Windsor, James and Winter, Othon and Winther, Mark Lykke and Winton, Alistair and Witteck, Ulrike and Witzke, Veronika and Woitke, Peter and Wolter, David and Wuchterl, Günther and Wyatt, Mark and Yang, Dan and Yu, Jie and Zanmar Sanchez, Ricardo and Zapatero Osorio, María Rosa and Zechmeister, Mathias and Zhou, Yixiao and Ziemke, Claas and Zwintz, Konstanze and Böhm, Torsten and Dansac, Léo Michel},
  doi = {10.1007/s10686-025-09985-9},
  fjournal = {Experimental Astronomy},
  issn = {1572-9508},
  journal = {Experimental Astronomy},
  month = {April},
  number = {3},
  publisher = {Springer Science and Business Media LLC},
  title = {The PLATO mission},
  url = {https://doi.org/10.1007/s10686-025-09985-9},
  volume = {59},
  year = {2025}
}

@article{auvergneCoRoTSatelliteFlight2009,
  author = {Auvergne, M. and Bodin, P. and Boisnard, L. and Buey, J.-T. and Chaintreuil, S. and Epstein, G. and Jouret, M. and Lam-Trong, T. and Levacher, P. and Magnan, A. and Perez, R. and Plasson, P. and Plesseria, J. and Peter, G. and Steller, M. and Tiphène, D. and Baglin, A. and Agogué, P. and Appourchaux, T. and Barbet, D. and Beaufort, T. and Bellenger, R. and Berlin, R. and Bernardi, P. and Blouin, D. and Boumier, P. and Bonneau, F. and Briet, R. and Butler, B. and Cautain, R. and Chiavassa, F. and Costes, V. and Cuvilho, J. and Cunha-Parro, V. and De Oliveira Fialho, F. and Decaudin, M. and Defise, J.-M. and Djalal, S. and Docclo, A. and Drummond, R. and Dupuis, O. and Exil, G. and Fauré, C. and Gaboriaud, A. and Gamet, P. and Gavalda, P. and Grolleau, E. and Gueguen, L. and Guivarc’h, V. and Guterman, P. and Hasiba, J. and Huntzinger, G. and Hustaix, H. and Imbert, C. and Jeanville, G. and Johlander, B. and Jorda, L. and Journoud, P. and Karioty, F. and Kerjean, L. and Lafond, L. and Lapeyrere, V. and Landiech, P. and Larqué, T. and Laudet, P. and Le Merrer, J. and Leporati, L. and Leruyet, B. and Levieuge, B. and Llebaria, A. and Martin, L. and Mazy, E. and Mesnager, J.-M. and Michel, J.-P. and Moalic, J.-P. and Monjoin, W. and Naudet, D. and Neukirchner, S. and Nguyen-Kim, K. and Ollivier, M. and Orcesi, J.-L. and Ottacher, H. and Oulali, A. and Parisot, J. and Perruchot, S. and Piacentino, A. and Pinheiro da Silva, L. and Platzer, J. and Pontet, B. and Pradines, A. and Quentin, C. and Rohbeck, U. and Rolland, G. and Rollenhagen, F. and Romagnan, R. and Russ, N. and Samadi, R. and Schmidt, R. and Schwartz, N. and Sebbag, I. and Smit, H. and Sunter, W. and Tello, M. and Toulouse, P. and Ulmer, B. and Vandermarcq, O. and Vergnault, E. and Wallner, R. and Waultier, G. and Zanatta, P.},
  doi = {10.1051/0004-6361/200810860},
  fjournal = {Astronomy \& Astrophysics},
  issn = {1432-0746},
  journal = {A\&A},
  month = {March},
  number = {1},
  pages = {411--424},
  publisher = {EDP Sciences},
  title = {The CoRoT satellite in flight: description and performance},
  url = {https://doi.org/10.1051/0004-6361/200810860},
  volume = {506},
  year = {2009}
}

@article{kochKEPLERMISSIONDESIGN2010,
  author = {Koch, David G. and Borucki, William J. and Basri, Gibor and Batalha, Natalie M. and Brown, Timothy M. and Caldwell, Douglas and Christensen-Dalsgaard, Jørgen and Cochran, William D. and DeVore, Edna and Dunham, Edward W. and Gautier, Thomas N. and Geary, John C. and Gilliland, Ronald L. and Gould, Alan and Jenkins, Jon and Kondo, Yoji and Latham, David W. and Lissauer, Jack J. and Marcy, Geoffrey and Monet, David and Sasselov, Dimitar and Boss, Alan and Brownlee, Donald and Caldwell, John and Dupree, Andrea K. and Howell, Steve B. and Kjeldsen, Hans and Meibom, Søren and Morrison, David and Owen, Tobias and Reitsema, Harold and Tarter, Jill and Bryson, Stephen T. and Dotson, Jessie L. and Gazis, Paul and Haas, Michael R. and Kolodziejczak, Jeffrey and Rowe, Jason F. and Van Cleve, Jeffrey E. and Allen, Christopher and Chandrasekaran, Hema and Clarke, Bruce D. and Li, Jie and Quintana, Elisa V. and Tenenbaum, Peter and Twicken, Joseph D. and Wu, Hayley},
  doi = {10.1088/2041-8205/713/2/l79},
  fjournal = {The Astrophysical Journal},
  issn = {2041-8213},
  journal = {ApJ},
  month = {March},
  number = {2},
  pages = {L79--L86},
  publisher = {American Astronomical Society},
  title = {KEPLER MISSION DESIGN, REALIZED PHOTOMETRIC PERFORMANCE, AND EARLY SCIENCE},
  url = {https://doi.org/10.1088/2041-8205/713/2/l79},
  volume = {713},
  year = {2010}
}

@article{Howell2014,
  author = {Howell, Steve B. and Sobeck, Charlie and Haas, Michael and Still, Martin and Barclay, Thomas and Mullally, Fergal and Troeltzsch, John and Aigrain, Suzanne and Bryson, Stephen T. and Caldwell, Doug and Chaplin, William J. and Cochran, William D. and Huber, Daniel and Marcy, Geoffrey W. and Miglio, Andrea and Najita, Joan R. and Smith, Marcie and Twicken, J. D. and Fortney, Jonathan J.},
  doi = {10.1086/676406},
  fjournal = {Publications of the Astronomical Society of the Pacific},
  issn = {1538-3873},
  journal = {PASP},
  month = {April},
  number = {938},
  pages = {398--408},
  publisher = {IOP Publishing},
  title = {The K2 Mission: Characterization and Early Results},
  url = {https://doi.org/10.1086/676406},
  volume = {126},
  year = {2014}
}

@article{rickerTransitingExoplanetSurvey2014,
  author = {Ricker, George R. and Winn, Joshua N. and Vanderspek, Roland and Latham, David W. and Bakos, Gáspár Á. and Bean, Jacob L. and Berta-Thompson, Zachory K. and Brown, Timothy M. and Buchhave, Lars and Butler, Nathaniel R. and Butler, R. Paul and Chaplin, William J. and Charbonneau, David and Christensen-Dalsgaard, Jørgen and Clampin, Mark and Deming, Drake and Doty, John and De Lee, Nathan and Dressing, Courtney and Dunham, Edward W. and Endl, Michael and Fressin, Francois and Ge, Jian and Henning, Thomas and Holman, Matthew J. and Howard, Andrew W. and Ida, Shigeru and Jenkins, Jon M. and Jernigan, Garrett and Johnson, John Asher and Kaltenegger, Lisa and Kawai, Nobuyuki and Kjeldsen, Hans and Laughlin, Gregory and Levine, Alan M. and Lin, Douglas and Lissauer, Jack J. and MacQueen, Phillip and Marcy, Geoffrey and McCullough, Peter R. and Morton, Timothy D. and Narita, Norio and Paegert, Martin and Palle, Enric and Pepe, Francesco and Pepper, Joshua and Quirrenbach, Andreas and Rinehart, Stephen A. and Sasselov, Dimitar and Sato, Bun’ei and Seager, Sara and Sozzetti, Alessandro and Stassun, Keivan G. and Sullivan, Peter and Szentgyorgyi, Andrew and Torres, Guillermo and Udry, Stephane and Villasenor, Joel},
  doi = {10.1117/1.jatis.1.1.014003},
  fjournal = {Journal of Astronomical Telescopes, Instruments, and Systems},
  issn = {2329-4124},
  journal = {JATIS},
  month = {October},
  number = {1},
  pages = {014003},
  publisher = {SPIE-Intl Soc Optical Eng},
  title = {Transiting Exoplanet Survey Satellite},
  url = {https://doi.org/10.1117/1.jatis.1.1.014003},
  volume = {1},
  year = {2014}
}

@article{catalaPLATOPLAnetaryTransits2009,
  title = {PLATO: PLAnetary Transits and Oscillations of Stars},
  author = {Catala, Claude and {The PLATO Consortium}},
  year = {2009},
  fjournal = {Experimental Astronomy},
  journal = {Exp Astron},
  volume = {23},
  pages = {329--356},
  doi = {10.1007/s10686-008-9122-9},
  url = {https://doi.org/10.1007/s10686-008-9122-9},
  langid = {english}
}

@article{Carter2013,
  author = {Carter, Joshua A. and Agol, Eric},
  doi = {10.1088/0004-637x/765/2/132},
  fjournal = {The Astrophysical Journal},
  issn = {1538-4357},
  journal = {ApJ},
  month = {February},
  number = {2},
  pages = {132},
  publisher = {American Astronomical Society},
  title = {THE QUASIPERIODIC AUTOMATED TRANSIT SEARCH ALGORITHM},
  url = {https://doi.org/10.1088/0004-637x/765/2/132},
  volume = {765},
  year = {2013}
}

@article{kovacsBoxfittingAlgorithmSearch2002,
  author = {Kovács, G. and Zucker, S. and Mazeh, T.},
  doi = {10.1051/0004-6361:20020802},
  fjournal = {Astronomy \& Astrophysics},
  issn = {1432-0746},
  journal = {A\&A},
  month = {July},
  number = {1},
  pages = {369--377},
  publisher = {EDP Sciences},
  title = {A box-fitting algorithm in the search for periodic transits},
  url = {https://doi.org/10.1051/0004-6361:20020802},
  volume = {391},
  year = {2002}
}

@article{hippkeOptimizedTransitDetection2019,
  author = {Hippke, Michael and Heller, René},
  doi = {10.1051/0004-6361/201834672},
  fjournal = {Astronomy \& Astrophysics},
  issn = {1432-0746},
  journal = {A\&A},
  month = {February},
  pages = {A39},
  publisher = {EDP Sciences},
  title = {Optimized transit detection algorithm to search for periodic transits of small planets},
  url = {https://doi.org/10.1051/0004-6361/201834672},
  volume = {623},
  year = {2019}
}

@article{thompsonMACHINELEARNINGTECHNIQUE2015,
  author = {Thompson, Susan E. and Mullally, Fergal and Coughlin, Jeff and Christiansen, Jessie L. and Henze, Christopher E. and Haas, Michael R. and Burke, Christopher J.},
  doi = {10.1088/0004-637x/812/1/46},
  fjournal = {The Astrophysical Journal},
  issn = {1538-4357},
  journal = {ApJ},
  month = {October},
  number = {1},
  pages = {46},
  publisher = {American Astronomical Society},
  title = {A MACHINE LEARNING TECHNIQUE TO IDENTIFY TRANSIT SHAPED SIGNALS},
  url = {https://doi.org/10.1088/0004-637x/812/1/46},
  volume = {812},
  year = {2015}
}

@article{Mullally2016,
  author = {Mullally, F. and Coughlin, Jeffery L. and Thompson, Susan E. and Christiansen, Jessie and Burke, Christopher and Clarke, Bruce D. and Haas, Michael R.},
  doi = {10.1088/1538-3873/128/965/074502},
  fjournal = {Publications of the Astronomical Society of the Pacific},
  issn = {1538-3873},
  journal = {PASP},
  month = {June},
  number = {965},
  pages = {074502},
  publisher = {IOP Publishing},
  title = {Identifying False Alarms in theKeplerPlanet Candidate Catalog},
  url = {https://doi.org/10.1088/1538-3873/128/965/074502},
  volume = {128},
  year = {2016}
}

@article{armstrongTransitShapesSelforganizing2017,
  author = {Armstrong, D. J. and Pollacco, D. and Santerne, A.},
  doi = {10.1093/mnras/stw2881},
  fjournal = {Monthly Notices of the Royal Astronomical Society},
  issn = {1365-2966},
  journal = {MNRAS},
  month = {November},
  number = {3},
  pages = {2634--2642},
  publisher = {Oxford University Press (OUP)},
  title = {Transit shapes and self-organizing maps as a tool for ranking planetary candidates: application toKeplerandK2},
  url = {https://doi.org/10.1093/mnras/stw2881},
  volume = {465},
  year = {2016}
}

@article{Mislis2018a,
  author = {Mislis, D and Pyrzas, S and Alsubai, K A},
  doi = {10.1093/mnras/sty2361},
  fjournal = {Monthly Notices of the Royal Astronomical Society},
  issn = {1365-2966},
  journal = {MNRAS},
  month = {September},
  number = {2},
  pages = {1624--1630},
  publisher = {Oxford University Press (OUP)},
  title = {TSARDI: a Machine Learning data rejection algorithm for transiting exoplanet light curves},
  url = {https://doi.org/10.1093/mnras/sty2361},
  volume = {481},
  year = {2018}
}

@article{Armstrong2021a,
  author = {Armstrong, David J and Gamper, Jevgenij and Damoulas, Theodoros},
  doi = {10.1093/mnras/staa2498},
  fjournal = {Monthly Notices of the Royal Astronomical Society},
  issn = {1365-2966},
  journal = {MNRAS},
  month = {August},
  number = {4},
  pages = {5327--5344},
  publisher = {Oxford University Press (OUP)},
  title = {Exoplanet validation with machine learning: 50 new validated Kepler planets},
  url = {https://doi.org/10.1093/mnras/staa2498},
  volume = {504},
  year = {2020}
}

@article{Jenkins2012,
  title = {Auto-Vetting Transiting Planet Candidates Identified by the Kepler Pipeline},
  author = {Jenkins, Jon M. and McCauliff, Sean and Burke, Christopher and Seader, Shawn and Twicken, Joseph and Klaus, Todd and Sanderfer, Dwight and Srivastava, Ashok and Haas, Michael R.},
  year = {2012},
  fjournal = {Proceedings of the International Astronomical Union},
  journal = {Proc. IAU},
  volume = {8},
  pages = {94--99},
  doi = {10.1017/S1743921313012611},
  url = {https://www.cambridge.org/core/product/identifier/S1743921313012611/type/journal_article},
  langid = {english}
}

@article{mccauliffAUTOMATICCLASSIFICATIONKEPLER2015,
  author = {McCauliff, Sean D. and Jenkins, Jon M. and Catanzarite, Joseph and Burke, Christopher J. and Coughlin, Jeffrey L. and Twicken, Joseph D. and Tenenbaum, Peter and Seader, Shawn and Li, Jie and Cote, Miles},
  doi = {10.1088/0004-637x/806/1/6},
  fjournal = {The Astrophysical Journal},
  issn = {1538-4357},
  journal = {ApJ},
  month = {June},
  number = {1},
  pages = {6},
  publisher = {American Astronomical Society},
  title = {AUTOMATIC CLASSIFICATION OFKEPLERPLANETARY TRANSIT CANDIDATES},
  url = {https://doi.org/10.1088/0004-637x/806/1/6},
  volume = {806},
  year = {2015}
}

@article{Mislis2016,
  title = {Sidra: A Blind Algorithm for Signal Detection in Photometric Surveys},
  author = {Mislis, D. and Bachelet, E. and Alsubai, K. A. and Bramich, D. M. and Parley, N.},
  year = {2016},
  fjournal = {Monthly Notices of the Royal Astronomical Society},
  journal = {MNRAS},
  volume = {455},
  pages = {626--633},
  doi = {10.1093/mnras/stv2333},
  url = {https://academic.oup.com/mnras/article-lookup/doi/10.1093/mnras/stv2333},
  langid = {english}
}

@article{Melton2024a,
  author = {Melton, Elizabeth J. and Feigelson, Eric D. and Montalto, Marco and Caceres, Gabriel A. and Rosenswie, Andrew W. and Abelson, Cullen S.},
  doi = {10.3847/1538-3881/ad29f0},
  fjournal = {The Astronomical Journal},
  issn = {1538-3881},
  journal = {AJ},
  month = {April},
  number = {5},
  pages = {202},
  publisher = {American Astronomical Society},
  title = {DIAmante TESS AutoRegressive Planet Search (DTARPS). I. Analysis of 0.9 Million Light Curves},
  url = {https://doi.org/10.3847/1538-3881/ad29f0},
  volume = {167},
  year = {2024}
}

@article{panahiDetectionTransitingExoplanets2022,
  author = {Panahi, Aviad and Zucker, Shay and Clementini, Gisella and Audard, Marc and Binnenfeld, Avraham and Cusano, Felice and Evans, Dafydd Wyn and Gomel, Roy and Holl, Berry and Ilyin, Ilya and de Fombelle, Grégory Jevardat and Mazeh, Tsevi and Mowlavi, Nami and Nienartowicz, Krzysztof and Rimoldini, Lorenzo and Shahaf, Sahar and Eyer, Laurent},
  doi = {10.1051/0004-6361/202243497},
  fjournal = {Astronomy \& Astrophysics},
  issn = {1432-0746},
  journal = {A\&A},
  month = {July},
  pages = {A101},
  publisher = {EDP Sciences},
  title = {The detection of transiting exoplanets by Gaia},
  url = {https://doi.org/10.1051/0004-6361/202243497},
  volume = {663},
  year = {2022}
}

@article{malikExoplanetDetectionUsing2021,
  author = {Malik, Abhishek and Moster, Benjamin P and Obermeier, Christian},
  doi = {10.1093/mnras/stab3692},
  fjournal = {Monthly Notices of the Royal Astronomical Society},
  issn = {1365-2966},
  journal = {MNRAS},
  month = {December},
  publisher = {Oxford University Press (OUP)},
  title = {Exoplanet detection using machine learning},
  url = {https://doi.org/10.1093/mnras/stab3692},
  year = {2021}
}

@misc{pratyushAutomationTransitingExoplanet,
      title={Automation Of Transiting Exoplanet Detection, Identification and Habitability Assessment Using Machine Learning Approaches}, 
      author={Pawel Pratyush and Akshata Gangrade},
      year={2021},
      eprint={2112.03298},
      archivePrefix={arXiv},
      primaryClass={astro-ph.EP},
      url={https://arxiv.org/abs/2112.03298}, 
}

@article{shallueIdentifyingExoplanetsDeep2018,
  author = {Shallue, Christopher J. and Vanderburg, Andrew},
  doi = {10.3847/1538-3881/aa9e09},
  fjournal = {The Astronomical Journal},
  issn = {1538-3881},
  journal = {AJ},
  month = {January},
  number = {2},
  pages = {94},
  publisher = {American Astronomical Society},
  title = {Identifying Exoplanets with Deep Learning: A Five-planet Resonant Chain around Kepler-80 and an Eighth Planet around Kepler-90},
  url = {https://doi.org/10.3847/1538-3881/aa9e09},
  volume = {155},
  year = {2018}
}

@article{Pearson2018,
  author = {Pearson, Kyle A. and Palafox, Leon and Griffith, Caitlin A.},
  doi = {10.1093/mnras/stx2761},
  fjournal = {Monthly Notices of the Royal Astronomical Society},
  issn = {1365-2966},
  journal = {MNRAS},
  month = {October},
  number = {1},
  pages = {478--491},
  publisher = {Oxford University Press (OUP)},
  title = {Searching for exoplanets using artificial intelligence},
  url = {https://doi.org/10.1093/mnras/stx2761},
  volume = {474},
  year = {2017}
}

@article{dattiloIdentifyingExoplanetsDeep2019,
  author = {Dattilo, Anne and Vanderburg, Andrew and Shallue, Christopher J. and Mayo, Andrew W. and Berlind, Perry and Bieryla, Allyson and Calkins, Michael L. and Esquerdo, Gilbert A. and Everett, Mark E. and Howell, Steve B. and Latham, David W. and Scott, Nicholas J. and Yu, Liang},
  doi = {10.3847/1538-3881/ab0e12},
  fjournal = {The Astronomical Journal},
  issn = {1538-3881},
  journal = {AJ},
  month = {April},
  number = {5},
  pages = {169},
  publisher = {American Astronomical Society},
  title = {Identifying Exoplanets with Deep Learning. II. Two New Super-Earths Uncovered by a Neural Network in K2 Data},
  url = {https://doi.org/10.3847/1538-3881/ab0e12},
  volume = {157},
  year = {2019}
}

@article{Yu2019,
  author = {Yu, Liang and Vanderburg, Andrew and Huang, Chelsea and Shallue, Christopher J. and Crossfield, Ian J. M. and Gaudi, B. Scott and Daylan, Tansu and Dattilo, Anne and Armstrong, David J. and Ricker, George R. and Vanderspek, Roland K. and Latham, David W. and Seager, Sara and Dittmann, Jason and Doty, John P. and Glidden, Ana and Quinn, Samuel N.},
  doi = {10.3847/1538-3881/ab21d6},
  fjournal = {The Astronomical Journal},
  issn = {1538-3881},
  journal = {AJ},
  month = {June},
  number = {1},
  pages = {25},
  publisher = {American Astronomical Society},
  title = {Identifying Exoplanets with Deep Learning. III. Automated Triage and Vetting of TESS Candidates},
  url = {https://doi.org/10.3847/1538-3881/ab21d6},
  volume = {158},
  year = {2019}
}

@article{Tey2023,
  author = {Tey, Evan and Moldovan, Dan and Kunimoto, Michelle and Huang, Chelsea X. and Shporer, Avi and Daylan, Tansu and Muthukrishna, Daniel and Vanderburg, Andrew and Dattilo, Anne and Ricker, George R. and Seager, S.},
  doi = {10.3847/1538-3881/acad85},
  fjournal = {The Astronomical Journal},
  issn = {1538-3881},
  journal = {AJ},
  month = {February},
  number = {3},
  pages = {95},
  publisher = {American Astronomical Society},
  title = {Identifying Exoplanets with Deep Learning. V. Improved Light-curve Classification for TESS Full-frame Image Observations},
  url = {https://doi.org/10.3847/1538-3881/acad85},
  volume = {165},
  year = {2023}
}

@article{Ansdell2018,
  author = {Ansdell, Megan and Ioannou, Yani and Osborn, Hugh P. and Sasdelli, Michele and Smith, Jeffrey C. and Caldwell, Douglas and Jenkins, Jon M. and Räissi, Chedy and Angerhausen, Daniel},
  doi = {10.3847/2041-8213/aaf23b},
  fjournal = {The Astrophysical Journal Letters},
  issn = {2041-8213},
  journal = {ApJL},
  month = {December},
  number = {1},
  pages = {L7},
  publisher = {American Astronomical Society},
  title = {Scientific Domain Knowledge Improves Exoplanet Transit Classification with Deep Learning},
  url = {https://doi.org/10.3847/2041-8213/aaf23b},
  volume = {869},
  year = {2018}
}

@article{valizadeganExoMinerHighlyAccurate2022,
  author = {Valizadegan, Hamed and Martinho, Miguel J. S. and Wilkens, Laurent S. and Jenkins, Jon M. and Smith, Jeffrey C. and Caldwell, Douglas A. and Twicken, Joseph D. and Gerum, Pedro C. L. and Walia, Nikash and Hausknecht, Kaylie and Lubin, Noa Y. and Bryson, Stephen T. and Oza, Nikunj C.},
  doi = {10.3847/1538-4357/ac4399},
  fjournal = {The Astrophysical Journal},
  issn = {1538-4357},
  journal = {ApJ},
  month = {February},
  number = {2},
  pages = {120},
  publisher = {American Astronomical Society},
  title = {ExoMiner: A Highly Accurate and Explainable Deep Learning Classifier That Validates 301 New Exoplanets},
  url = {https://doi.org/10.3847/1538-4357/ac4399},
  volume = {926},
  year = {2022}
}

@article{Valizadegan2023a,
  author = {Valizadegan, Hamed and Martinho, Miguel J. S. and Jenkins, Jon M. and Caldwell, Douglas A. and Twicken, Joseph D. and Bryson, Stephen T.},
  doi = {10.3847/1538-3881/acd344},
  fjournal = {The Astronomical Journal},
  issn = {1538-3881},
  journal = {AJ},
  month = {June},
  number = {1},
  pages = {28},
  publisher = {American Astronomical Society},
  title = {Multiplicity Boost of Transit Signal Classifiers: Validation of 69 New Exoplanets using the Multiplicity Boost of ExoMiner},
  url = {https://doi.org/10.3847/1538-3881/acd344},
  volume = {166},
  year = {2023}
}

@article{Zucker2018,
  author = {Zucker, Shay and Giryes, Raja},
  doi = {10.3847/1538-3881/aaae05},
  fjournal = {The Astronomical Journal},
  issn = {1538-3881},
  journal = {AJ},
  month = {March},
  number = {4},
  pages = {147},
  publisher = {American Astronomical Society},
  title = {Shallow Transits—Deep Learning. I. Feasibility Study of Deep Learning to Detect Periodic Transits of Exoplanets},
  url = {https://doi.org/10.3847/1538-3881/aaae05},
  volume = {155},
  year = {2018}
}

@article{Alvarez2023,
  author = {Iglesias Álvarez, Santiago and Díez Alonso, Enrique and Sánchez Rodríguez, María Luisa and Rodríguez Rodríguez, Javier and Sánchez Lasheras, Fernando and de Cos Juez, Francisco Javier},
  doi = {10.3390/axioms12040348},
  fjournal = {Axioms},
  issn = {2075-1680},
  journal = {Axioms},
  month = {March},
  number = {4},
  pages = {348},
  publisher = {MDPI AG},
  title = {One-Dimensional Convolutional Neural Networks for Detecting Transiting Exoplanets},
  url = {https://doi.org/10.3390/axioms12040348},
  volume = {12},
  year = {2023}
}

@article{osbornRapidClassificationTESS2020,
  author = {Osborn, H. P. and Ansdell, M. and Ioannou, Y. and Sasdelli, M. and Angerhausen, D. and Caldwell, D. and Jenkins, J. M. and Räissi, C. and Smith, J. C.},
  doi = {10.1051/0004-6361/201935345},
  fjournal = {Astronomy \& Astrophysics},
  issn = {1432-0746},
  journal = {A\&A},
  month = {January},
  pages = {A53},
  publisher = {EDP Sciences},
  title = {Rapid classification of TESS planet candidates with convolutional neural networks},
  url = {https://doi.org/10.1051/0004-6361/201935345},
  volume = {633},
  year = {2020}
}

@article{Li2019,
  title = {\mkbibemph{Kepler} Data Validation II–Transit Model Fitting and Multiple-Planet Search},
  author = {Li, Jie and Tenenbaum, Peter and Twicken, Joseph D. and Burke, Christopher J. and Jenkins, Jon M. and Quintana, Elisa V. and Rowe, Jason F. and Seader, Shawn E.},
  year = {2019},
  fjournal = {Publications of the Astronomical Society of the Pacific},
  journal = {PASP},
  volume = {131},
  pages = {024506},
  doi = {10.1088/1538-3873/aaf44d},
  url = {https://iopscience.iop.org/article/10.1088/1538-3873/aaf44d},
  langid = {english}
}

@article{yehSearchingPossibleExoplanet2021,
  author = {Yeh, Li-Chin and Jiang, Ing-Guey},
  doi = {10.1088/1538-3873/abbb24},
  fjournal = {Publications of the Astronomical Society of the Pacific},
  issn = {1538-3873},
  journal = {PASP},
  month = {December},
  number = {1019},
  pages = {014401},
  publisher = {IOP Publishing},
  title = {Searching for Possible Exoplanet Transits from BRITE Data through a Machine Learning Technique},
  url = {https://doi.org/10.1088/1538-3873/abbb24},
  volume = {133},
  year = {2020}
}

@article{cuellarDeepLearningExoplanets2022,
  title = {Deep Learning Exoplanets Detection by Combining Real and Synthetic Data},
  author = {Cuéllar, Sara and Granados, Paulo and Fabregas, Ernesto and Curé, Michel and Vargas, Héctor and Dormido-Canto, Sebastián and Farias, Gonzalo},
  editor = {V E, Sathishkumar},
  year = {2022},
  fjournal = {PLOS ONE},
  journal = {PLoS ONE},
  volume = {17},
  pages = {e0268199},
  doi = {10.1371/journal.pone.0268199},
  url = {https://dx.plos.org/10.1371/journal.pone.0268199},
  langid = {english}
}

@article{chintarungruangchaiDetectingExoplanetTransits2019a,
  author = {Chintarungruangchai, Pattana and Jiang, Ing-Guey},
  doi = {10.1088/1538-3873/ab13d3},
  fjournal = {Publications of the Astronomical Society of the Pacific},
  issn = {1538-3873},
  journal = {PASP},
  month = {May},
  number = {1000},
  pages = {064502},
  publisher = {IOP Publishing},
  title = {Detecting Exoplanet Transits through Machine-learning Techniques with Convolutional Neural Networks},
  url = {https://doi.org/10.1088/1538-3873/ab13d3},
  volume = {131},
  year = {2019}
}

@article{cuiIdentifyLightcurveSignals2022,
  author = {Cui, Kaiming and Liu, Junjie and Feng, Fabo and Liu, Jifeng},
  doi = {10.3847/1538-3881/ac3482},
  fjournal = {The Astronomical Journal},
  issn = {1538-3881},
  journal = {AJ},
  month = {December},
  number = {1},
  pages = {23},
  publisher = {American Astronomical Society},
  title = {Identify Light-curve Signals with Deep Learning Based Object Detection Algorithm. I. Transit Detection},
  url = {https://doi.org/10.3847/1538-3881/ac3482},
  volume = {163},
  year = {2021}
}

@article{Brown_2011,
	author = {Brown, Timothy M. and Latham, David W. and Everett, Mark E. and Esquerdo, Gilbert A.},
	doi = {10.1088/0004-6256/142/4/112},
	fjournal = {The Astronomical Journal},
    journal = {AJ},
	month = {sep},
	number = {4},
	pages = {112},
	publisher = {The American Astronomical Society},
	title = {KEPLER INPUT CATALOG: PHOTOMETRIC CALIBRATION AND STELLAR CLASSIFICATION},
	url = {https://doi.org/10.1088/0004-6256/142/4/112},
	volume = {142},
	year = {2011}
}

@article{Dvash2022,
  author = {Dvash, Elad and Peleg, Yam and Zucker, Shay and Giryes, Raja},
  doi = {10.3847/1538-3881/ac5ea2},
  fjournal = {The Astronomical Journal},
  issn = {1538-3881},
  journal = {AJ},
  month = {April},
  number = {5},
  pages = {237},
  publisher = {American Astronomical Society},
  title = {Shallow Transits—Deep Learning. II. Identify Individual Exoplanetary Transits in Red Noise using Deep Learning},
  url = {https://doi.org/10.3847/1538-3881/ac5ea2},
  volume = {163},
  year = {2022}
}

@misc{aydoganExoplanetDetectionMachine2022,
      title={Exoplanet Detection by Machine Learning with Data Augmentation}, 
      author={Koray Aydoğan},
      year={2022},
      eprint={2211.15577},
      archivePrefix={arXiv},
      primaryClass={astro-ph.EP},
      url={https://arxiv.org/abs/2211.15577}, 
}

@article{salinasDistinguishingPlanetaryTransit2023,
  author = {Salinas, Helem and Pichara, Karim and Brahm, Rafael and Pérez-Galarce, Francisco and Mery, Domingo},
  doi = {10.1093/mnras/stad1173},
  fjournal = {Monthly Notices of the Royal Astronomical Society},
  issn = {1365-2966},
  journal = {MNRAS},
  month = {April},
  number = {3},
  pages = {3201--3216},
  publisher = {Oxford University Press (OUP)},
  title = {Distinguishing a planetary transit from false positives: a Transformer-based classification for planetary transit signals},
  url = {https://doi.org/10.1093/mnras/stad1173},
  volume = {522},
  year = {2023}
}

@misc{G2023,
      title={Identification and Classification of Exoplanets Using Machine Learning Techniques}, 
      author={Prithivraj G and Alka Kumari},
      year={2023},
      eprint={2305.09596},
      archivePrefix={arXiv},
      primaryClass={astro-ph.EP},
      url={https://arxiv.org/abs/2305.09596}, 
}

@article{Wang2024,
  author = {Wang, Kaitlyn and Ge, Jian and Willis, Kevin and Wang, Kevin and Zhao, Yinan and Hu, Quanquan},
  doi = {10.1093/mnras/stae2155},
  fjournal = {Monthly Notices of the Royal Astronomical Society},
  issn = {1365-2966},
  journal = {MNRAS},
  month = {September},
  number = {3},
  pages = {1913--1927},
  publisher = {Oxford University Press (OUP)},
  title = {Discovery of small ultra-short-period planets orbiting Kepler KG dwarfs with GPU phase folding and deep learning},
  url = {https://doi.org/10.1093/mnras/stae2155},
  volume = {534},
  year = {2024}
}

@article{Wang2024a,
  author = {Wang, Kaitlyn and Ge, Jian and Willis, Kevin and Wang, Kevin and Zhao, Yinan},
  doi = {10.1093/mnras/stae245},
  fjournal = {Monthly Notices of the Royal Astronomical Society},
  issn = {1365-2966},
  journal = {MNRAS},
  month = {January},
  number = {3},
  pages = {4053--4067},
  publisher = {Oxford University Press (OUP)},
  title = {The GPU phase folding and deep learning method for detecting exoplanet transits},
  url = {https://doi.org/10.1093/mnras/stae245},
  volume = {528},
  year = {2024}
}

@inproceedings{vaswaniAttentionAllYou2017,
  author={Vaswani, Ashish and Shazeer, Noam and Parmar, Niki and Uszkoreit, Jakob and Jones, Llion and Gomez, Aidan N. and Kaiser, {\L}ukasz and Polosukhin, Illia},
  title={Attention Is All You Need},
  booktitle={Advances in Neural Information Processing Systems},
  volume={30},
  year={2017}
}

@inproceedings{Longa,
  author = {Long, Jonathan and Shelhamer, Evan and Darrell, Trevor},
  booktitle = {Proceedings of the IEEE Conference on Computer Vision and Pattern Recognition (CVPR)},
  doi = {10.1109/CVPR.2015.7298965},
  langid = {english},
  pages = {3431--3440},
  publisher = {IEEE},
  title = {Fully Convolutional Networks for Semantic Segmentation},
  year = {2015}
}

@misc{Howard2017,
      title={MobileNets: Efficient Convolutional Neural Networks for Mobile Vision Applications}, 
      author={Andrew G. Howard and Menglong Zhu and Bo Chen and Dmitry Kalenichenko and Weijun Wang and Tobias Weyand and Marco Andreetto and Hartwig Adam},
      year={2017},
      eprint={1704.04861},
      archivePrefix={arXiv},
      primaryClass={cs.CV},
      url={https://arxiv.org/abs/1704.04861}, 
}

@inproceedings{Tan2020,
  title = 	 {{E}fficient{N}et: Rethinking Model Scaling for Convolutional Neural Networks},
  author =       {Tan, Mingxing and Le, Quoc},
  booktitle = 	 {Proceedings of the 36th International Conference on Machine Learning},
  pages = 	 {6105--6114},
  year = 	 {2019},
  editor = 	 {Chaudhuri, Kamalika and Salakhutdinov, Ruslan},
  volume = 	 {97},
  series = 	 {Proceedings of Machine Learning Research},
  month = 	 {09--15 Jun},
  publisher =    {PMLR},
  url = 	 {https://proceedings.mlr.press/v97/tan19a.html}
}

@inproceedings{Chollet2017,
  author = {Chollet, Fran{\c{c}}ois},
  booktitle = {Proceedings of the IEEE Conference on Computer Vision and Pattern Recognition (CVPR)},
  doi = {10.1109/CVPR.2017.195},
  langid = {english},
  pages = {1251--1258},
  publisher = {IEEE},
  title = {Xception: Deep Learning with Depthwise Separable Convolutions},
  year = {2017}
}

@article{Srivastava2014,
  title = {Dropout: A Simple Way to Prevent Neural Networks from Overfitting},
  author = {Srivastava, Nitish and Hinton, Geoffrey and Krizhevsky, Alex and Sutskever, Ilya and Salakhutdinov, Ruslan},
  journal = {Journal of Machine Learning Research},
  year = {2014},
  volume = {15},
  pages = {1929--1958}
}

@inproceedings{Ioffe2015,
  title = 	 {Batch Normalization: Accelerating Deep Network Training by Reducing Internal Covariate Shift},
  author = 	 {Ioffe, Sergey and Szegedy, Christian},
  booktitle = 	 {Proceedings of the 32nd International Conference on Machine Learning},
  pages = 	 {448--456},
  year = 	 {2015},
  editor = 	 {Bach, Francis and Blei, David},
  volume = 	 {37},
  series = 	 {Proceedings of Machine Learning Research},
  address = 	 {Lille, France},
  month = 	 {07--09 Jul},
  publisher =    {PMLR},
  url = 	 {https://proceedings.mlr.press/v37/ioffe15.html}
}

@inproceedings{loshchilovDecoupledWeightDecay2019,
  title={Decoupled Weight Decay Regularization},
  author={Ilya Loshchilov and Frank Hutter},
  booktitle={International Conference on Learning Representations (ICLR)},
  year={2019}
}

@inproceedings{Kingma2017,
  title={Adam: A Method for Stochastic Optimization},
  author={Diederik P. Kingma and Jimmy Ba},
  booktitle={International Conference on Learning Representations (ICLR)},
  year={2015},
}

@article{gondhalekarTCFPeriodogramHigh2023,
	author = {Gondhalekar, Yash and Feigelson, Eric D. and Caceres, Gabriel A. and Montalto, Marco and Saha, Snehanshu},
	doi = {10.3847/2041-8213/ad0844},
	fjournal = {The Astrophysical Journal Letters},
  journal = {ApJL},
	month = {dec},
	number = {2},
	pages = {L16},
	publisher = {The American Astronomical Society},
	title = {A Study of Two Periodogram Algorithms for Improving the Detection of Small Transiting Planets},
	url = {https://doi.org/10.3847/2041-8213/ad0844},
	volume = {959},
	year = {2023}
}

@article{Petigura2013,
  author = {Petigura, Erik A. and Marcy, Geoffrey W. and Howard, Andrew W.},
  doi = {10.1088/0004-637x/770/1/69},
  fjournal = {The Astrophysical Journal},
  issn = {1538-4357},
  journal = {ApJ},
  month = {May},
  number = {1},
  pages = {69},
  publisher = {American Astronomical Society},
  title = {A PLATEAU IN THE PLANET POPULATION BELOW TWICE THE SIZE OF EARTH},
  url = {https://doi.org/10.1088/0004-637x/770/1/69},
  volume = {770},
  year = {2013}
}

@misc{HuGPUTLS,
  title={GTLS: A GPU-accelerated method for periodic transit detection}, 
  author={Quanquan Hu and Jian Ge and Luoxi Jin and Kevin Willis},
  year={2026},
  eprint={2607.00348},
  archivePrefix={arXiv},
  primaryClass={astro-ph.IM},
  url={https://arxiv.org/abs/2607.00348}, 
}

@unpublished{Telesco2026,
  author = {Michael Telesco and Jian Ge and Kevin Willis and
            Chuanfei Dong and Jun Yang and Beibei Liu and Luoxi Jin},
  title  = {A scaled analog of the Solar System around a Sun-like star},
  year   = {2026},
  note   = {Submitted to Science}
}

@article{jenkinsOVERVIEWKEPLERSCIENCE2010,
  author = {Jenkins, Jon M. and Caldwell, Douglas A. and Chandrasekaran, Hema and Twicken, Joseph D. and Bryson, Stephen T. and Quintana, Elisa V. and Clarke, Bruce D. and Li, Jie and Allen, Christopher and Tenenbaum, Peter and Wu, Hayley and Klaus, Todd C. and Middour, Christopher K. and Cote, Miles T. and McCauliff, Sean and Girouard, Forrest R. and Gunter, Jay P. and Wohler, Bill and Sommers, Jeneen and Hall, Jennifer R. and Uddin, AKM K. and Wu, Michael S. and Bhavsar, Paresh A. and Van Cleve, Jeffrey and Pletcher, David L. and Dotson, Jessie A. and Haas, Michael R. and Gilliland, Ronald L. and Koch, David G. and Borucki, William J.},
  doi = {10.1088/2041-8205/713/2/l87},
  fjournal = {The Astrophysical Journal},
  issn = {2041-8213},
  journal = {ApJ},
  month = {March},
  number = {2},
  pages = {L87--L91},
  publisher = {American Astronomical Society},
  title = {OVERVIEW OF THE KEPLER SCIENCE PROCESSING PIPELINE},
  url = {https://doi.org/10.1088/2041-8205/713/2/l87},
  volume = {713},
  year = {2010}
}

@inproceedings{jenkinsTransitingPlanetSearch2010,
  title = {Transiting Planet Search in the Kepler Pipeline},
  booktitle = {SPIE Astronomical Telescopes + Instrumentation},
  author = {Jenkins, Jon M. and Chandrasekaran, Hema and McCauliff, Sean D. and Caldwell, Douglas A. and Tenenbaum, Peter and Li, Jie and Klaus, Todd C. and Cote, Miles T. and Middour, Christopher},
  editor = {Radziwill, Nicole M. and Bridger, Alan},
  year = {2010},
  pages = {77400D},
  address = {San Diego, California, USA},
  doi = {10.1117/12.856764},
  langid = {english}
}

@article{rauerPLATOMission2014,
	author = {Rauer, H. and Catala, C. and Aerts, C. and Appourchaux, T. and Benz, W. and Brandeker, A. and Christensen-Dalsgaard, J. and Deleuil, M. and Gizon, L. and Goupil, M. -J. and G{\"u}del, M. and Janot-Pacheco, E. and Mas-Hesse, M. and Pagano, I. and Piotto, G. and Pollacco, D. and Santos, {\.C}. and Smith, A. and Su{\'a}rez, J. -C. and Szab{\'o}, R. and Udry, S. and Adibekyan, V. and Alibert, Y. and Almenara, J. -M. and Amaro-Seoane, P. and Eiff, M. Ammler-von and Asplund, M. and Antonello, E. and Barnes, S. and Baudin, F. and Belkacem, K. and Bergemann, M. and Bihain, G. and Birch, A. C. and Bonfils, X. and Boisse, I. and Bonomo, A. S. and Borsa, F. and Brand{\~a}o, I. M. and Brocato, E. and Brun, S. and Burleigh, M. and Burston, R. and Cabrera, J. and Cassisi, S. and Chaplin, W. and Charpinet, S. and Chiappini, C. and Church, R. P. and Csizmadia, Sz. and Cunha, M. and Damasso, M. and Davies, M. B. and Deeg, H. J. and D{\'\i}az, R. F. and Dreizler, S. and Dreyer, C. and Eggenberger, P. and Ehrenreich, D. and Eigm{\"u}ller, P. and Erikson, A. and Farmer, R. and Feltzing, S. and Oliveira Fialho, F. de and Figueira, P. and Forveille, T. and Fridlund, M. and Garc{\'\i}a, R. A. and Giommi, P. and Giuffrida, G. and Godolt, M. and da Silva, J. Gomes and Granzer, T. and Grenfell, J. L. and Grotsch-Noels, A. and G{\"u}nther, E. and Haswell, C. A. and Hatzes, A. P. and H{\'e}brard, G. and Hekker, S. and Helled, R. and Heng, K. and Jenkins, J. M. and Johansen, A. and Khodachenko, M. L. and Kislyakova, K. G. and Kley, W. and Kolb, U. and Krivova, N. and Kupka, F. and Lammer, H. and Lanza, A. F. and Lebreton, Y. and Magrin, D. and Marcos-Arenal, P. and Marrese, P. M. and Marques, J. P. and Martins, J. and Mathis, S. and Mathur, S. and Messina, S. and Miglio, A. and Montalban, J. and Montalto, M. and P. F. G. Monteiro, M. J. and Moradi, H. and Moravveji, E. and Mordasini, C. and Morel, T. and Mortier, A. and Nascimbeni, V. and Nelson, R. P. and Nielsen, M. B. and Noack, L. and Norton, A. J. and Ofir, A. and Oshagh, M. and Ouazzani, R. -M. and P{\'a}pics, P. and Parro, V. C. and Petit, P. and Plez, B. and Poretti, E. and Quirrenbach, A. and Ragazzoni, R. and Raimondo, G. and Rainer, M. and Reese, D. R. and Redmer, R. and Reffert, S. and Rojas-Ayala, B. and Roxburgh, I. W. and Salmon, S. and Santerne, A. and Schneider, J. and Schou, J. and Schuh, S. and Schunker, H. and Silva-Valio, A. and Silvotti, R. and Skillen, I. and Snellen, I. and Sohl, F. and Sousa, S. G. and Sozzetti, A. and Stello, D. and Strassmeier, K. G. and {\v S}vanda, M. and Szab{\'o}, Gy. M. and Tkachenko, A. and Valencia, D. and Van Grootel, V. and Vauclair, S. D. and Ventura, P. and Wagner, F. W. and Walton, N. A. and Weingrill, J. and Werner, S. C. and Wheatley, P. J. and Zwintz, K.},
	doi = {10.1007/s10686-014-9383-4},
	id = {Rauer2014},
	isbn = {1572-9508},
	journal = {Experimental Astronomy},
	number = {1},
	pages = {249--330},
	title = {The PLATO 2.0 mission},
	url = {https://doi.org/10.1007/s10686-014-9383-4},
	volume = {38},
	year = {2014}
}

@article{Christiansen2025,
  author = {Christiansen, Jessie L. and McElroy, Douglas L. and Harbut, Marcy and Ciardi, David R. and Crane, Megan and Good, John and Hardegree-Ullman, Kevin K. and Kesseli, Aurora Y. and Lund, Michael B. and Lynn, Meca and Muthiar, Ananda and Nilsson, Ricky and Oluyide, Toba and Papin, Michael and Rivera, Amalia and Swain, Melanie and Susemiehl, Nicholas D. and Tam, Raymond and van Eyken, Julian and Beichman, Charles},
  doi = {10.3847/psj/ade3c2},
  fjournal = {The Planetary Science Journal},
  issn = {2632-3338},
  journal = {PSJ},
  month = {August},
  number = {8},
  pages = {186},
  publisher = {American Astronomical Society},
  title = {The NASA Exoplanet Archive and Exoplanet Follow-up Observing Program: Data, Tools, and Usage},
  url = {https://doi.org/10.3847/psj/ade3c2},
  volume = {6},
  year = {2025}
}

@article{Valizadegan2025,
  title = {ExoMiner++: Enhanced Transit Classification and a New Vetting Catalog for 2-Minute TESS Data},
  author = {Valizadegan, Hamed and Martinho, Miguel J.~S. and Jenkins, Jon M. and Twicken, Joseph D. and Caldwell, Douglas A. and Maynard, Patrick and Wei, Hongbo and Zhong, William and Yates, Charles and Donald, Sam and Collins, Karen A. and others},
  year = {2025},
  journal = {AJ},
  fjournal = {The Astronomical Journal},
  volume = {170},
  number = {5},
  pages = {287},
  doi = {10.3847/1538-3881/ae03a4},
  eprint = {2502.09790},
  eprinttype = {arxiv},
  primaryClass = {astro-ph.EP},
  langid = {english}
}

@misc{Martinho2026ExoMinerPP20,
  author = {Martinho, Miguel J.~S. and Valizadegan, Hamed and Jenkins, Jon M. and Caldwell, Douglas A. and Twicken, Joseph D. and Tofflemire, Ben and Jafariyazani, Marziye},
  title = {{ExoMiner++} 2.0: Vetting {TESS} Full-Frame Image Transit Signals},
  year = {2026},
  eprint = {2601.14877},
  eprinttype = {arxiv},
  primaryClass = {astro-ph.EP},
  howpublished = {\url{https://arxiv.org/abs/2601.14877}},
  note = {preprint},
  langid = {english}
}

@article{Salinas2025,
  author  = {Salinas, Helem and Brahm, Rafael and Olmschenk, Greg and Barry, Richard K. and Pichara, Karim and Ishitani Silva, Stela and Araujo, Vladimir},
  title   = {Exoplanet Transit Candidate Identification in {TESS} Full-Frame Images via a Transformer-Based Algorithm},
  fjournal = {Monthly Notices of the Royal Astronomical Society},
  journal = {MNRAS},
  volume  = {538},
  number  = {3},
  pages   = {2031--2049},
  year    = {2025},
  doi     = {10.1093/mnras/staf347}
}

@article{Vivien2025,
  author = {Vivien, H. G. and Deleuil, M. and Jannsen, N. and De Ridder, J. and Seynaeve, D. and Carpine, M.-A. and Zerah, Y.},
  doi = {10.1051/0004-6361/202452124},
  fjournal = {Astronomy \& Astrophysics},
  issn = {1432-0746},
  journal = {A\&A},
  month = {February},
  pages = {A293},
  publisher = {EDP Sciences},
  title = {PANOPTICON: A novel deep learning model to detect single transit events with no prior data filtering in PLATO light curves},
  url = {https://doi.org/10.1051/0004-6361/202452124},
  volume = {694},
  year = {2025}
}

@misc{Paetzold2025TRANSCENDENCE,
  author = {P{\"a}tzold, Martin and Grziwa, Sascha and Hribar, Rok and Schmerling, Hendrik},
  title = {{TRANSCENDENCE} --- {A} {TRANS}it {CaptureENgine} for {DE}tection and {N}eural network {C}haracterization of {E}xoplanets},
  howpublished = {EPSC--DPS Joint Meeting 2025},
  year = {2025},
  doi = {10.5194/epsc-dps2025-1430},
  langid = {english}
}

@misc{Thomas2025KeplerHybrid,
  author = {Thomas, Bibin and Bhat, Vittal M. and Mohammed, Salman Arafath and Mohammed, Abdul Wase and Dessalegn, Adis Abebaw and Mittal, Mohit},
  title = {Identifying Exoplanets with Deep Learning: {A} {CNN} and {RNN} Classifier for {Kepler} {DR25} and Candidate Vetting},
  year = {2025},
  eprint = {2509.04793},
  eprinttype = {arxiv},
  primaryClass = {astro-ph.EP},
  howpublished = {\url{https://arxiv.org/abs/2509.04793}},
  note = {preprint},
  langid = {english}
}

@article{Choudhary2025ViTGAFRP,
	author = {Choudhary, Anupma and Bandari, Sohith and Kushvah, B. S. and Swastik, C.},
	doi = {10.3847/1538-3881/ade99c},
	journal = {The Astronomical Journal},
	month = {jul},
	number = {2},
	pages = {120},
	publisher = {The American Astronomical Society},
	title = {Exoplanet Classification Through Vision Transformers with Temporal Image Analysis},
	url = {https://doi.org/10.3847/1538-3881/ade99c},
	volume = {170},
	year = {2025}
}

@article{Xie2025RAAKeplerTESS,
  author = {Xie, Dapeng and Wang, Ying and Liu, Fuyao and Sun, Wei},
  title = {Deep Learning to Classify Exoplanet Light Curves in {Kepler} and {TESS}},
  journal = {Research in Astronomy and Astrophysics},
  fjournal = {Research in Astronomy and Astrophysics},
  year = {2025},
  volume = {25},
  number = {10},
  eid = {104004},
  pages = {104004},
  doi = {10.1088/1674-4527/adf70e},
  langid = {english}
}

@article{Fiscale2025CFM,
  author = {Fiscale, Stefano and Ferone, Alessio and Ciaramella, Angelo and Inno, Laura and Giordano Orsini, Massimiliano and Covone, Giovanni and Rotundi, Alessandra},
  doi = {10.3390/electronics14091738},
  fjournal = {Electronics},
  issn = {2079-9292},
  journal = {Electronics},
  month = {April},
  number = {9},
  pages = {1738},
  publisher = {MDPI AG},
  title = {Detection of Exoplanets in Transit Light Curves with Conditional Flow Matching and XGBoost},
  url = {https://doi.org/10.3390/electronics14091738},
  volume = {14},
  year = {2025}
}

@article{Christiansen2012,
  author = {Christiansen, Jessie L. and Jenkins, Jon M. and Caldwell, Douglas A. and Burke, Christopher J. and Tenenbaum, Peter and Seader, Shawn and Thompson, Susan E. and Barclay, Thomas S. and Clarke, Bruce D. and Li, Jie and Smith, Jeffrey C. and Stumpe, Martin C. and Twicken, Joseph D. and Van Cleve, Jeffrey},
	doi = {10.1086/668847},
	fjournal = {Publications of the Astronomical Society of the Pacific},
  journal = {PASP},
	month = {nov},
	number = {922},
	pages = {1279},
	publisher = {University of Chicago Press},
	title = {The Derivation, Properties, and Value of Kepler's Combined Differential Photometric Precision},
	url = {https://doi.org/10.1086/668847},
	volume = {124},
	year = {2012}
}

@misc{cuvarbase,
  author = {Hoffman, John},
  title = {cuvarbase: fast period finding utilities for GPUs},
  howpublished = {Astrophysics Source Code Library},
  year = {2022},
  eid = {ascl:2210.030}
}

@article{astropy:2013,
  Adsurl = {https://adsabs.harvard.edu/abs/2013A%26A...558A..33A},
  Archiveprefix = {arXiv},
  Author = {{Astropy Collaboration} and {Robitaille}, T.~P. and {Tollerud}, E.~J. and {Greenfield}, P. and {Droettboom}, M. and {Bray}, E. and {Aldcroft}, T. and {Davis}, M. and {Ginsburg}, A. and {Price-Whelan}, A.~M. and {Kerzendorf}, W.~E. and {Conley}, A. and {Crighton}, N. and {Barbary}, K. and {Muna}, D. and {Ferguson}, H. and {Grollier}, F. and {Parikh}, M.~M. and {Nair}, P.~H. and {Unther}, H.~M. and {Deil}, C. and {Woillez}, J. and {Conseil}, S. and {Kramer}, R. and {Turner}, J.~E.~H. and {Singer}, L. and {Fox}, R. and {Weaver}, B.~A. and {Zabalza}, V. and {Edwards}, Z.~I. and {Azalee Bostroem}, K. and {Burke}, D.~J. and {Casey}, A.~R. and {Crawford}, S.~M. and {Dencheva}, N. and {Ely}, J. and {Jenness}, T. and {Labrie}, K. and {Lim}, P.~L. and {Pierfederici}, F. and {Pontzen}, A. and {Ptak}, A. and {Refsdal}, B. and {Servillat}, M. and {Streicher}, O.},
  Doi = {10.1051/0004-6361/201322068},
  Eid = {A33},
  Eprint = {1307.6212},
  Journal = {A\&A},
  Month = oct,
  Pages = {A33},
  Primaryclass = {astro-ph.IM},
  Title = {{Astropy: A community Python package for astronomy}},
  Volume = 558,
  Year = 2013
}

@ARTICLE{astropy:2018,
        author = {{Astropy Collaboration} and {Price-Whelan}, A.~M. and
          {Sip{\H{o}}cz}, B.~M. and {G{\"u}nther}, H.~M. and {Lim}, P.~L. and
          {Crawford}, S.~M. and {Conseil}, S. and {Shupe}, D.~L. and
          {Craig}, M.~W. and {Dencheva}, N. and {Ginsburg}, A. and {Vand
        erPlas}, J.~T. and {Bradley}, L.~D. and {P{\'e}rez-Su{\'a}rez}, D. and
          {de Val-Borro}, M. and {Aldcroft}, T.~L. and {Cruz}, K.~L. and
          {Robitaille}, T.~P. and {Tollerud}, E.~J. and {Ardelean}, C. and
          {Babej}, T. and {Bach}, Y.~P. and {Bachetti}, M. and {Bakanov}, A.~V. and
          {Bamford}, S.~P. and {Barentsen}, G. and {Barmby}, P. and
          {Baumbach}, A. and {Berry}, K.~L. and {Biscani}, F. and {Boquien}, M. and
          {Bostroem}, K.~A. and {Bouma}, L.~G. and {Brammer}, G.~B. and
          {Bray}, E.~M. and {Breytenbach}, H. and {Buddelmeijer}, H. and
          {Burke}, D.~J. and {Calderone}, G. and {Cano Rodr{\'\i}guez}, J.~L. and
          {Cara}, M. and {Cardoso}, J.~V.~M. and {Cheedella}, S. and {Copin}, Y. and
          {Corrales}, L. and {Crichton}, D. and {D'Avella}, D. and {Deil}, C. and
          {Depagne}, {\'E}. and {Dietrich}, J.~P. and {Donath}, A. and
          {Droettboom}, M. and {Earl}, N. and {Erben}, T. and {Fabbro}, S. and
          {Ferreira}, L.~A. and {Finethy}, T. and {Fox}, R.~T. and
          {Garrison}, L.~H. and {Gibbons}, S.~L.~J. and {Goldstein}, D.~A. and
          {Gommers}, R. and {Greco}, J.~P. and {Greenfield}, P. and
          {Groener}, A.~M. and {Grollier}, F. and {Hagen}, A. and {Hirst}, P. and
          {Homeier}, D. and {Horton}, A.~J. and {Hosseinzadeh}, G. and {Hu}, L. and
          {Hunkeler}, J.~S. and {Ivezi{\'c}}, {\v{Z}}. and {Jain}, A. and
          {Jenness}, T. and {Kanarek}, G. and {Kendrew}, S. and {Kern}, N.~S. and
          {Kerzendorf}, W.~E. and {Khvalko}, A. and {King}, J. and {Kirkby}, D. and
          {Kulkarni}, A.~M. and {Kumar}, A. and {Lee}, A. and {Lenz}, D. and
          {Littlefair}, S.~P. and {Ma}, Z. and {Macleod}, D.~M. and
          {Mastropietro}, M. and {McCully}, C. and {Montagnac}, S. and
          {Morris}, B.~M. and {Mueller}, M. and {Mumford}, S.~J. and {Muna}, D. and
          {Murphy}, N.~A. and {Nelson}, S. and {Nguyen}, G.~H. and
          {Ninan}, J.~P. and {N{\"o}the}, M. and {Ogaz}, S. and {Oh}, S. and
          {Parejko}, J.~K. and {Parley}, N. and {Pascual}, S. and {Patil}, R. and
          {Patil}, A.~A. and {Plunkett}, A.~L. and {Prochaska}, J.~X. and
          {Rastogi}, T. and {Reddy Janga}, V. and {Sabater}, J. and
          {Sakurikar}, P. and {Seifert}, M. and {Sherbert}, L.~E. and
          {Sherwood-Taylor}, H. and {Shih}, A.~Y. and {Sick}, J. and
          {Silbiger}, M.~T. and {Singanamalla}, S. and {Singer}, L.~P. and
          {Sladen}, P.~H. and {Sooley}, K.~A. and {Sornarajah}, S. and
          {Streicher}, O. and {Teuben}, P. and {Thomas}, S.~W. and
          {Tremblay}, G.~R. and {Turner}, J.~E.~H. and {Terr{\'o}n}, V. and
          {van Kerkwijk}, M.~H. and {de la Vega}, A. and {Watkins}, L.~L. and
          {Weaver}, B.~A. and {Whitmore}, J.~B. and {Woillez}, J. and
          {Zabalza}, V. and {Astropy Contributors}},
        title = "{The Astropy Project: Building an Open-science Project and Status of the v2.0 Core Package}",
      journal = {AJ},
          year = 2018,
        month = sep,
        volume = {156},
        number = {3},
          eid = {123},
        pages = {123},
          doi = {10.3847/1538-3881/aabc4f},
archivePrefix = {arXiv},
        eprint = {1801.02634},
  primaryClass = {astro-ph.IM},
        adsurl = {https://ui.adsabs.harvard.edu/abs/2018AJ....156..123A}
}

@ARTICLE{astropy:2022,
        author = {{Astropy Collaboration} and {Price-Whelan}, Adrian M. and {Lim}, Pey Lian and {Earl}, Nicholas and {Starkman}, Nathaniel and {Bradley}, Larry and {Shupe}, David L. and {Patil}, Aarya A. and {Corrales}, Lia and {Brasseur}, C.~E. and {N{"o}the}, Maximilian and {Donath}, Axel and {Tollerud}, Erik and {Morris}, Brett M. and {Ginsburg}, Adam and {Vaher}, Eero and {Weaver}, Benjamin A. and {Tocknell}, James and {Jamieson}, William and {van Kerkwijk}, Marten H. and {Robitaille}, Thomas P. and {Merry}, Bruce and {Bachetti}, Matteo and {G{"u}nther}, H. Moritz and {Aldcroft}, Thomas L. and {Alvarado-Montes}, Jaime A. and {Archibald}, Anne M. and {B{'o}di}, Attila and {Bapat}, Shreyas and {Barentsen}, Geert and {Baz{'a}n}, Juanjo and {Biswas}, Manish and {Boquien}, M{'e}d{'e}ric and {Burke}, D.~J. and {Cara}, Daria and {Cara}, Mihai and {Conroy}, Kyle E. and {Conseil}, Simon and {Craig}, Matthew W. and {Cross}, Robert M. and {Cruz}, Kelle L. and {D'Eugenio}, Francesco and {Dencheva}, Nadia and {Devillepoix}, Hadrien A.~R. and {Dietrich}, J{"o}rg P. and {Eigenbrot}, Arthur Davis and {Erben}, Thomas and {Ferreira}, Leonardo and {Foreman-Mackey}, Daniel and {Fox}, Ryan and {Freij}, Nabil and {Garg}, Suyog and {Geda}, Robel and {Glattly}, Lauren and {Gondhalekar}, Yash and {Gordon}, Karl D. and {Grant}, David and {Greenfield}, Perry and {Groener}, Austen M. and {Guest}, Steve and {Gurovich}, Sebastian and {Handberg}, Rasmus and {Hart}, Akeem and {Hatfield-Dodds}, Zac and {Homeier}, Derek and {Hosseinzadeh}, Griffin and {Jenness}, Tim and {Jones}, Craig K. and {Joseph}, Prajwel and {Kalmbach}, J. Bryce and {Karamehmetoglu}, Emir and {Ka{l}uszy{'n}ski}, Miko{l}aj and {Kelley}, Michael S.~P. and {Kern}, Nicholas and {Kerzendorf}, Wolfgang E. and {Koch}, Eric W. and {Kulumani}, Shankar and {Lee}, Antony and {Ly}, Chun and {Ma}, Zhiyuan and {MacBride}, Conor and {Maljaars}, Jakob M. and {Muna}, Demitri and {Murphy}, N.~A. and {Norman}, Henrik and {O'Steen}, Richard and {Oman}, Kyle A. and {Pacifici}, Camilla and {Pascual}, Sergio and {Pascual-Granado}, J. and {Patil}, Rohit R. and {Perren}, Gabriel I. and {Pickering}, Timothy E. and {Rastogi}, Tanuj and {Roulston}, Benjamin R. and {Ryan}, Daniel F. and {Rykoff}, Eli S. and {Sabater}, Jose and {Sakurikar}, Parikshit and {Salgado}, Jes{'u}s and {Sanghi}, Aniket and {Saunders}, Nicholas and {Savchenko}, Volodymyr and {Schwardt}, Ludwig and {Seifert-Eckert}, Michael and {Shih}, Albert Y. and {Jain}, Anany Shrey and {Shukla}, Gyanendra and {Sick}, Jonathan and {Simpson}, Chris and {Singanamalla}, Sudheesh and {Singer}, Leo P. and {Singhal}, Jaladh and {Sinha}, Manodeep and {Sip{H{o}}cz}, Brigitta M. and {Spitler}, Lee R. and {Stansby}, David and {Streicher}, Ole and {{\v{S}}umak}, Jani and {Swinbank}, John D. and {Taranu}, Dan S. and {Tewary}, Nikita and {Tremblay}, Grant R. and {Val-Borro}, Miguel de and {Van Kooten}, Samuel J. and {Vasovi{'c}}, Zlatan and {Verma}, Shresth and {de Miranda Cardoso}, Jos{'e} Vin{'i}cius and {Williams}, Peter K.~G. and {Wilson}, Tom J. and {Winkel}, Benjamin and {Wood-Vasey}, W.~M. and {Xue}, Rui and {Yoachim}, Peter and {Zhang}, Chen and {Zonca}, Andrea and {Astropy Project Contributors}},
        title = "{The Astropy Project: Sustaining and Growing a Community-oriented Open-source Project and the Latest Major Release (v5.0) of the Core Package}",
      journal = {ApJ},
          year = 2022,
        month = aug,
        volume = {935},
        number = {2},
          eid = {167},
        pages = {167},
          doi = {10.3847/1538-4357/ac7c74},
archivePrefix = {arXiv},
        eprint = {2206.14220},
  primaryClass = {astro-ph.IM},
        adsurl = {https://ui.adsabs.harvard.edu/abs/2022ApJ...935..167A}
}

@inproceedings{Paszke2017,
	author = {Paszke, Adam and Gross, Sam and Massa, Francisco and Lerer, Adam and Bradbury, James and Chanan, Gregory and Killeen, Trevor and Lin, Zeming and Gimelshein, Natalia and Antiga, Luca and Desmaison, Alban and Kopf, Andreas and Yang, Edward and DeVito, Zachary and Raison, Martin and Tejani, Alykhan and Chilamkurthy, Sasank and Steiner, Benoit and Fang, Lu and Bai, Junjie and Chintala, Soumith},
	booktitle = {Advances in Neural Information Processing Systems},
	editor = {H. Wallach and H. Larochelle and A. Beygelzimer and F. d\textquotesingle Alch\'{e}-Buc and E. Fox and R. Garnett},
	publisher = {Curran Associates, Inc.},
	title = {PyTorch: An Imperative Style, High-Performance Deep Learning Library},
	url = {https://proceedings.neurips.cc/paper_files/paper/2019/file/bdbca288fee7f92f2bfa9f7012727740-Paper.pdf},
	volume = {32},
	year = {2019}
  }

@inproceedings{Kluyver2016,
  title={Jupyter Notebooks - a publishing format for reproducible computational workflows},
  author={Thomas Kluyver and Benjamin Ragan-Kelley and Fernando P{\'e}rez and Brian E. Granger and Matthias Bussonnier and Jonathan Frederic and Kyle Kelley and Jessica B. Hamrick and Jason Grout and Sylvain Corlay and Paul Ivanov and Dami{\'a}n Encalada Avila and Safia Abdalla and Carol Willing and Jupyter Development Team},
  booktitle={International Conference on Electronic Publishing},
  year={2016},
  url={https://api.semanticscholar.org/CorpusID:36928206}
}

@Misc{Jones2001,
  author  = {Virtanen, Pauli and Gommers, Ralf and Oliphant, Travis E. and
            Haberland, Matt and Reddy, Tyler and Cournapeau, David and
            Burovski, Evgeni and Peterson, Pearu and Weckesser, Warren and
            Bright, Jonathan and {van der Walt}, St{\'e}fan J. and
            Brett, Matthew and Wilson, Joshua and Millman, K. Jarrod and
            Mayorov, Nikolay and Nelson, Andrew R. J. and Jones, Eric and
            Kern, Robert and Larson, Eric and Carey, C J and
            Polat, {\.I}lhan and Feng, Yu and Moore, Eric W. and
            {VanderPlas}, Jake and Laxalde, Denis and Perktold, Josef and
            Cimrman, Robert and Henriksen, Ian and Quintero, E. A. and
            Harris, Charles R. and Archibald, Anne M. and
            Ribeiro, Ant{\^o}nio H. and Pedregosa, Fabian and
            {van Mulbregt}, Paul and {SciPy 1.0 Contributors}},
  title   = {{{SciPy} 1.0: Fundamental Algorithms for Scientific
            Computing in Python}},
  journal = {Nature Methods},
  year    = {2020},
  volume  = {17},
  pages   = {261--272},
  adsurl  = {https://rdcu.be/b08Wh},
  doi     = {10.1038/s41592-019-0686-2},
}

@article{scikit-learn,
  title={Scikit-learn: Machine Learning in {P}ython},
  author={Pedregosa, F. and Varoquaux, G. and Gramfort, A. and Michel, V.
          and Thirion, B. and Grisel, O. and Blondel, M. and Prettenhofer, P.
          and Weiss, R. and Dubourg, V. and Vanderplas, J. and Passos, A. and
          Cournapeau, D. and Brucher, M. and Perrot, M. and Duchesnay, E.},
  journal={Journal of Machine Learning Research},
  volume={12},
  pages={2825--2830},
  year={2011}
}

@inproceedings{sklearn_api,
  author    = {Lars Buitinck and Gilles Louppe and Mathieu Blondel and
                Fabian Pedregosa and Andreas Mueller and Olivier Grisel and
                Vlad Niculae and Peter Prettenhofer and Alexandre Gramfort
                and Jaques Grobler and Robert Layton and Jake VanderPlas and
                Arnaud Joly and Brian Holt and Ga{\"{e}}l Varoquaux},
  title     = {{API} design for machine learning software: experiences from the scikit-learn
                project},
  booktitle = {ECML PKDD Workshop: Languages for Data Mining and Machine Learning},
  year      = {2013},
  pages = {108--122},
}

@article{Hunter2007,
  Author    = {Hunter, J. D.},
  Title     = {Matplotlib: A 2D graphics environment},
  Journal   = {Computing in Science \& Engineering},
  Volume    = {9},
  Number    = {3},
  Pages     = {90--95},
  publisher = {IEEE COMPUTER SOC},
  doi       = {10.1109/MCSE.2007.55},
  year      = 2007
}
\begin{appendix}
% \FloatBarrier

\section{Validation of periodic chunk permutation}
\label{app:pcp}
To evaluate the effectiveness of PCP in suppressing residual transit signatures, we randomly selected 1000 ATLCs from the Cross-KIC Recovery Set for analysis. BLS, TLS, and \TransitNet were applied to both the original ATLCs and their PCP-transformed counterparts, and the resulting detection scores at the true transit periods were compared (Fig.~\ref{fig:pcp_tp_score_ridge}).

The results show that transit signals at the true periods are substantially weakened after applying PCP. This effect is reflected by the clear separation between the score distributions of the original ATLCs and PCP-transformed ATLCs across all three methods. In particular, the PCP-transformed samples are concentrated in the low-score regime, indicating that PCP effectively disrupts the periodic transit structure. 

Consequently, when the light curves are phase-folded at the true period, the transit signatures can no longer be coherently aligned and are therefore strongly attenuated. These results support the use of PCP-transformed TMLCs as non-transit samples, as they preserve the underlying noise characteristics of real light curves while substantially mitigating the influence of residual or undiscovered transit signals.

\begin{figure}[H]
  \centering
  \includegraphics[width=0.98\hsize]{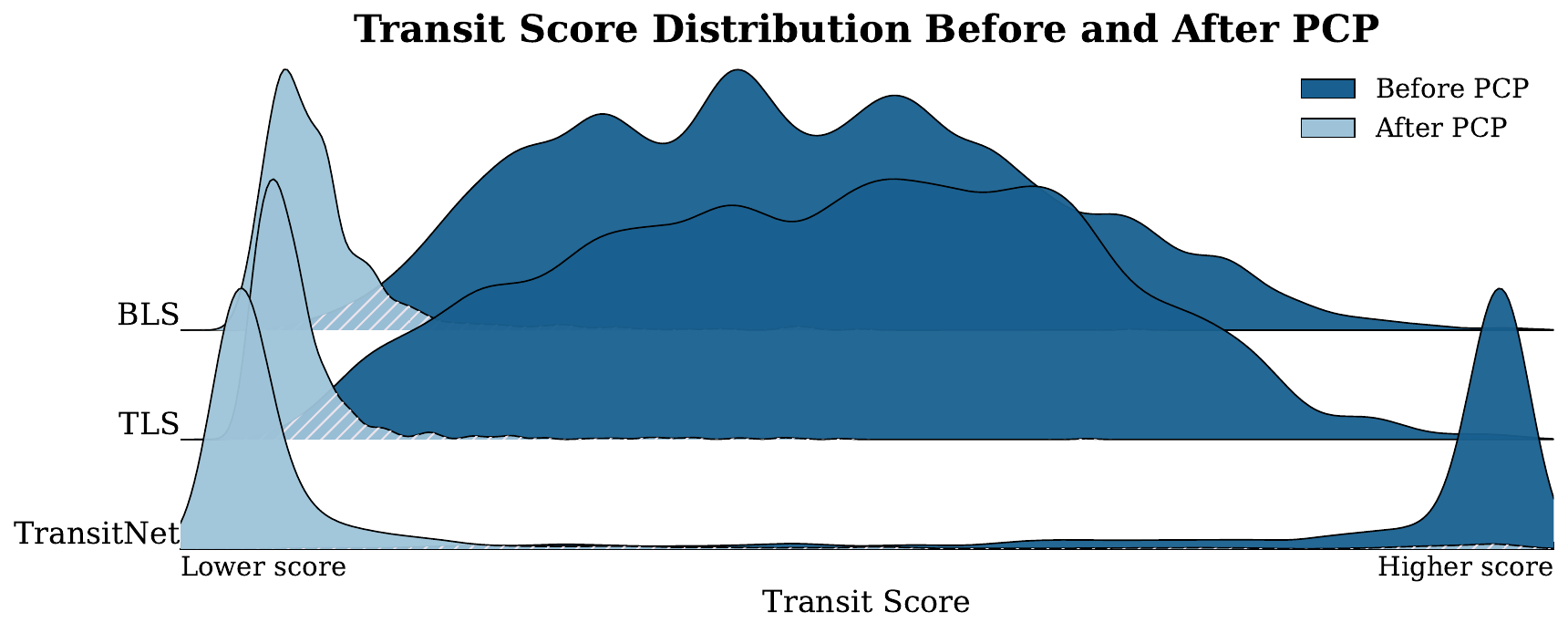}
  \caption{Kernel-density distributions of transit detection scores at the injected orbital period for \textsc{BLS}, \textsc{TLS}, and \TransitNet. All methods show substantially lower scores after PCP, demonstrating effective attenuation of injected periodic transit signals.}
  % Alt text: Ridge density plots of detection scores at the true period before and after periodic chunk permutation for BLS, TLS, and TransitNet, with post-permutation scores shifted to much lower values.
  \label{fig:pcp_tp_score_ridge}
\end{figure}

\section{Threshold selection criteria}
\label{app:thres_select}
To obtain a unified operating threshold under realistic transit blind-search conditions, detection-score distributions are constructed from both ATLCs and PCP-transformed TMLCs across multiple target KICs. The ATLCs provide transit samples embedded in realistic stellar and instrumental noise environments, while the PCP-transformed TMLCs serve as non-transit controls that preserve the statistical properties of the original observations. 

Rather than determining a threshold separately for each target, we seek a single operating threshold that generalizes across heterogeneous stellar environments. Let $\mathcal{D}$ denote the scored samples produced by a given transit-search algorithm. Candidate thresholds $\mathcal{T}$ are derived from the pooled score distribution across all KICs. For each $\tau\in\mathcal{T}$, TPR and FPR are computed separately for each KIC containing both classes. The macro-averaged Youden statistic is then defined as
\begin{equation}
J(\tau)=\frac{1}{|\mathcal{K}|}
\sum_{k\in\mathcal{K}}
\left[\mathrm{TPR}_k(\tau)-\mathrm{FPR}_k(\tau)\right],
\end{equation}
where $\mathcal{K}$ denotes the set of eligible KICs. The shared operating threshold $\tau^\ast$ is selected to maximize $J(\tau)$, with ties resolved in favor of the larger threshold \citep{youden1950}. The complete procedure is summarized in Algorithm~\ref{alg:threshold}. This approach assigns equal weight to each eligible KIC, reducing sensitivity to differences in sample counts and target-specific noise characteristics.

\begin{algorithm}[H]
\caption{Macro-Averaged Youden Threshold Selection}
\label{alg:threshold}
\begin{algorithmic}[1]
    \Require Scored samples $\mathcal{D}$; candidate count $N_\tau$
    \Ensure Optimal shared threshold $\tau^\ast$

    \State Construct $\mathcal{T}$ from the empirical quantiles
    of pooled scores across all KICs
    \State $\mathcal{K}\gets$ KICs containing both classes
    \State $\tau^\ast\gets-\infty$, $J^\ast\gets-\infty$

    \For{each $\tau\in\mathcal{T}$}
        \State Compute $\mathrm{TPR}_k(\tau)$ and
        $\mathrm{FPR}_k(\tau)$ for each $k\in\mathcal{K}$
        \State $J(\tau)\gets
        \frac{1}{|\mathcal{K}|}
        \sum_{k\in\mathcal{K}}
        [\mathrm{TPR}_k(\tau)-\mathrm{FPR}_k(\tau)]$
        \If{$J(\tau)>J^\ast$ or
        $(J(\tau)=J^\ast$ and $\tau>\tau^\ast)$}
            \State $\tau^\ast\gets\tau$
            \State $J^\ast\gets J(\tau)$
        \EndIf
    \EndFor

    \State \Return $\tau^\ast$
\end{algorithmic}
\end{algorithm}

The equivalence between maximizing the macro-averaged Youden statistic and minimizing the macro-averaged balanced classification error can be shown as follows. 
Since
\begin{equation}
\overline{\mathrm{FNR}}(\theta) = 1- \overline{\mathrm{TPR}} (\theta),
\end{equation}
we have
\begin{equation}
    E(\theta) =
    \frac{1}{2}
    \left[
    \overline{\mathrm{FPR}}(\theta) +
    1-\overline{\mathrm{TPR}}(\theta)
    \right] = \frac{1}{2}\left[1-J(\theta)\right].
\end{equation}
Therefore,
\begin{equation}
J(\theta)=1-2E(\theta),
\end{equation}
and hence $\arg\max_{\theta}J(\theta)=\arg\min_{\theta}E(\theta).$

Applying the operating threshold selected via Algorithm~\ref{alg:threshold}, the resulting confusion matrices for all transit blind-search algorithms on the full Low-\snr{} Transit Recovery Set are presented in Fig.~\ref{fig:cnf_m_overall}.

\begin{figure*}[!tp]
  \centering
  \includegraphics[width=0.96\textwidth]{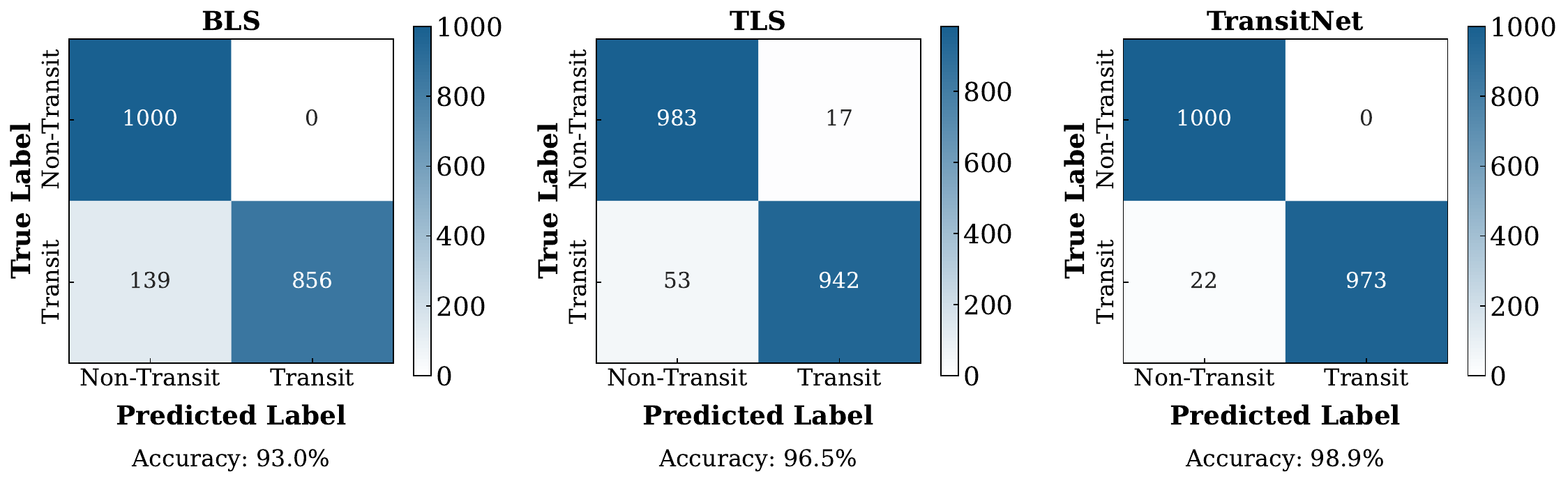}
  \caption{
    Overall confusion matrices showing the classification performance of BLS, TLS, and \TransitNet on the Low-\snr{} Transit Recovery Set. 
    Classification is performed using the operating thresholds selected from the macro-averaged threshold optimization analysis shown in Fig.~\ref{fig:threshold_macro_metrics_sweep}. 
    Among the three methods, \TransitNet demonstrates the best overall performance, achieving the highest accuracy (98.9\%), compared to TLS (96.5\%) and BLS (93.0\%).
  }
  % Alt text: Three confusion matrices on the full low-signal recovery set for BLS, TLS, and TransitNet, reporting overall accuracies of 93.0, 96.5, and 98.9 percent at the selected thresholds.
  \label{fig:cnf_m_overall}
\end{figure*}

\section{Detection score distribution by KIC}
\label{app:dsd_kic}
Per-KIC score distributions for the Cross-KIC Recovery Set are shown in Figs.~\ref{fig:kic_score_ridge_dlm}--\ref{fig:kic_score_ridge_bls}. Dark blue and light blue curves represent the detection-score distributions. TransitNet achieves a clearer separation between transit and non-transit score distributions in blind searches of previously unseen KICs, indicating stronger discrimination between the two classes without target-specific optimization.

\begin{figure*}[!tp]
  \centering
  \includegraphics[width=\linewidth]{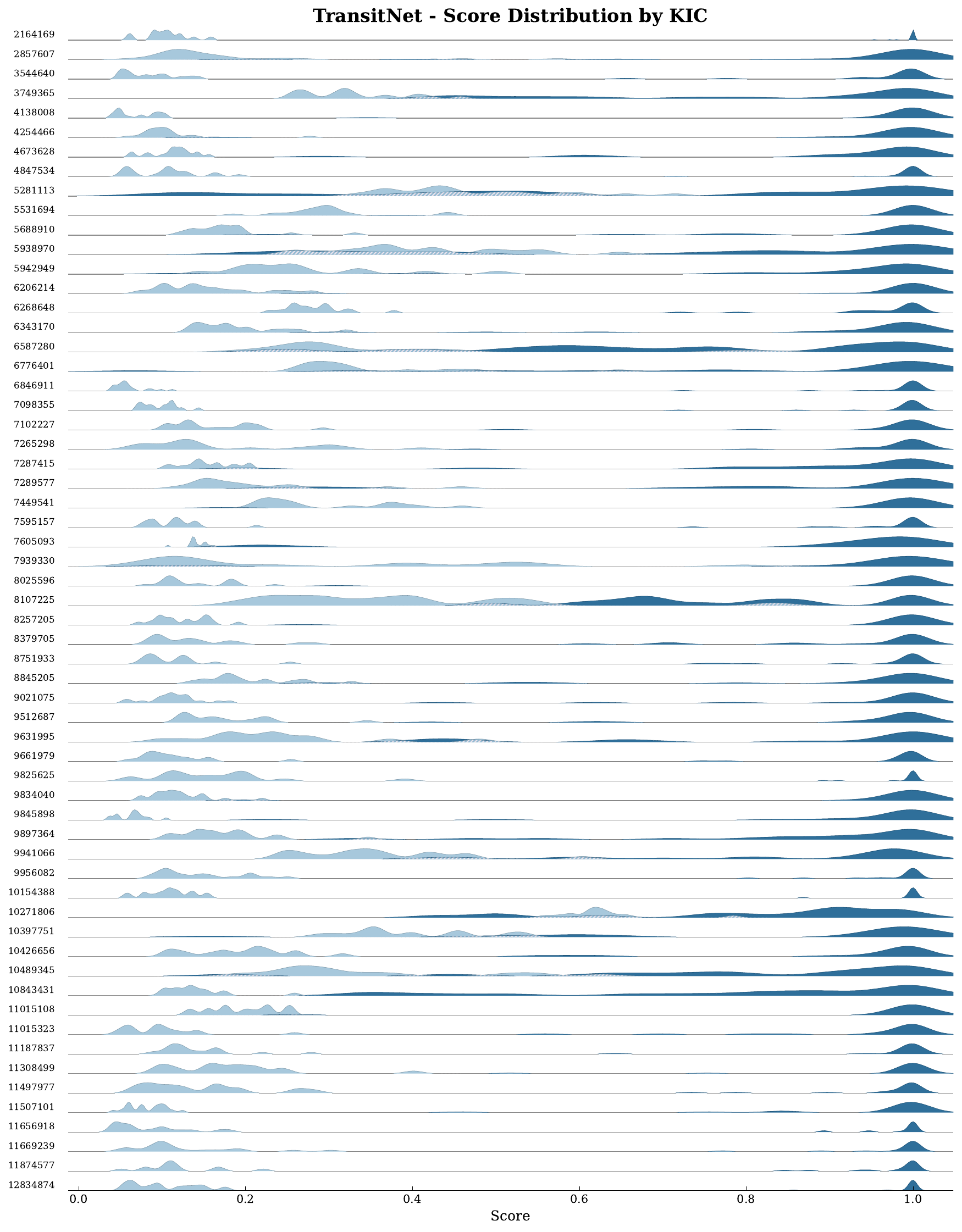}
  % \hfill
  \caption{Per-KIC distribution of \TransitNet detection scores on the Cross-KIC Recovery Set. 
  ATLC samples (with injected transit signals) and PCP-TMLC samples (without transit signals) are generally well separated in score space, indicating robust discrimination between transit and non-transit light curves.
  }
  % Alt text: Per-target ridge plots of TransitNet scores for injected transit and permuted non-transit samples across sixty Kepler targets, with transit scores generally higher than non-transit scores.
  \label{fig:kic_score_ridge_dlm}
\end{figure*}

\begin{figure*}[!tp]
  \centering
  \includegraphics[width=\linewidth]{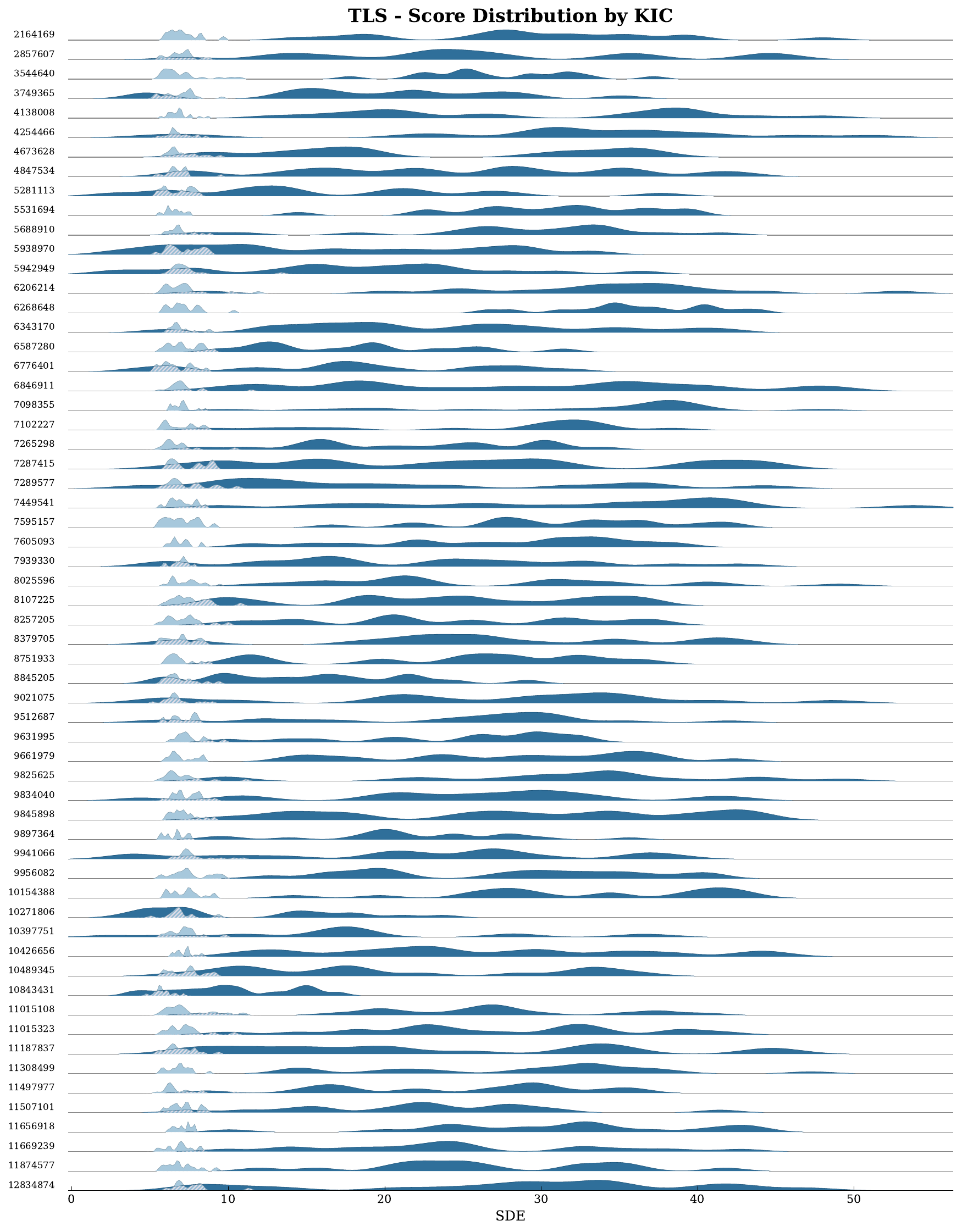}
  % \hfill
  \caption{
  Per-KIC distribution of TLS detection scores on the Cross-KIC Recovery Set. 
  Compared to \TransitNet, TLS exhibits greater overlap between the score distributions of ATLC and PCP-TMLC samples. These overlapping regions correspond to false positives and missed transit detections, indicating a weaker separation between transit and non-transit light curves.
  }
  % Alt text: Per-target ridge plots of TLS scores for transit and non-transit samples across sixty Kepler targets, showing broader overlap between the two classes than for TransitNet.
  \label{fig:kic_score_ridge_tls}
\end{figure*}

\begin{figure*}[!tp]
  \centering
  \includegraphics[width=\linewidth]{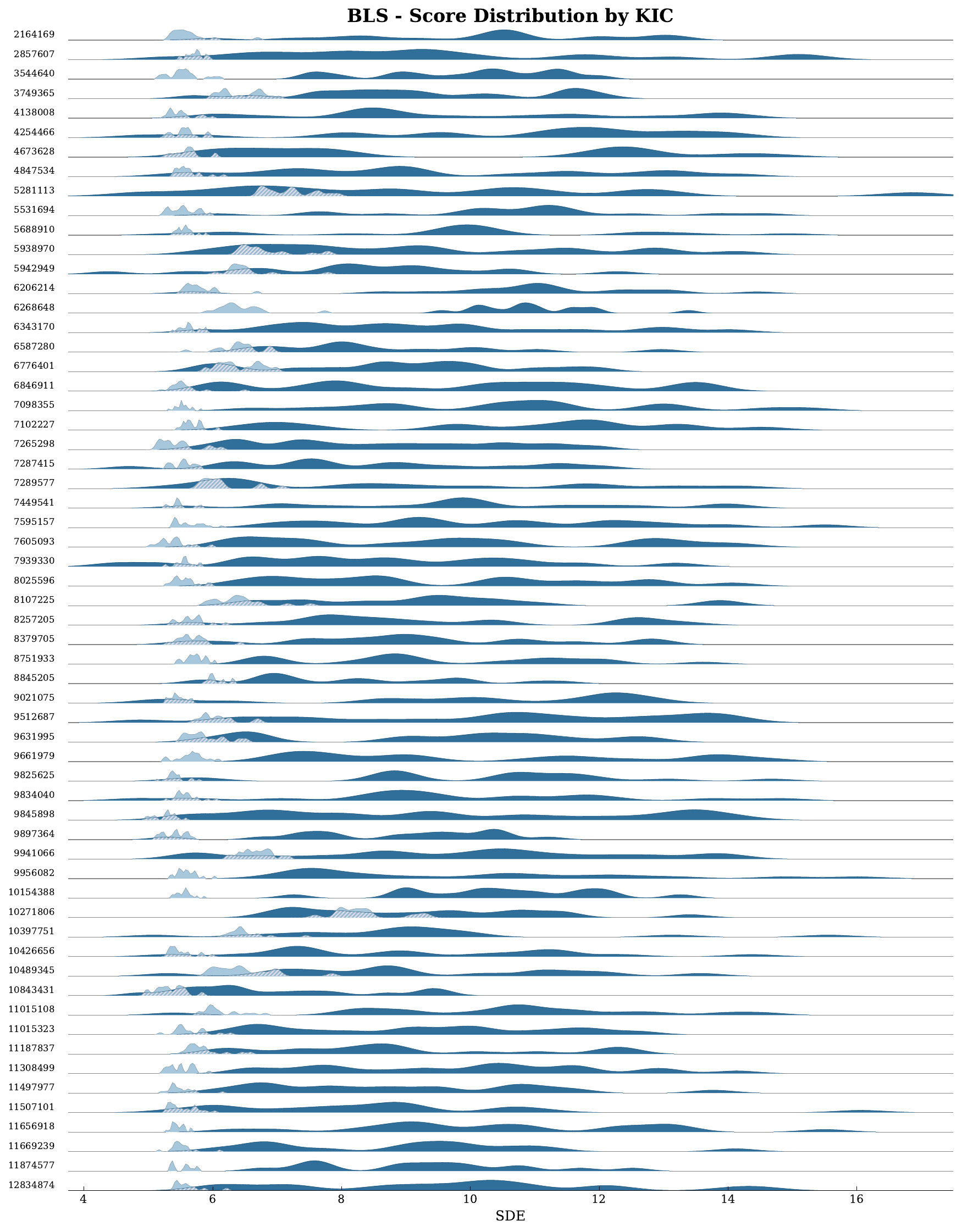}
  % \hfill
  \caption{
    Per-KIC distribution of BLS detection scores on the Cross-KIC Recovery Set. 
  Compared to TLS, BLS produces a broader and more variable false-positive score distribution, resulting in substantially greater overlap between ATLC and PCP-TMLC samples. The enlarged overlap region reflects reduced separability between transit and non-transit signals.
  }
  % Alt text: Per-target ridge plots of BLS scores for transit and non-transit samples across sixty Kepler targets, with the widest overlap and most variable non-transit scores of the three methods.
  \label{fig:kic_score_ridge_bls}
\end{figure*}

\section{Recovered confirmed Kepler low-\snr{} planets}
\label{app:real}
Unlike the three benchmark sets described in Section~\ref{sec:apply}, this evaluation involves neither signal injection nor permutation. Instead, we conduct a blind search on detrended Kepler light curves of confirmed planets, using the same period grid and operating threshold. This provides an end-to-end evaluation on genuine, limb-darkened transit signals beyond the semi-synthetic benchmarks. The successful recovery of these signals provides empirical support for both the effectiveness of the proposed dataset-construction strategy and TransitNet's ability to generalize from synthetic training data to real exoplanet observations.

\begin{table*}[!tp]
  \centering
  \caption{
Summary of known I2LP low-\snr{} transit signals recovered by \TransitNet.
Targets were drawn from the KOI catalog hosted by the NASA Exoplanet Archive \citep{Christiansen2025}, following \citet{Thompson_2018}. We consider confirmed Kepler planets with $30 \leq P \leq 60$\,d and $6 \leq \SNR{} \leq 15$. The catalog transit midpoint is defined as $\tau_0 = T_0 \bmod P$, where $T_0$ is the KOI transit epoch (\texttt{koi\_time0bk}, BKJD). The corresponding \TransitNet prediction is denoted by $\hat{\tau}_0$; the mean absolute timing error is $\langle|\hat{\tau}_0 - \tau_0|\rangle = 1.24$\,h (median $0.62$\,h). The Score column lists the detection-spectrum value at the catalog period, obtained from a search over approximately $3\times10^4$ trial periods. All recovered targets have scores above the adopted recovery threshold of 0.54.
  }
  \label{tab:transitnet_results}
  \begin{tabular*}{\textwidth}{@{\extracolsep{\fill}}lccccccc@{}}
  \hline
  \hline
  Kepler Name & Period $P$ (d) & \snr{} & Depth $\delta$ (ppm) & $\tau_0$ (d) & $\hat{\tau}_0$ (d) & $|\hat{\tau}_0-\tau_0|$ (h) & Score \\
  \hline
  $\text{[1]}$ \object{Kepler-1162~c} & 59.284 & 13.1 & 424.4 & 30.186 & 30.156 & 0.72 & 0.93 \\
  $\text{[2]}$ \object{Kepler-1178~b} & 31.806 & 10.9 & 141.7 & 23.457 & 23.493 & 0.86 & 0.97 \\
  $\text{[3]}$ \object{Kepler-1251~b} & 45.091 & 14.1 & 436.3 & 43.370 & 43.335 & 0.84 & 1.00 \\
  $\text{[4]}$ \object{Kepler-1419~b} & 42.522 & 14.7 & 685.6 & 8.562 & 8.487 & 1.80 & 1.00 \\
  $\text{[5]}$ \object{Kepler-1440~b} & 39.860 & 13.9 & 155.5 & 35.183 & 35.242 & 1.41 & 0.87 \\
  $\text{[6]}$ \object{Kepler-1444~b} & 33.420 & 8.9 & 436.6 & 25.459 & 25.453 & 0.14 & 1.00 \\
  $\text{[7]}$ \object{Kepler-1451~b} & 35.623 & 9.6 & 651.1 & 34.584 & 34.583 & 0.02 & 1.00 \\
  $\text{[8]}$ \object{Kepler-1453~b} & 47.161 & 13.6 & 723.2 & 8.776 & 8.802 & 0.62 & 1.00 \\
  $\text{[9]}$ \object{Kepler-1454~b} & 47.032 & 13.7 & 367.8 & 34.377 & 34.384 & 0.17 & 1.00 \\
  $\text{[10]}$ \object{Kepler-1472~b} & 38.130 & 11.6 & 180.4 & 6.866 & 6.744 & 2.93 & 1.00 \\
  $\text{[11]}$ \object{Kepler-1610~c} & 44.985 & 12.7 & 617.8 & 15.615 & 15.623 & 0.19 & 0.96 \\
  $\text{[12]}$ \object{Kepler-1697~b} & 33.497 & 13.2 & 163.5 & 29.776 & 29.788 & 0.29 & 1.00 \\
  \hline
  $\text{[13]}$ \object{Kepler-1703~c} & 31.825 & 7.7 & 79.0 & 17.231 & 17.198 & 0.79 & 1.00 \\
  $\text{[14]}$ \object{Kepler-176~e} & 51.166 & 12.2 & 259.2 & 18.812 & 19.218 & 9.74 & 1.00 \\
  $\text{[15]}$ \object{Kepler-1760~b} & 38.326 & 15.0 & 392.1 & 27.105 & 27.112 & 0.17 & 1.00 \\
  $\text{[16]}$ \object{Kepler-1853~b} & 48.888 & 13.5 & 462.8 & 7.169 & 7.179 & 0.24 & 1.00 \\
  $\text{[17]}$ \object{Kepler-1914~b} & 30.828 & 13.9 & 641.3 & 19.834 & 19.866 & 0.77 & 0.99 \\
  $\text{[18]}$ \object{Kepler-1916~b} & 31.254 & 14.7 & 438.0 & 30.030 & 29.930 & 2.40 & 1.00 \\
  $\text{[19]}$ \object{Kepler-1918~b} & 47.056 & 12.1 & 762.0 & 32.796 & 32.839 & 1.03 & 1.00 \\
  $\text{[20]}$ \object{Kepler-1919~b} & 37.886 & 14.4 & 672.2 & 19.373 & 19.373 & 0.00 & 1.00 \\
  $\text{[21]}$ \object{Kepler-1920~b} & 30.254 & 12.3 & 576.3 & 15.622 & 15.626 & 0.10 & 1.00 \\
  $\text{[22]}$ \object{Kepler-1926~b} & 42.877 & 14.8 & 1068.8 & 27.437 & 27.411 & 0.62 & 0.99 \\
  $\text{[23]}$ \object{Kepler-1965~b} & 41.868 & 12.5 & 150.7 & 21.372 & 21.757 & 9.24 & 1.00 \\
  $\text{[24]}$ \object{Kepler-1980~b} & 33.026 & 12.2 & 507.1 & 28.907 & 28.958 & 1.22 & 0.95 \\
  \hline
  $\text{[25]}$ \object{Kepler-263~c} & 47.333 & 14.0 & 931.5 & 9.417 & 9.412 & 0.12 & 0.97 \\
  $\text{[26]}$ \object{Kepler-265~d} & 43.131 & 11.8 & 447.8 & 36.867 & 36.850 & 0.41 & 0.82 \\
  $\text{[27]}$ \object{Kepler-276~d} & 48.648 & 12.2 & 641.8 & 26.184 & 26.159 & 0.60 & 1.00 \\
  $\text{[28]}$ \object{Kepler-296~e} & 34.142 & 13.3 & 788.0 & 33.610 & 33.613 & 0.07 & 0.99 \\
  $\text{[29]}$ \object{Kepler-299~e} & 38.286 & 15.0 & 322.8 & 28.441 & 28.541 & 2.40 & 0.96 \\
  $\text{[30]}$ \object{Kepler-324~d} & 34.206 & 11.6 & 183.2 & 11.982 & 11.971 & 0.26 & 1.00 \\
  $\text{[31]}$ \object{Kepler-331~d} & 32.134 & 11.8 & 1262.0 & 13.740 & 13.764 & 0.58 & 1.00 \\
  $\text{[32]}$ \object{Kepler-383~c} & 31.201 & 13.7 & 360.1 & 15.266 & 15.307 & 0.98 & 1.00 \\
  $\text{[33]}$ \object{Kepler-395~c} & 34.990 & 11.5 & 498.5 & 4.250 & 4.250 & 0.00 & 0.95 \\
  $\text{[34]}$ \object{Kepler-438~b} & 35.233 & 14.5 & 352.8 & 29.666 & 29.646 & 0.48 & 1.00 \\
  \hline
  \hline
  \end{tabular*}
\end{table*}
\end{appendix}
\end{document}